\documentclass[a4paper,fleqn,usenatbib]{mnras}

\usepackage{newtxtext,newtxmath}

\usepackage[T1]{fontenc}
\usepackage{ae,aecompl}
\usepackage{multirow}

\usepackage{graphicx,layout}	
\usepackage{amsmath}	
\usepackage{amssymb}	
\usepackage{lscape}
\usepackage{longtable}
\usepackage{color}
\usepackage{mathtools}
\usepackage{xcolor}
\usepackage{amssymb}

\newcommand{\hst}{{\it Hubble Space Telescope}}
\def\farcs{\hbox{$.\!\!^{\prime\prime}$}}
\def\re{\hbox{$R_{\rm e}$}}

\def\msun{\hbox{$M_{\odot}$}}

\catcode`\@=11
\def\gsim{\ifmmode{\mathrel{\mathpalette\@versim>}}
    \else{$\mathrel{\mathpalette\@versim>}$}\fi}
\def\lsim{\ifmmode{\mathrel{\mathpalette\@versim<}}
    \else{$\mathrel{\mathpalette\@versim<}$}\fi}
\def\@versim#1#2{\lower 2.9truept \vbox{\baselineskip 0pt \lineskip
    0.5truept \ialign{$\m@th#1\hfil##\hfil$\crcr#2\crcr\sim\crcr}}}
\catcode`\@=12

\title[Rejuvenated galaxies with very old bulges]{Rejuvenated galaxies with very old bulges at the origin of the bending of the main sequence and of the ``green valley''}
\author[C. Mancini et al.]{
Chiara Mancini,$^{1,2}$\thanks{E-mail: chiara.mancini@unipd.it}
Emanuele Daddi,$^{3}$
St\'ephanie Juneau,$^{4}$
Alvio Renzini,$^{2}$
\newauthor
Giulia Rodighiero,$^{1}$
Michele Cappellari,$^{5}$
Luc\'{\i}a Rodr\'{\i}guez-Mu\~noz,$^{1}$
Daizhong Liu,$^{6}$
\newauthor
Maurilio Pannella,$^{7}$
Ivano Baronchelli,$^{1}$
Alberto Franceschini,$^{1}$
Pietro Bergamini,$^{1}$
\newauthor
Chiara D'Eugenio,$^{3}$
Annagrazia Puglisi$^{3}$
\\
$^{1}$Department of Physics and Astronomy, University of Padova, Vicolo dell'Osservatorio, 3, I-35122, Padova, Italy\\
$^{2}$INAF Padova Astronomical Observatory, Vicolo dell'Osservatorio, 5, I-35122, Padova, Italy\\
$^{3}$CEA, IRFU, DAp, AIM, Universit\'e Paris-Saclay, Universit\'e Paris Diderot, Sorbonne Paris Cit\'e, CNRS, F-91191 Gif-sur-Yvette, France\\
$^{4}$National Optical Astronomy Observatory, 950 N Cherry Ave, Tucson AZ 85719, USA\\
$^{5}$Sub-department of Astrophysics, Department of Physics, University of Oxford, Denys Wilkinson Building, Keble Road, Oxford OX1 3RH, UK\\
$^{6}$Max Planck Institute for Astronomy, Konigstuhl 17, D-69117 Heidelberg, Germany\\
$^{7}$Faculty of Physics, Ludwig-Maximilians Universit\"at, Scheinerstr. 1, 81679 Munich, German
}

\date{Accepted XXX. Received YYY; in original form ZZZ}

\pubyear{2017}

\begin{document}
\label{firstpage}
\pagerange{\pageref{firstpage}--\pageref{lastpage}}
\maketitle

\begin{abstract}
We investigate the nature of star-forming galaxies with reduced specific star formation rate (sSFR) and high stellar masses, those ``green valley'' objects that seemingly cause a reported bending, or flattening, of the star-forming main sequence. The fact that such objects host large bulges recently led some to suggest that the internal formation of bulges was a late event that induced the sSFRs of massive galaxies to drop in a slow downfall, and thus the main sequence to bend. We have studied in detail a sample of 10 galaxies at $0.45<z<1$ with secure SFR from Herschel, deep Keck optical spectroscopy, and HST imaging from CANDELS allowing us to perform multi-wavelength bulge to disc  decomposition, and to derive star formation histories for the separated bulge and disc components. We find that the bulges hosted in these systems below main sequence are virtually all maximally old, with ages approaching the age of the Universe at the time of observation, while discs are young ($\langle$ T$_{50}\rangle \sim 1.5$ Gyr). We conclude that, at least based on our sample, the bending of the main sequence is, for a major part, due to rejuvenation, and we disfavour mechanisms that postulate the internal formation of bulges at late times. 
The very old stellar ages of our bulges suggest a number density of Early Type galaxies at $z=1-3$ higher than actually observed. If confirmed, this might represent one of the first direct validations of hierarchical assembly of bulges at high redshifts.
\end{abstract}




\section{Introduction}
The existence and universality of the {\it main sequence} (MS) of star-forming galaxies can be regarded as one of the most important findings in the field of galaxy evolution, over the last decade. The term {\it main sequence} was adopted for the first time by \citet{2007ApJ...660L..43N} to indicate the tight correlation found at $z=0-1$ between galaxy stellar mass ($M_*$) and star formation rate (SFR)  in the form of a power law (SFR $\propto M_*^{\beta}$). As confirmed by many works, this correlation holds over an extensive redshift range with a similar slope and scatter, but different normalisation, which strongly evolves with cosmic time \citep[e.g., decreasing by a factor of $\sim 20$ from $z=2$ to $z=0$, ][]{2007A&A...468...33E,2007ApJ...670..156D,2009ApJ...698L.116P,2014ApJS..214...15S,2012ApJ...754L..29W,2014MNRAS.443...19R,2014ApJ...793...19S,2016ApJ...817..118T}. This has opened a new paradigm in galaxy evolution, in which galaxies preferentially form stars in a quasi-steady mode, fuelled by smooth and continuous cold gas accretion, rather than in a starburst (SB) mode, triggered by stochastic and short-timescale events like major mergers \citep{2011ApJ...739L..40R}.  
Such scenario is further supported by detailed studies based on integral-field spectroscopy, both in the local Universe \citep{2010ApJ...716..198B,2011MNRAS.413..813C}, and up to redshift about 2.5 \citep{2006ApJ...645.1062F,2009ApJ...706.1364F,2018ApJS..238...21F,2009ApJ...697.2057L,2011ApJ...743...86M,2017MNRAS.466..892M}, showing that the majority of star-forming galaxies above the dwarfs regime ($M_*\geq 2\times 10^9 M_{\odot}$) appear to be rotationally-supported discs \citep[see review by ][]{2016ARA&A..54..597C}.
 Hence, the MS trend is driven essentially by disc  galaxies, with a smooth gas accretion  sustaining elevated SFRs over much longer timescales than major mergers could, while preserving and increasing the angular momentum and therefore allowing the disc  structure to survive and grow.  
Passively evolving galaxies, i.e., those in which star formation has been quenched, are also predominantly fast rotators, both at high \citep{2011ApJ...730...38V} and low redshift \citep{2007MNRAS.379..418C, 2007MNRAS.379..401E,2011MNRAS.414..888E}.
 
Many studies focused on the derivation of the best functional form (shape, slope, normalisation, scatter) to describe the MS, which however can depend on several factors, differing from one study to another, such as the sample selection criteria, with or without preselection of star-forming galaxies, SFR and stellar mass indicators, treatment of dust attenuation, etc. \citep{2015ApJ...801L..29R}. The precise form of the MS relation has critical implications for the rate of galaxy growth, and eventually for quenching \citep{2009MNRAS.398L..58R,2010ApJ...721..193P,2014ApJS..214...15S,2017A&A...608A..41C}. A general consensus has been reached on the evolution of the MS normalisation with redshift, $\propto$ (1+z)$^{2.8}$ at least up to $z\sim 2.5$ \citep[e.g.][]{2012ApJ...747L..31S,2015A&A...579A...2I}, reflecting the fact that galaxies in the past had larger reservoir of molecular gas, sustaining higher SFR \citep{2008ApJ...673L..21D, 2010ApJ...713..686D,2010Natur.463..781T,2018ApJ...853..179T}. Its narrowness at all redshifts (scatter of 0.2$-$0.3 dex) is also widely documented, confirming that at any epoch galaxies spend most of their lifetime on the MS  \citep[see][and references therein]{2014ApJS..214...15S}. On the other hand, different results have been obtained so far for the MS slope ($\beta$), which regulates the mass growth rate (i.e., expressed in terms of specific SFR, sSFR$\equiv$ SFR/$M_*$) of high-mass versus low-mass galaxies. 
Some studies found a slope close to unity \citep{2007A&A...468...33E,2009ApJ...698L.116P,2010ApJ...721..193P}, while many authors agreed about slope shallower than unity, between 0.6 and 0.8 \citep{2007ApJ...670..156D,2014MNRAS.443...19R,2014ApJ...793...19S,2014ApJS..214...15S,2015ApJ...801L..29R,2015ApJ...807..141P}. \citet{2012ApJ...754L..14S} and \citet{2012ApJ...754L..29W} found that when only disc-dominated galaxies (S\'ersic index $n< 1.5$) are considered, their MS slope  is close to $\sim 1$, although there is not a general agreement on this issue \citep[e.g.,][hereafter S16]{2016A&A...589A..35S}. 

Nonetheless,  other studies have advocated that a more complex function than a single power law is needed to properly describe the data, suggesting that while at lower masses the relation has $\beta\sim 1$, it bends at the high-mass end. This was parametrized with either a broken power-law or a smooth function flattening with increasing mass. Moreover, some of these studies derived a high-mass end slope constant at all epochs, i.e., a MS constant in shape and evolving only in normalisation \citep{2012ApJ...754L..29W,2014ApJ...795..104W,2015ApJ...801...80L}, while others found a MS that  becomes steeper with redshift, so that a single power law would be appropriate to fit the MS at $z\gtrsim 2$ \citep[][hereafter, S15]{2016ApJ...817..118T,2015A&A...575A..74S}. All these studies agree with the bending of the MS becoming more pronounced with decreasing redshift, whereas \citet{2015ApJ...801L..29R} find no evidence for bending among local Sloan Digital Sky Survey (SDSS) data\footnote{But note that in the SDSS towards the high mass end  the SFR of even star-forming galaxies is inferred predominantly from the D4000 \citep{2016MNRAS.462.2355M}, a rather poor SFR indicator.}.

Despite the conflicting results about the value of $\beta$, one thing seems to be well established: at least at $z<2$ many among the most massive star-forming galaxies have a lower sSFR compared to the less massive ones, thus falling below a linear MS extrapolated from lower stellar masses. 
Hence, the crucial question is understanding whether this observational evidence, referred to as the {\it bending of the MS}, is directly related to physical processes happening in all the most massive galaxies. 
Some studies have interpreted this MS bending as due to the quenching of star formation  driving high-mass galaxies (the first ones to be quenched) to depart from a reference single-slope MS  \citep[][S15]{2015ApJ...801...80L,2015Sci...348..314T}. In this scenario the quenching mechanisms should be able to maintain the galaxies as star-forming for longer time (slow quenching) compared to what expected for ``fast quenching'' processes \citep[e.g.][]{2010ApJ...721..193P,2013A&A...556A..55I}.

On the other hand,  it has been extensively shown that the most massive galaxies tend to have larger Bulge-to-Total (B/T) mass ratios, and steeper mass profiles, compared to the lower-mass ones, both in the local Universe (see \citealt{2016ARA&A..54..597C} for a review), and at high-z \citep{2012MNRAS.427.1666B,2015MNRAS.450..763M,2015Sci...348..314T,2017arXiv170400733T}.  Building on this, \citet{2014ApJ...785L..36A} suggested that the MS deviation from linearity ($\beta<1$, plus possible bending) may be caused by the inclusion of the mass of the {\it quiescent} bulge at the denominator of the sSFR=SFR/($M_{*~\rm disc }+M*_{*~\rm bulge}$). Using 2D bulge/disc (B/D) decomposition analysis in the SDSS data, they showed that the MS slope indeed approaches unity at all masses when the total $M_*$ is replaced with just the mass of the star-forming disc , i.e., a better proxy of the available gas mass. Based on these results, they concluded that the mass growth  rate (sSFR) is nearly the same {\it in the  star-forming regions} of high-mass and low-mass galaxies, but see \citet[]{2015ApJ...808L..49G}, and S16 for a different view.  
In particular, S16 found a high-mass end slope of the M$_{*~\rm disc }-$SFR relation  at $z\sim 1$ well below  unity ($\sim0.6$), and concluded that the bending of the MS is instead due to a gradual reduction of the sSFR occurring in the discs themselves of the most massive galaxies. Based on stacking analysis in {\it Herschel} Far-IR data, they also suggested that such reduction is not caused by a lesser gas abundance, but by a progressive {\it slow downfall} of the star formation efficiency (SFE=SFR/M$_{\rm gas}$) at the MS high-mass end. 
However, it is not obvious from their Figure 8 whether they recover the trend of B/T with stellar mass and their data are quite sparse above $10^{11}\;M_\odot$. In a similar mood, some authors proposed that the MS bending at late times (low redshift) could be linked to a late growth of massive bulges \citep[e.g.][]{2014ApJ...788...11L,2015ApJ...811L..12W}, both as a direct effect on the sSFR as in \cite{2014ApJ...785L..36A}, and for an effect of the bulge on the SFE of the disc. The latter suggestion finds theoretical support in the so-called ``morphological'' (or gravitational) quenching \citep{2009ApJ...707..250M,2014ApJ...796....7G}, according to which the formation of a massive, compact bulge would stabilise the disc  against fragmentation into giant actively star-forming clumps, then smoothly reducing the SFR of the disc. The debate is still open on whether the deviation from linearity of the MS, and its eventual bending, are due only to the presence of a quiescent bulge reducing the global sSFR, or whether also the sSFR of the disc decreases with increasing disc mass.

The existence of such a relatively well constrained parametrization of the MS as a function of the stellar mass and redshift (at least up to $z\sim 4$) gives the opportunity of building the star formation history (SFH) of star-forming galaxies (i.e., derive their SFR at each cosmic time), by assuming that they stay on the MS all their lives. Several authors used this approach to study the ability of classical analytical SFHs used in the spectral energy distributions (SED) fitting analysis to recover the SFR of MS galaxies \citep{2009MNRAS.398L..58R,2011ApJ...738..106W,2012ApJ...745..149L,2014ApJS..214...15S,2017A&A...608A..41C}.

There are various possible reasons why a galaxy may be found somewhat below the MS: one is just measurement error, another is that it is on its way to be quenched, or in a temporary minimum SFR, after which it will return towards the MS, and finally that it was quenched and it is being rejuvenated by reactivating star formation in the disc. In this paper we investigate the nature of the MS bending, by studying in detail the SFHs of a sample of massive galaxies with reduced sSFR at $0.45<z<1$ (hereafter called ``bending galaxies''), and comparing them to the SFHs expected for galaxies evolving along a bending MS. 
This allows us to investigate whether, or not, the MS bending represents the last evolutionary phase experienced by all the star-forming galaxies before quenching, once they reach a critical mass. As we will show in the following, our results suggest that the bending of the MS is not only due to most massive galaxies having started to quench, but also to other quiescent galaxies having acquired a star-forming disc ({\it rejuvenation}).

Several studies have been performed so far on galaxy SFHs, based on spectroscopic, or photometric data, in the local Universe \citep{2005ApJ...619L.135J,2005ApJ...621..673T,2010MNRAS.404.1775T,2012ApJ...745..149L,2016MNRAS.463.2799I,2016A&A...592A..19C}, even with spatially resolved spectroscopy \citep{2018MNRAS.480.2544R,2019A&A...621A.120G}, and at high redshift \citep{2011ApJ...738..106W,2014ApJ...788...72G,2018ApJ...855...85W,2018ApJ...861...13C,2019arXiv190311082C}. 
Here we propose a quite new approach, which consists in deriving the SFHs for the morphologically separated bulge and disc components of our galaxies, based on multi-band bulge/disc decomposition. This technique is particularly effective for the sample analysed in this work, made by bulge-dominated galaxies with low sSFR, since it limits the systematics that would be introduced by deriving the SFHs from the total integrated SEDs, (where the properties of the passively evolving bulge, and the active star-forming disc are averaged out).
The analysis is performed in the North field of the Great Observatory Origins Deep Survey (GOODS-N), including a large amount of high-quality spectroscopic and photometric data (from UV to Far-IR) which are essential to accomplish our goal.
The ``bending galaxies'' are selected among massive galaxies ($M_*>2\times 10^{10} M_{\odot}$) with substantially reduced sSFR compared to what expected for galaxies following a single power-law MS \citep[][]{2014ApJ...793...19S}.  
We use B/D decomposition techniques in multi-band CANDELS/3DHST images to derive accurate spectral energy distributions (SED) separately for the bulges and discs. The SFH of each component is then reconstructed from the age distribution of its stellar populations. The bulge stellar ages are robustly constrained exploiting deep long-slit spectra fully sampling the central part of these galaxies, while the disc mean ages are derived from the broad-band SED fitting. 
Last but not least, by construction, the choice of an intermediate redshift sample allows us to better constrain the age of the old galaxy bulges, compared to what can be done in the local Universe, where the similarity of stellar spectra with ages $>$5 Gyr makes more difficult the age determination for old stellar populations \citep{2005MNRAS.362...41G,2018ApJ...861...13C}.

The paper is organised as follows. In Section \ref{sec:sample} we describe the data and sample selection, while the B/D decomposition is presented in Section \ref{sec:bd}. In Section \ref{sec:seds} we illustrate the SED fitting results, and the related estimate of the disc and bulge stellar ages.
Section \ref{sec:b_spec} describes the methodology used to derive the bulge mass-weighted ages from the galaxy spectra. 
In Section \ref{sec:sfhs} we derive the galaxy SFHs, and compare them with that predicted for a galaxy evolving along a bending MS in the slow downfall scenario of S16.  
The results are discussed in Section \ref{sec:disc}, while Section \ref{sec:concl} presents the conclusions. Some further analysis, simulations, and tests performed to derive the disc and bulge ages, SFHs, and the relative uncertainties, are described in detail in appendices \ref{app:disc_sfh}-\ref{app:med_sed}. Throughout the paper, we assume a cold dark matter cosmology with $H_0 =$ 70 km s$^{-1}$ Mpc$^{-1}$, $\Omega_M = 0.3$, and $\Omega_{\Lambda}= 0.7$ \citep{2011ApJS..192...18K}. All stellar masses and SFR are quoted for a \citet{2003PASP..115..763C} initial mass function (IMF), and magnitudes are given in the AB photometric system \citep{1974ApJS...27...21O}, unless explicitly stated otherwise.

\begin{figure}
\includegraphics[width=0.5\textwidth]{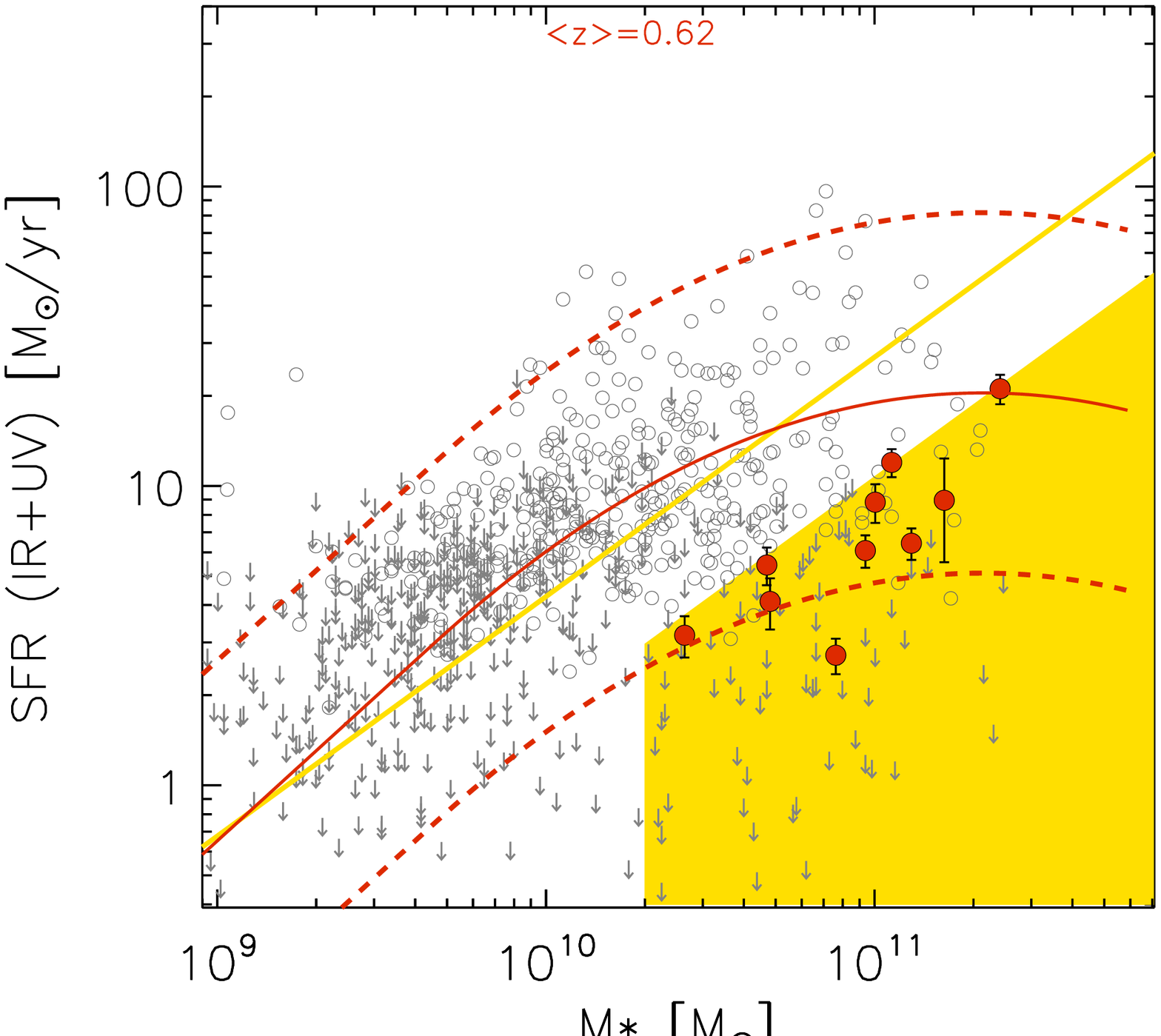}\\
\caption{Our sample galaxies (red filled circles) were selected among the parent sample of MIPS/VLA-detected objects with S/N$_{\rm FIR}>5$ from the \citet[][see the text for details]{2018ApJ...853..172L} catalogue (grey open circles) on the $M_*$-SFR plane, in the yellow shaded region, so to have $M_*\geq 2\times 10^{10}~M_{\odot}$ and SFR more that 2.5 times below the linear MS of \citet{2014ApJ...793...19S} (yellow solid line). The red solid and dashed lines represent the bending MS of S15, and a factor of 4 ($ \sim 2.5\sigma$) above and below the ridge line.}
    \label{fig:sample}
\end{figure}

\section{Data and Sample Selection}\label{sec:sample}
Our sample of ``bending galaxies'' at $\langle z\rangle= 0.62$ was selected in the GOODS-North field \citep[GOODS-N, ][]{2004ApJ...600L..93G} from the {\it Super-Deblended} FIR-to-mm catalogue of \citet[][hereafter, L18]{2018ApJ...853..172L}. As extensively described by the authors, this catalogue is built by using the MIPS/24$\mu$m and VLA/20 cm detections as priors for source deblending, providing reliable photometry, and achieving deeper detection limits in the more confused far-infrared and millimetric bands. This is crucial for the main purpose of this work, allowing us to pick up bona-fide objects with suppressed SFR (as derived from their IR SED) compared to main sequence galaxies with similar mass, and to avoid including blended contaminants in the sample \citep[see][]{2015MNRAS.450..763M}. In practice, this is a SFR selected sample, picking up galaxies which are still star-forming (hence are detected with MIPS and VLA), but do so well below the MS level.
The FIR information from the L18 catalogue was complemented with UV/optical/near-IR photometry, photometric redshifts and stellar masses from \citet{2015ApJ...807..141P}, and 3DHST \citep{2014ApJS..214...24S}.

The multi-band optical/near-IR \hst\ (HST) images from the GOODS \citep{2004ApJ...600L..93G}, CANDELS \citep{2011ApJS..197...35G,2011ApJS..197...36K}, and 3D-HST \citep{2014ApJS..214...24S} programs were exploited to perform the bulge/disc decomposition (cf. Section \ref{sec:bd}). We used the 3D-HST version of the images at all wavelengths, that were drizzled to the same plate scale ($0\farcs06$).     

Last but not least, we took advantage of deep spectra ($R_{AB}<24.4$) from the Team Keck Redshift Survey\footnote{Publicly available at http://tkserver.keck.hawaii.edu/tksurvey/} \citep[TKRS, ][]{2004AJ....127.3121W}, completed using DEIMOS/Keck II multi-slit spectrograph \citep{2003SPIE.4841.1657F}, which provide a spectral coverage of the range 4600-9800~\AA. The spatial and spectral resolution of the TKRS spectra are FWHM=$0\farcs7-1\farcs2$, and 3.5~\AA\ at 7200~\AA\ reference wavelength (corresponding to $\sim$146 km/s), respectively.
As described in \citet{2004AJ....127.3121W}, the spectroscopic target selection was magnitude-limited at R$_{AB}<$24.4 with no colour or shape selection applied to pre-select on galaxies or certain redshift range.
Slits were designed following pre-imaging with DEIMOS in the R-band. Elongated objects had the slit positioned along the semi-major axis (within a $\pm30\textdegree$ tilt from the position angle of the slit mask). The slit masks were observed with 3 exposures of 1200 seconds each, yielding a total integration time of 3600 seconds per target. The TKRS spectra were reduced using the pipeline\footnote{http://deep.ps.uci.edu/spec2d/} from the DEEP2 team \citep{2013ApJS..208....5N,2012ascl.soft03003C}.
The TKRS team released a redshift and photometric catalogue, extracted one-dimensional (1D) spectra as well as sky-subtracted two-dimensional spectra. More details on the TKRS data acquisition and reduction are given by \citet[][their sections 3.1-3.3]{2004AJ....127.3121W}. The public TKRS 1-D spectra were then flux-calibrated by \citet{2011ApJ...736..104J}, who used a throughput curve from the DEEP2 team  along with observations of standard stars with the same observing set-up as TKRS, and applied atmospheric absorption correction. They checked the resulting spectrophotometry and estimated slit-loss correction by computing synthetic magnitudes in the HST bands closest to the observed wavelength [OIII]5007 emission line (usually ACS F775w), and comparing to the observed HST photometry from GOODS-N.

The sample studied in this work includes 10 galaxies selected to match the following criteria :

\begin{itemize}
\item [$\bullet$] Availability of a DEIMOS/TKRS spectrum with a S/N$>3$ per pixel in the range 3800-4200~\AA, to guarantee an adequate sampling of the spectral region including the 4000~\AA\  break, i.e. the most age-sensitive spectral feature \citep[e.g.][]{2003MNRAS.344.1000B}. The S/N of the full spectra in our sample spans a range from 3 to 11, being $\sim 4.5$ on average.

\item [$\bullet$] Spectroscopic redshift in the range $0.45<z<1$. We limited our study in this redshift range for two main reasons. On the one hand, at $z>1-1.5$ the MS may become linear also at the high-masses, as suggested by some authors \citep[e.g., S15 and][]{2016ApJ...817..118T}. On the other hand, as already mentioned, at lower redshifts it becomes increasingly difficult to accurately reconstruct the SFH of old galaxy bulges, due to the convergence of stellar population spectra with ages $> 5$ Gyr. 
\item [$\bullet$] A total FIR S/N ratio S/N$_{\rm FIR}=\sqrt{\Sigma_{\rm i} \left(\frac{\rm S}{\rm N}\right)_{\rm i}^2} > 5$, with i=100$\mu$m, 160$\mu$m, 250$\mu$m, 350$\mu$m, 500$\mu$m (see Table~\ref{tab:ir_fluxes}), to exclude quiescent galaxies from the sample. In principle, part of the FIR emission could be due to diffused cirrus dust heated by old stellar population, but the following analysis based on bulge/disc  decomposition confirms that all the selected galaxies own star-forming discs. 
\item [$\bullet$] A stellar mass $M_*>2\times 10^{10} \msun$ (corresponding to the turnover mass where the MS starts to bend, as derived by S15\footnote{We refer to this work since it is based on a mass-complete sample of galaxies with SFR from FIR, complemented with stacking analysis.}), and a sSFR more than 2.5 times lower compared to similarly massive galaxies lying on the single power-law MS defined by \citet[][]{2014ApJ...793...19S}. This allows us to select our sample among those massive galaxies, lying more than 1.5$\sigma$ below the MS (the average scatter being $\sim 0.25$ dex at $z=0.5-1$, e.g., \citealt{2016ApJ...820L...1K}), whose presence contributes to flatten the MS slope at the high-mass end. 
\end{itemize}
The contribution to the MS bending by objects matching the last criterion can be roughly quantified as follows. Our parent sample of FIR-detected objects (S/N$_{\rm FIR}> 5$) at the same redshift in GOODS-N includes 157 galaxies with $M_*>2\times 10^{10} \msun$, 25 of which ($\sim 16\%$) are located more than 1.5$\sigma$ below the single power-law MS. This number is more than two times larger than what expected for a gaussian tail ($\sim 6.6\%$). As also shown by previous studies \citep[e.g.,][S15]{2012ApJ...754L..29W,2014ApJ...795..104W}, the fraction of outliers (e.g., 1.5$\sigma$) below the single power-law MS becomes larger and larger with increasing stellar mass. In fact, in our parent sample it increases from $\sim 6\%$ at $M_*=2-5\times 10^{10} \msun$, to $\sim 37\%$ at higher masses.    
In Figure \ref{fig:sample} we show the position of our sample (red filled circles) on the $M_*$-SFR plane compared to the single power-law MS of \citet[][yellow solid line]{2014ApJ...793...19S}, and to the MS rendition of S15 (red solid line, their Eq. 9). The shaded yellow area marks the region in which our sample galaxies were selected. Most of them have $M_*> 5\times 10^{10}\msun$, being located where the MS bending is more pronounced. 
The open grey circles represent the parent sample of MIPS/VLA-detected objects at $0.45<z<1$ from the L18 catalogue, while objects with S/N$_{\rm FIR}<5$ are shown as upper limits. The stellar masses are from \citet{2015ApJ...807..141P}, and the SFRs are inferred by summing the UV and IR contributions, the latter being derived from FIR-to-mm SED fitting (cf., L18). 

We cross-matched the selected galaxies with the public Chandra 2Ms catalogue in GOODS-N ~\citep{2003AJ....126..632B}, to check for the presence of X-ray detected AGNs, and did not find any counterpart. We can also exclude a significant contribution of obscured AGNs to the mid-/FAR-IR emission of our objects, based on IR SED fitting, including dusty-torus components, by L18.

\begin{table*}
\begin{center}
\begin{scriptsize}
\caption{Mid- and Far-IR fluxes, corresponding S/N, and SFR(IR+UV) for the 10 sample galaxies, taken from the publicly available catalogue in L18.}\label{tab:ir_fluxes}
\begin{tabular}{rrrrrrrrrrrrrr}
\hline
\hline
ID &$z_{\rm spec}$ &   F16[$\mu$Jy] &      	F24[$\mu$Jy] &      F100[mJy] &	       F160[mJy] &	     F250[mJy] &          F350[mJy] &		  S/N$_{100}$ &   S/N$_{160}$ &   S/N$_{250}$ &   S/N$_{350}$ &   S/N$_{\rm FIR}$   & SFR(IR+UV)\\
\hline
  2202  & 1.0159 & 128.90$\pm$  9.09 &  178.90$\pm$  7.14 &  1.43 $\pm$ 0.32 &  2.15$\pm$ 1.35 & 13.39$\pm$ 6.19  &  $<$2.072		   & 4.44   &  1.59   &  2.16	&   $-$  &  5.19  &16.51 $\pm$  6.26  \\ 
  4267  & 0.4547 & 151.80$\pm$  7.24 &  224.60$\pm$  9.54 &  3.61 $\pm$ 0.45 &  7.62$\pm$ 0.99 &  4.44$\pm$ 3.58  &  $<$2.072		   & 8.00   &  7.67   &  1.24	&   $-$  &  11.15 &6.52	 $\pm$  0.96  \\
  4751  & 0.6816 &   4.70$\pm$ 10.51 &   29.68$\pm$  2.64 &  0.95 $\pm$ 0.27 &  2.14$\pm$ 0.58 &  1.55$\pm$ 2.10  &  $<$2.072		   & 3.48   &  3.70   &  0.74	&   $-$  &  5.14  &4.56	 $\pm$  0.89 \\
  5138  & 0.7041 &  74.49$\pm$  7.07 &  100.80$\pm$  4.41 &  1.12 $\pm$ 0.30 &  2.23$\pm$ 0.55 &  4.29$\pm$ 0.84  &  2.88 $\pm$ 1.31	   & 3.73   &  4.06   &  5.11	&   2.20 &  7.84  &7.00	 $\pm$  0.87  \\ 
  8099  & 0.6419 &  86.03$\pm$  7.33 &   84.57$\pm$  4.96 &  1.36 $\pm$ 0.30 &  3.11$\pm$ 0.68 &  4.94$\pm$ 0.77  &  3.97 $\pm$ 1.71	   & 4.47   &  4.59   &  6.39	&   2.31 &  9.35  &6.68	 $\pm$  0.80  \\ 
  9514  & 0.5172 & 103.50$\pm$ 10.35 &  115.20$\pm$  7.31 &  1.55 $\pm$ 0.48 &  6.06$\pm$ 1.22 &  5.87$\pm$ 3.22  &  $<$2.072		   & 3.22   &  4.97   &  1.82	&   $-$  &  6.20  &4.53	 $\pm$  0.65 \\ 
  11900 & 0.5178 &  38.43$\pm$  7.75 &   24.59$\pm$  2.43 &  0.64 $\pm$ 0.25 &  2.06$\pm$ 0.51 &  1.72$\pm$ 1.10  &  $<$2.072		   & 2.53   &  4.06   &  1.57	&   $-$  &  5.03  &2.26	 $\pm$  0.31 \\ 
  12465 & 0.5625 & 311.10$\pm$ 10.50 &   34.60$\pm$ 10.19 &  6.47 $\pm$ 0.33 & 19.33$\pm$ 0.76 & 16.33$\pm$ 1.90  &  8.67 $\pm$ 3.06	   & 9.47   &  25.51  &  8.58	&   2.84 &  33.3  &19.08 $\pm$ 2.15  \\
  17219 & 0.5189 &  61.98$\pm$  7.42 &   73.80$\pm$  5.59 &  1.04 $\pm$ 0.33 &  2.46$\pm$ 0.51 &  2.52$\pm$ 1.41  &  $<$2.072		   & 3.20   &  4.82   &  1.79	&   $-$  &  6.05  &2.65	 $\pm$ 0.42  \\
  17320 & 0.5603 & 160.60$\pm$  9.40 &  167.50$\pm$  6.01 &  2.09 $\pm$ 0.33 &  7.76$\pm$ 0.62 &  8.10$\pm$ 1.04  &  2.36 $\pm$2.08	   & 6.38   &  12.56  &  7.78	&   1.13 & 16.13  &10.78 $\pm$ 1.16  \\

\hline
\hline
\end{tabular}
\end{scriptsize}
\end{center}
\end{table*}

\section{Bulge-disc decomposition}\label{sec:bd}
We performed multi-band bulge/disc decomposition in the CANDELS HST images, from the $B$ to the $H$ band (i.e., F435W, F606W, F775W, F850LP, F125W, F140W, and F160W), in order to derive the photometry and reconstruct the best-fit SEDs for the two galaxy components separately. 
Many softwares have been developed so far to perform two-dimensional photometric decomposition, such as GIM2D \citep{1998ASPC..145..108S},  GALFIT \citep{2002AJ....124..266P}, BUDDA \citep{2004ApJS..153..411D}, GASP2D \citep{2008A&A...478..353M}, etc. Here we used the {\it galfitm} software \citep{2013MNRAS.430..330H,2013MNRAS.435..623V}, i.e., a multi-wavelength version of {\it GALFIT3} \citep{2010AJ....139.2097P} which enables the automated measurement of wavelength-dependent S\'ersic profile parameters. The code uses a series of Chebyshev polynomials (with a user-specified degree) to set the variation of each parameter as a function of wavelength. 
In this work we fit the disc components with S\'ersic profiles by making the S\'ersic index to vary linearly as a function of wavelength, in the allowed range of $n=0.3-1.5$. 
For the bulge components we constrained the S\'ersic index to be in the range $n=2.5-8$, and to remain constant at all wavelengths. In both cases, magnitudes are left as free parameters, and effective semi-major axis (\re), minor to major axis ratios (q=b/a), and position angles (PA) are set to vary linearly with the wavelength. To avoid contamination from neighbours, we used appropriate masks for the fainter objects, while fitting the closest and brightest objects together with the central galaxy. 
The {\it galfitm} output parameters obtained in the F160W band for bulges and discs are shown in Table \ref{tab:sers_param}. 
Most of the galaxies have a relevant (if not dominant) bulge component, the average bulge-to-total {\it flux} ratio of the sample being $\langle$B/T$_{\rm flux}\rangle\sim0.5$ in H-band (F160W). This becomes even more evident when considering the B/T {\it mass} ratio of the sample, derived based on the SED fitting (in Section~\ref{sec:seds}, ranging from 0.3 to 0.9 and $\langle$B/T$_{\rm mass}\rangle\sim0.67$ on average).  Most discs are fitted with S\'ersic index $n<1$, which in some case could be due to the presence of a bar, or spiral-like structures \citep[e.g., ][]{2011MNRAS.415.3308G}. Although the bulge S\'ersic indices are more spread in the range $n=2.5-6.2$, most of them are $n \gtrsim 4$.

\begin{figure}
\includegraphics[width=\columnwidth]{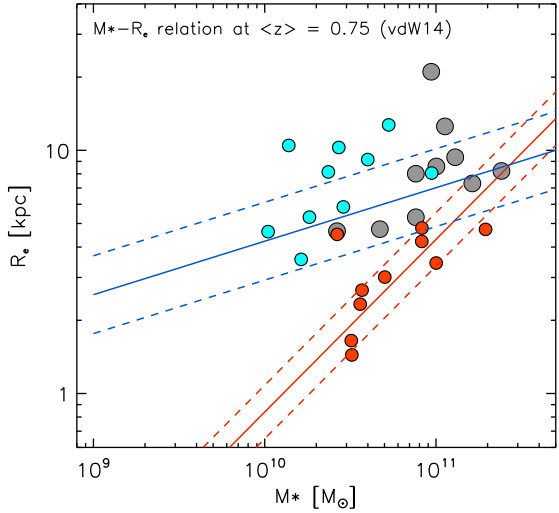}
  \caption{Stellar mass {\it vs} effective semi-major axis (\re) for the global galaxies (grey filled circles), bulges (red circles), and discs (cyan circles) compared with the Mass-size relation, and the 1$\sigma$ dispersion, derived by \citet[][i.e., vdW14]{2014ApJ...788...28V} for quiescent (red solid/dashed lines), and star-forming galaxies (blue solid/dashed lines) at $\langle z\rangle$=0.75 . }
  \label{fig:sizes}
\end{figure}

\begin{figure*}
  \includegraphics[width=0.23\textwidth]{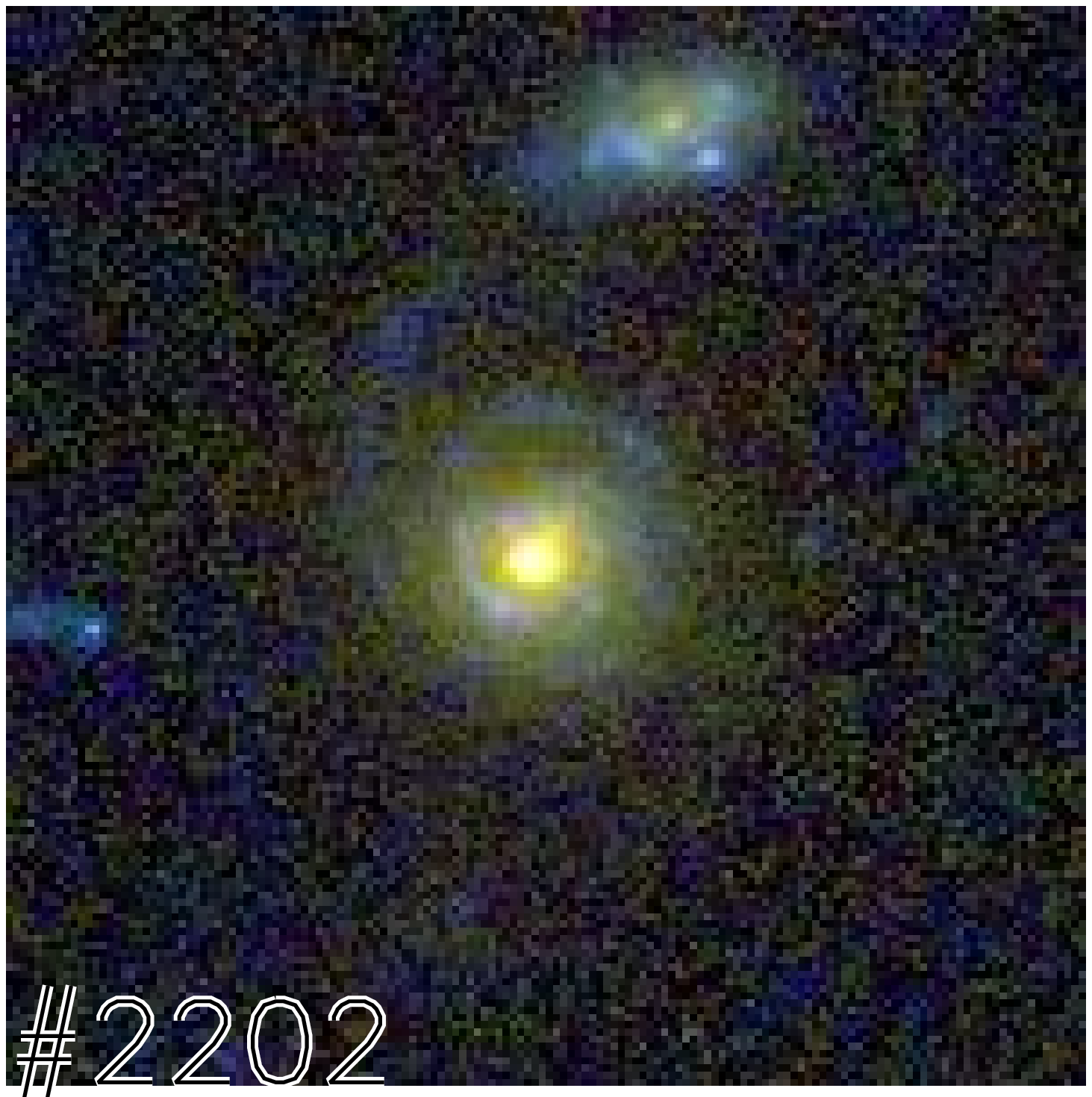}\includegraphics[width=0.23\textwidth]{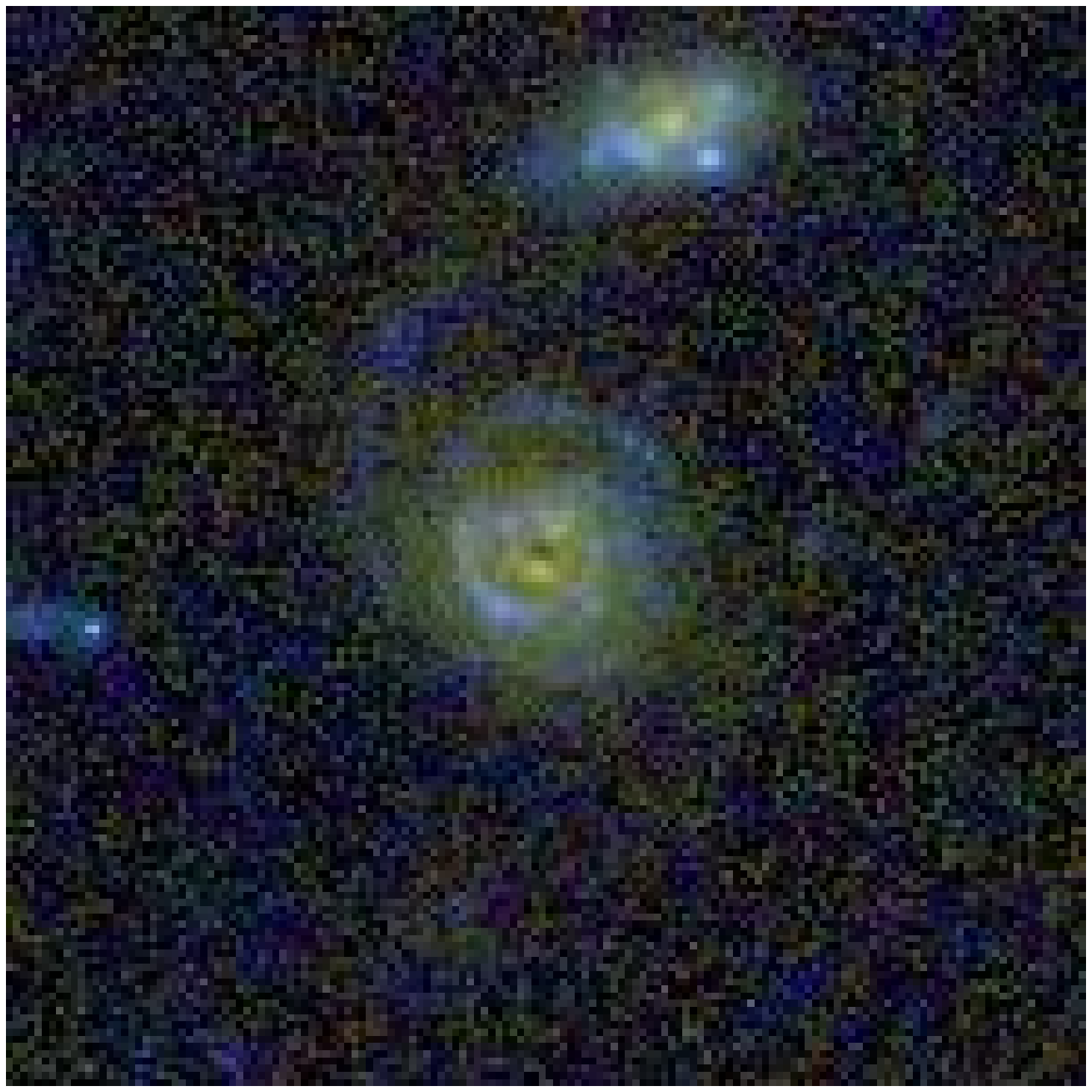} \hfill \includegraphics[width=0.23\textwidth]{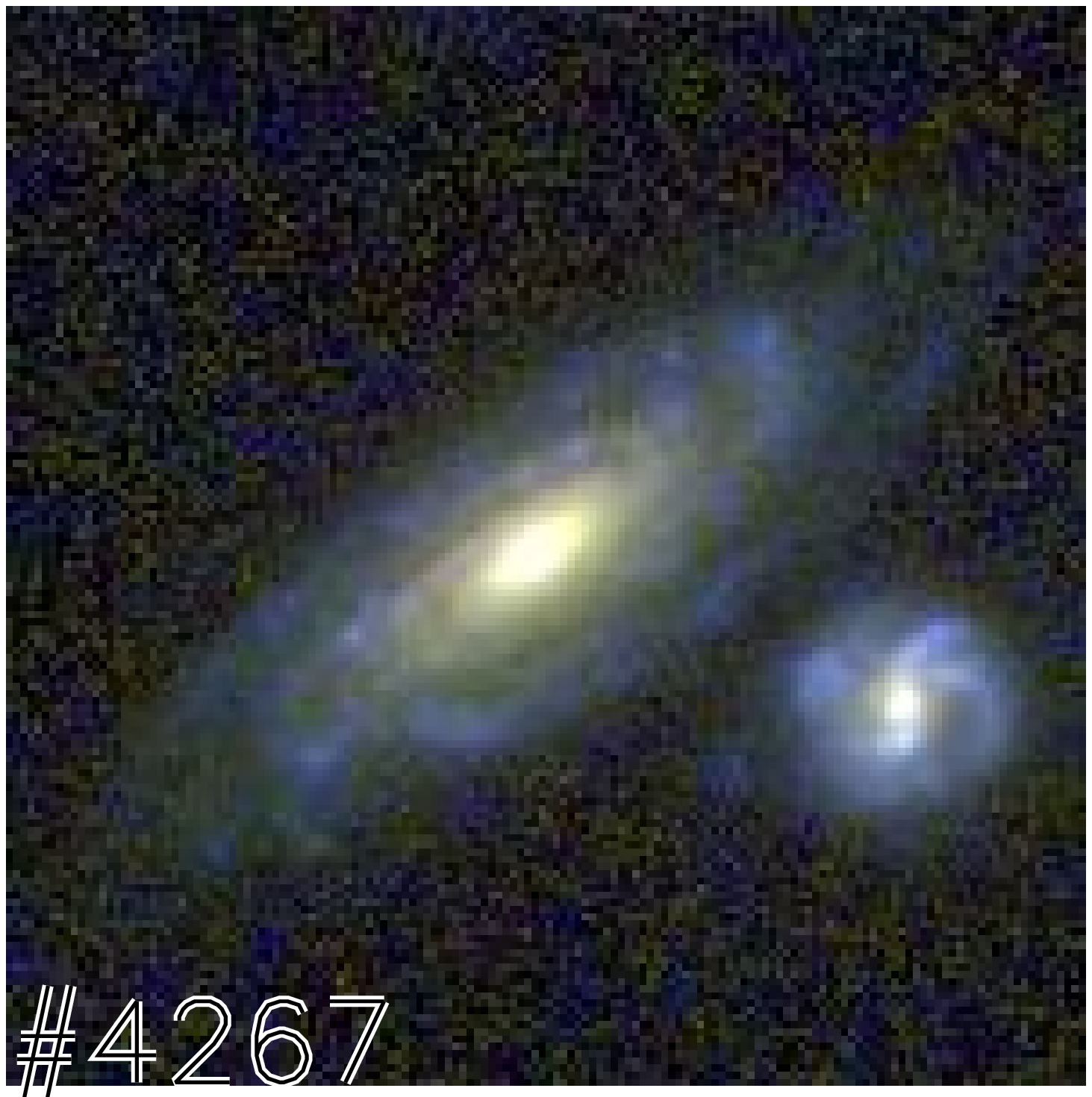}\includegraphics[width=0.23\textwidth]{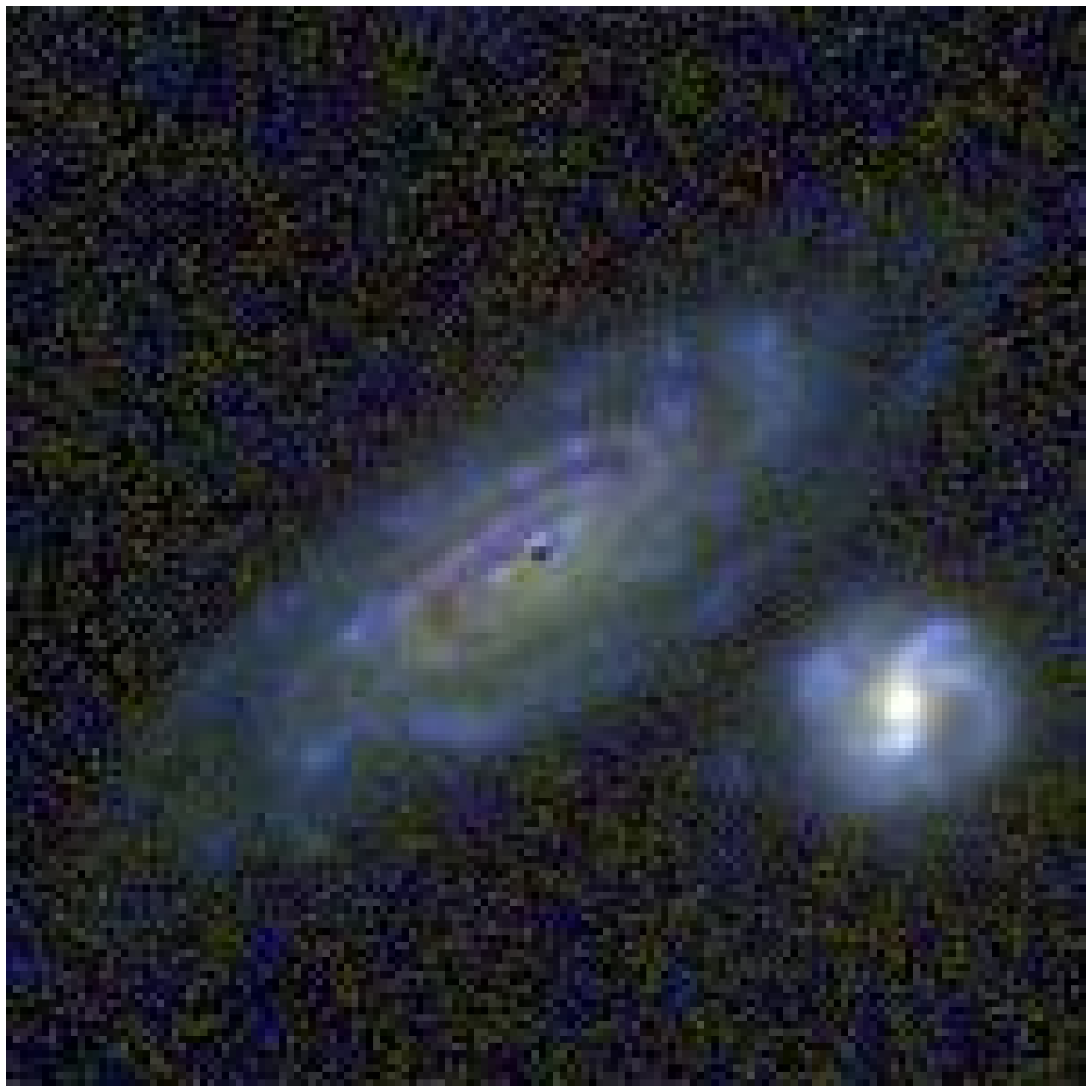}\\
\vfill

\includegraphics[width=0.23\textwidth]{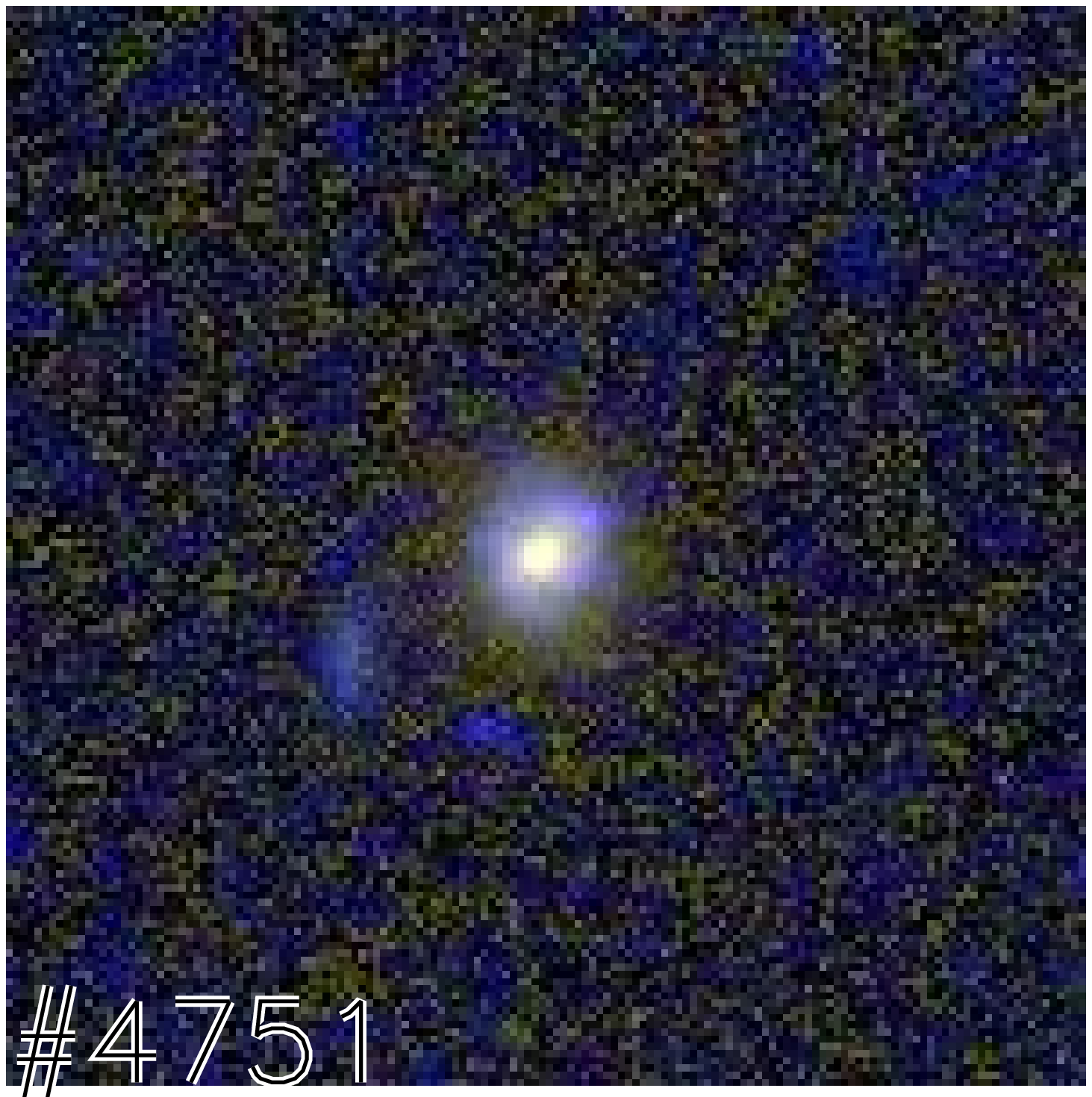}\includegraphics[width=0.23\textwidth]{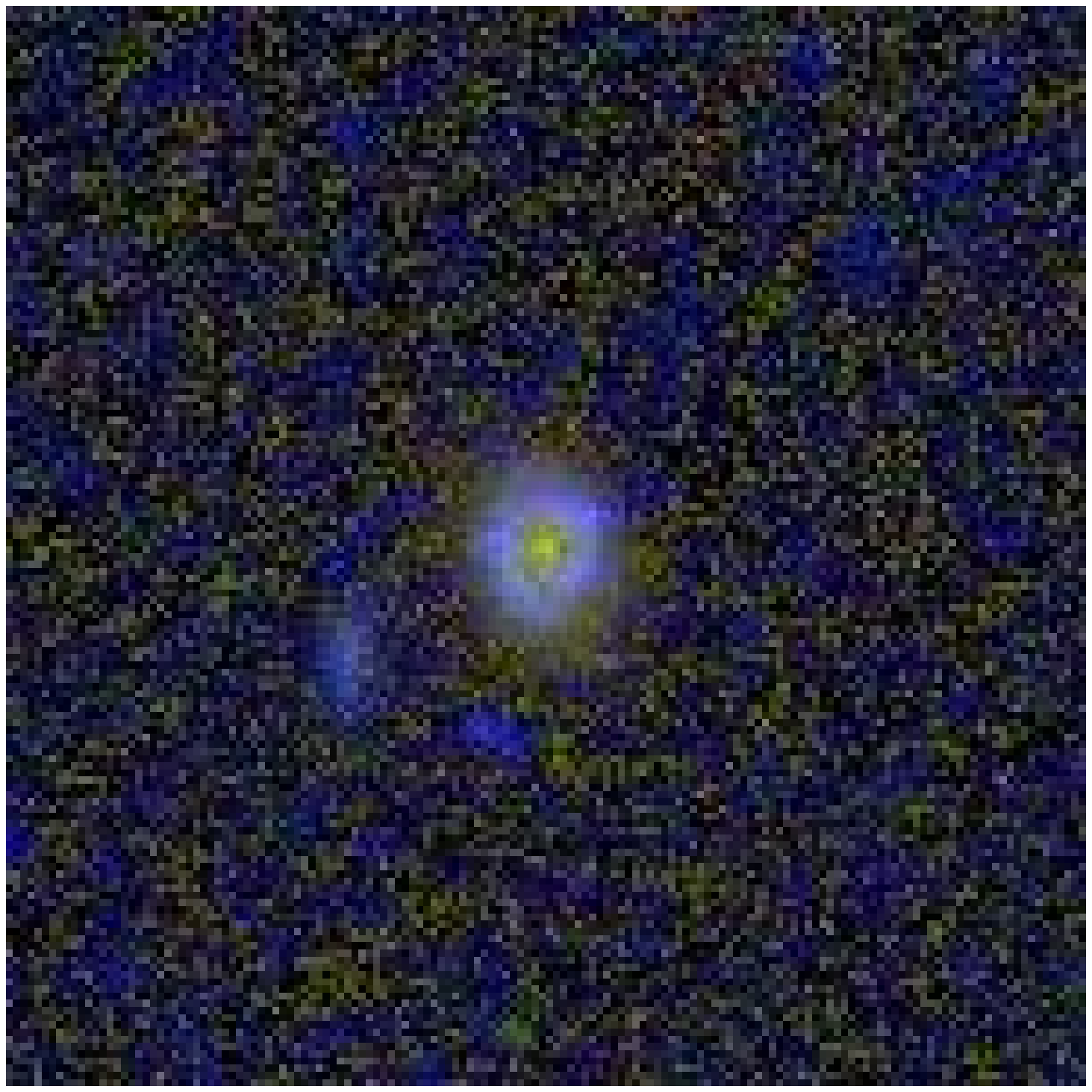}\hfill   \includegraphics[width=0.23\textwidth]{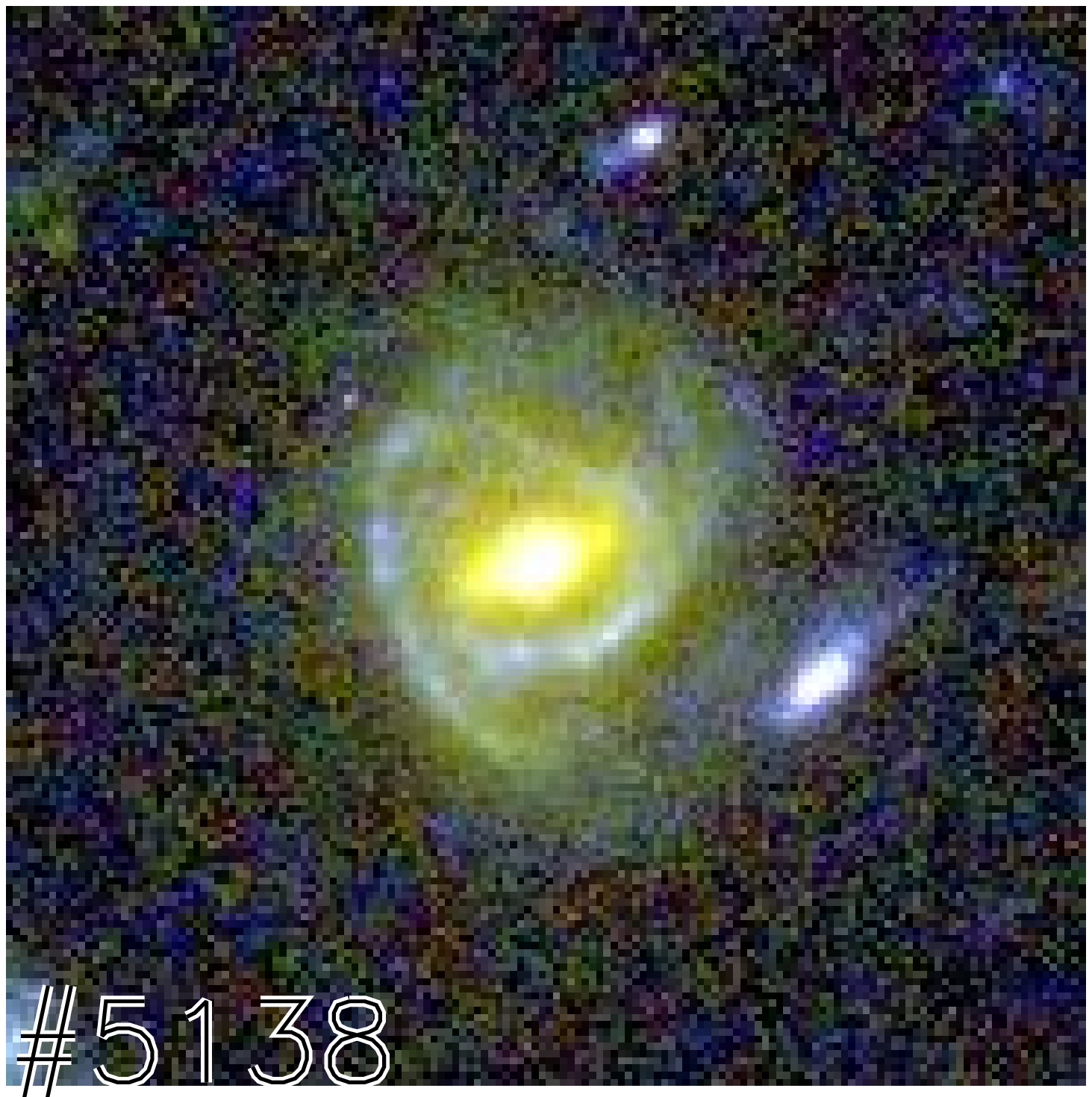}\includegraphics[width=0.23\textwidth]{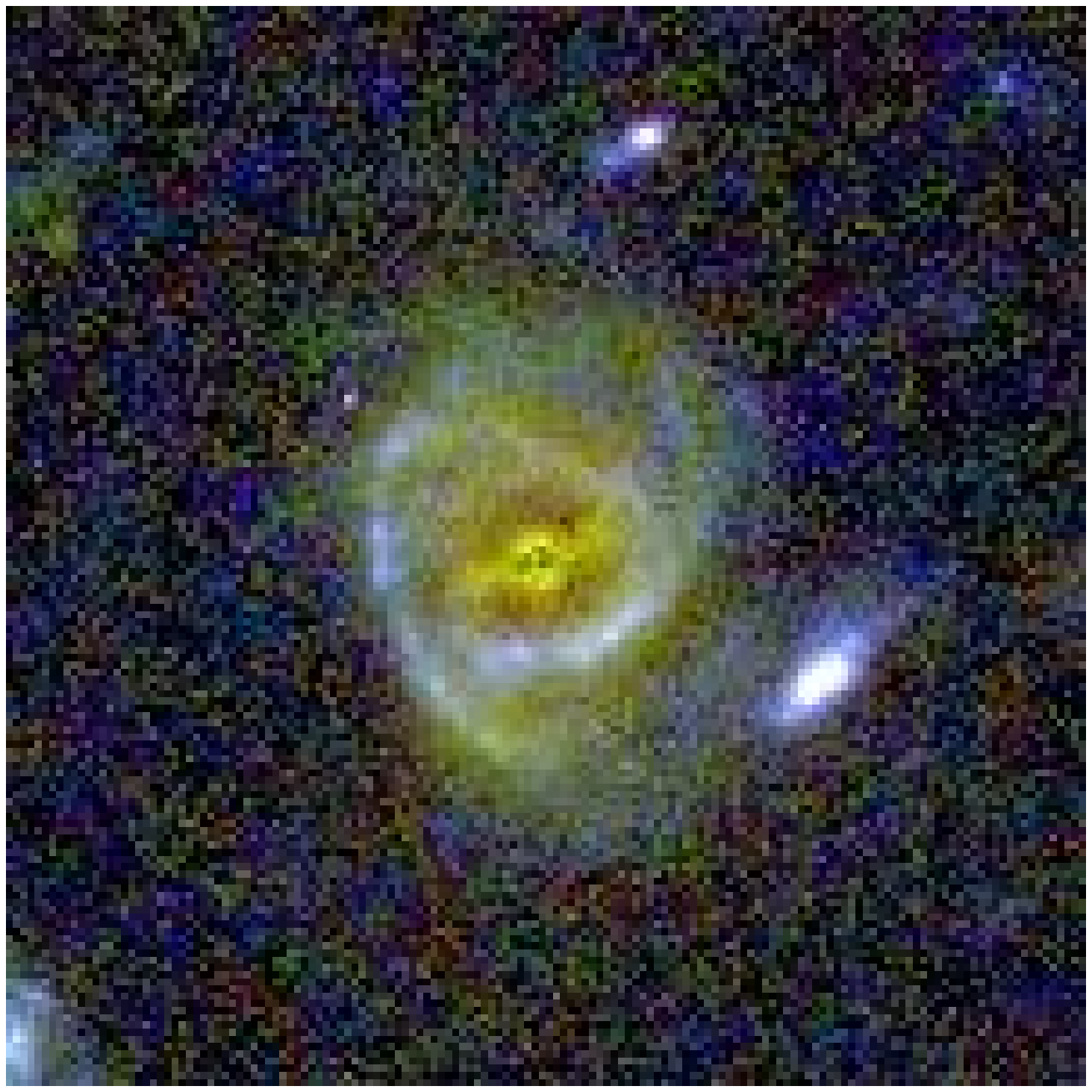}\\
\vfill

\includegraphics[width=0.23\textwidth]{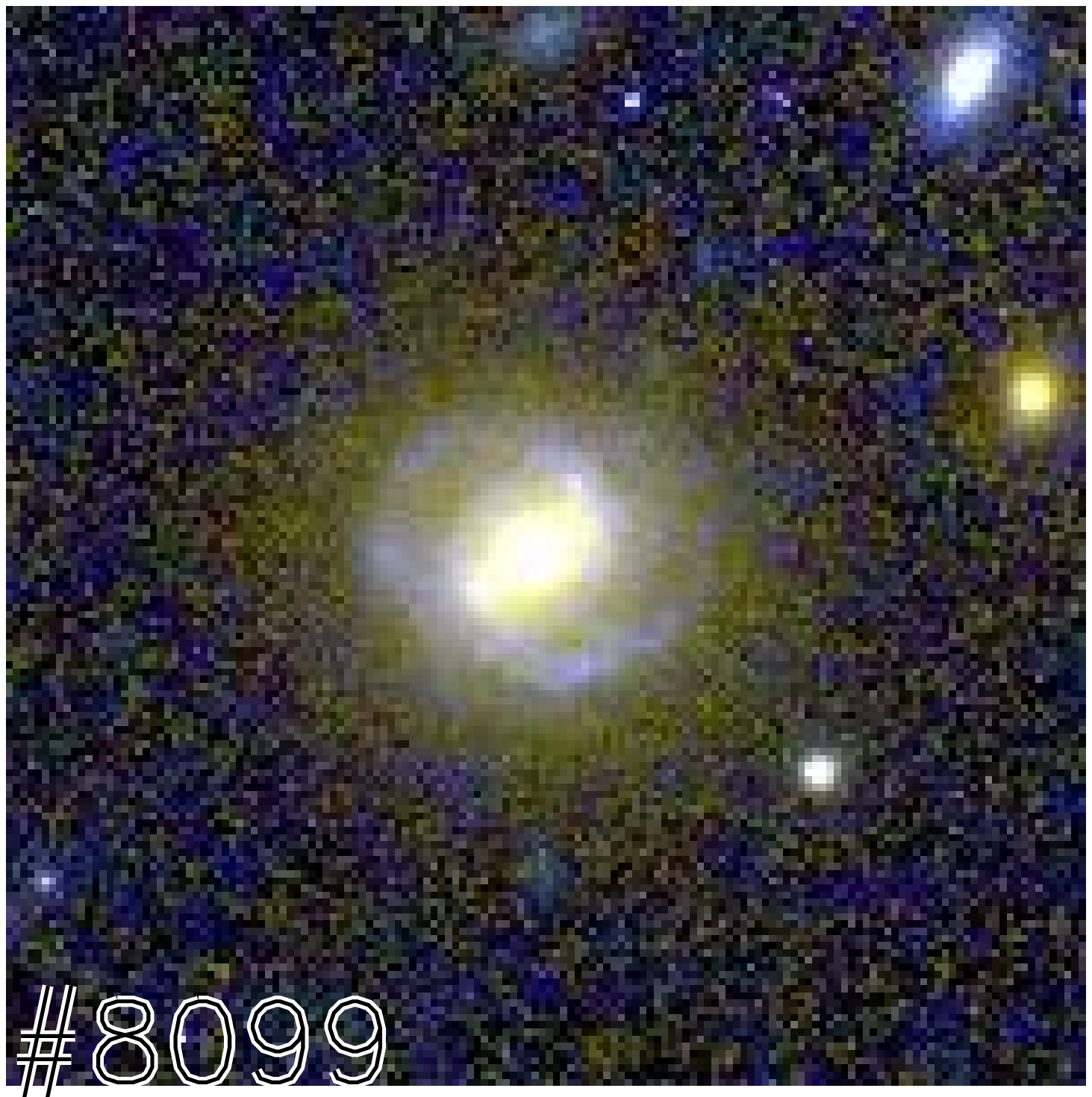}\includegraphics[width=0.23\textwidth]{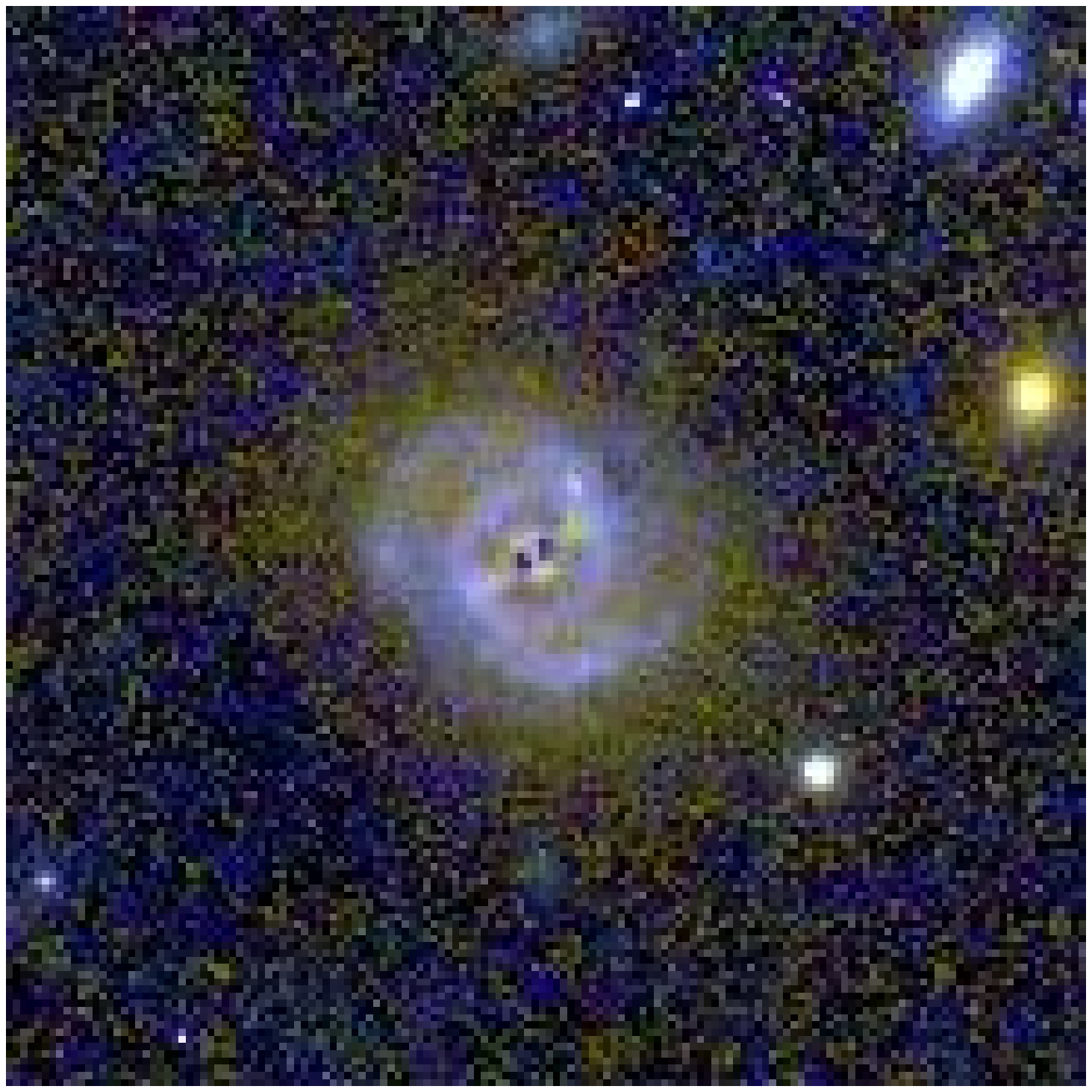}\hfill \includegraphics[width=0.23\textwidth]{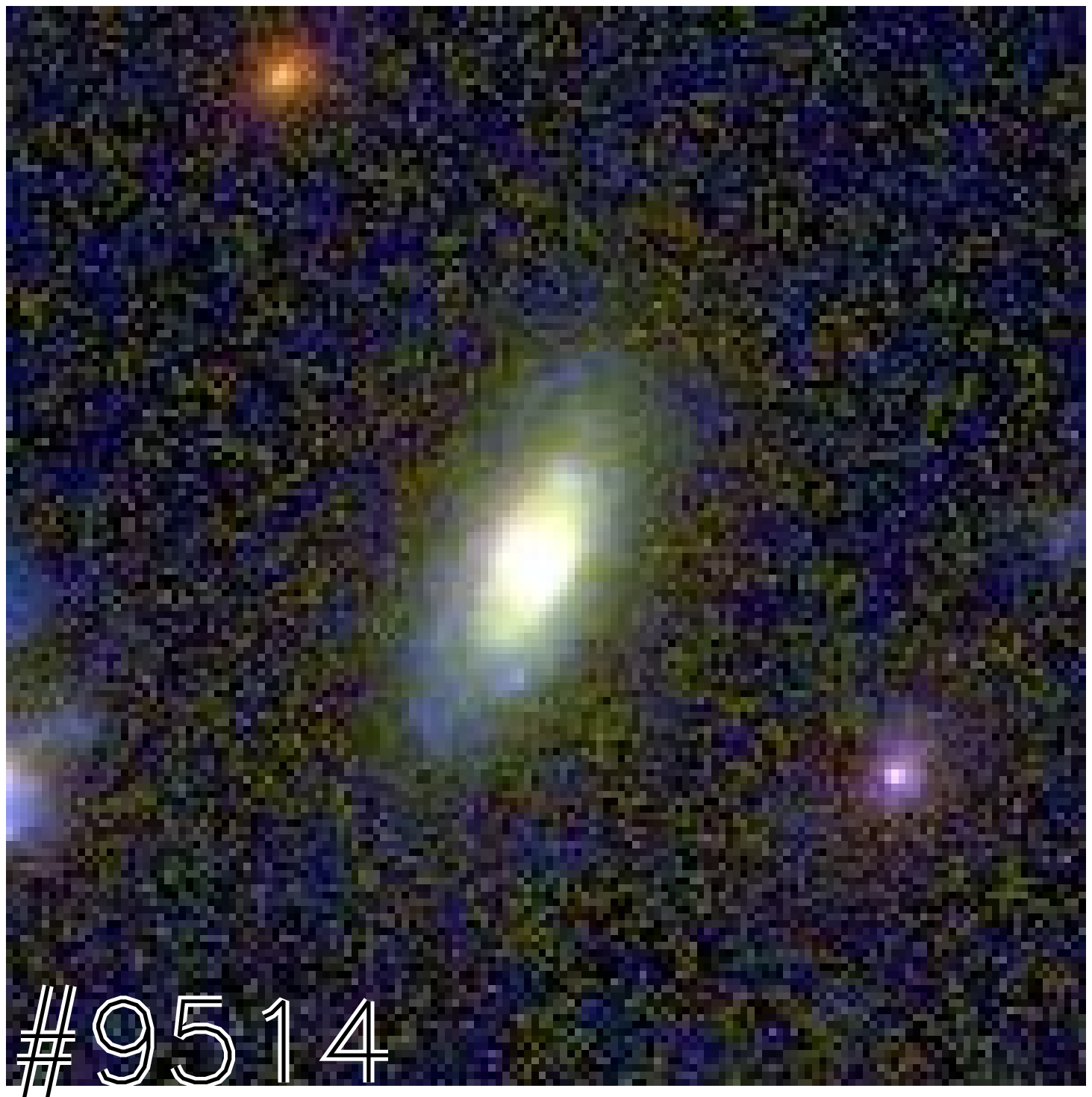}\includegraphics[width=0.23\textwidth]{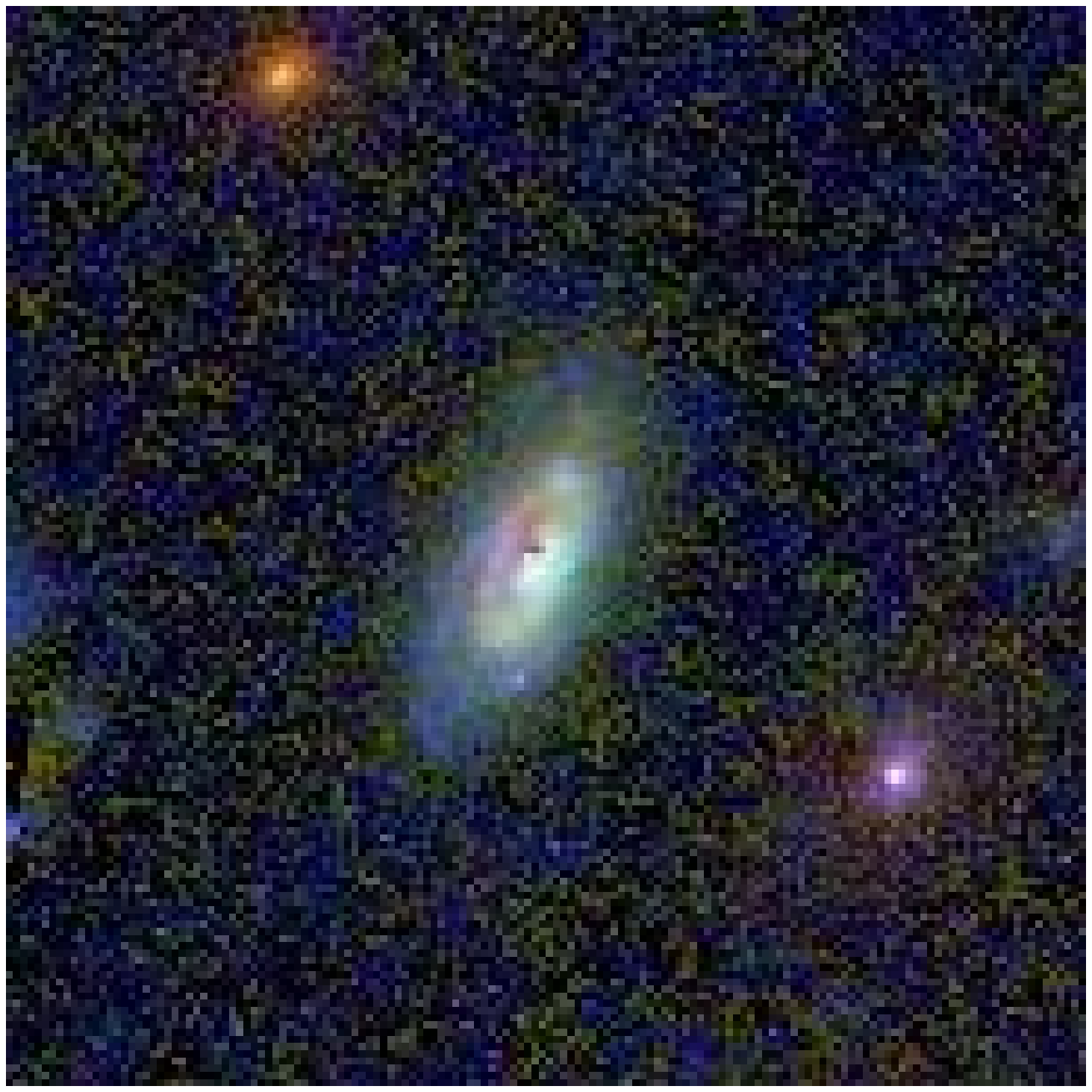}\\
\vfill

\includegraphics[width=0.23\textwidth]{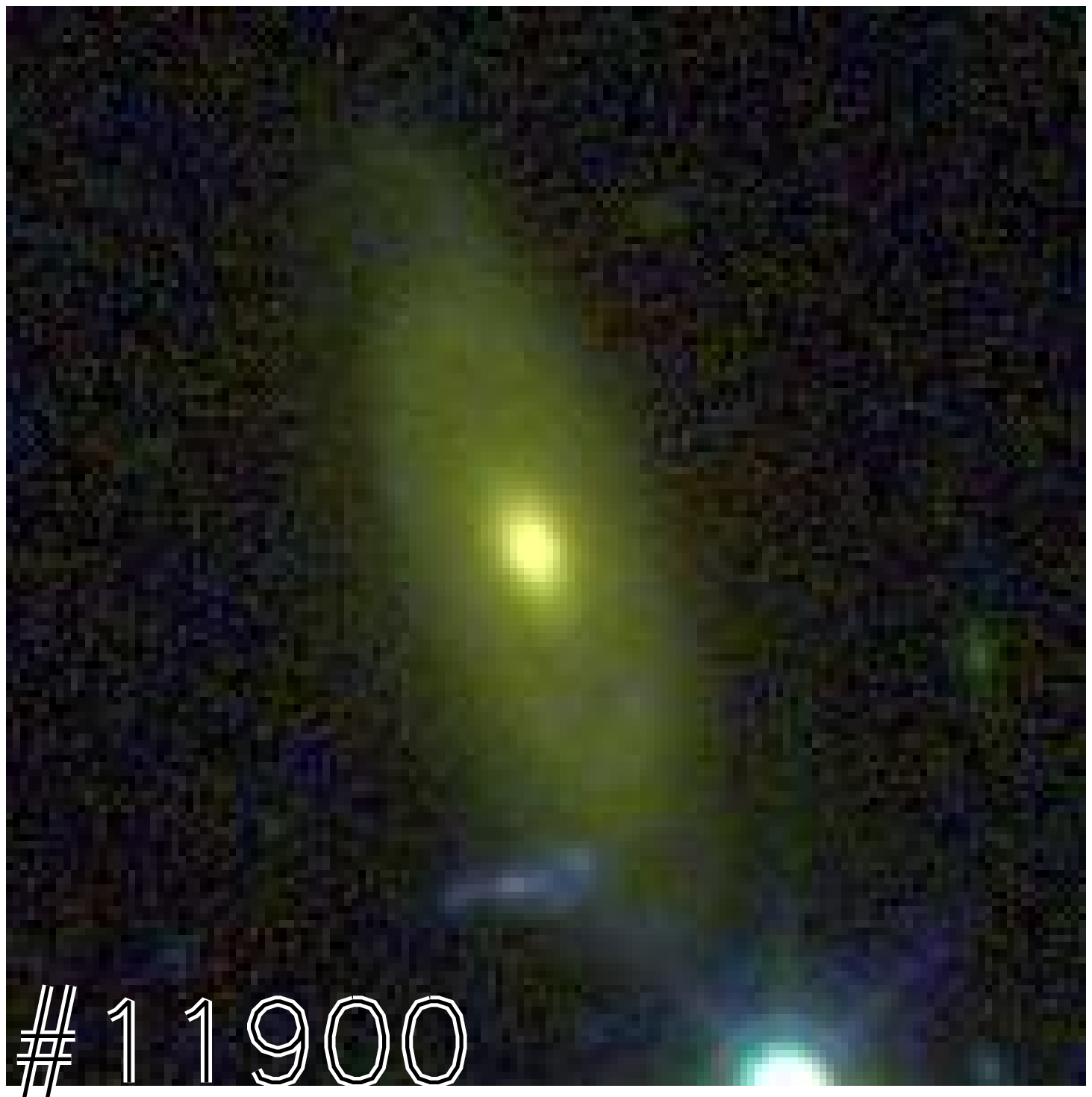}\includegraphics[width=0.23\textwidth]{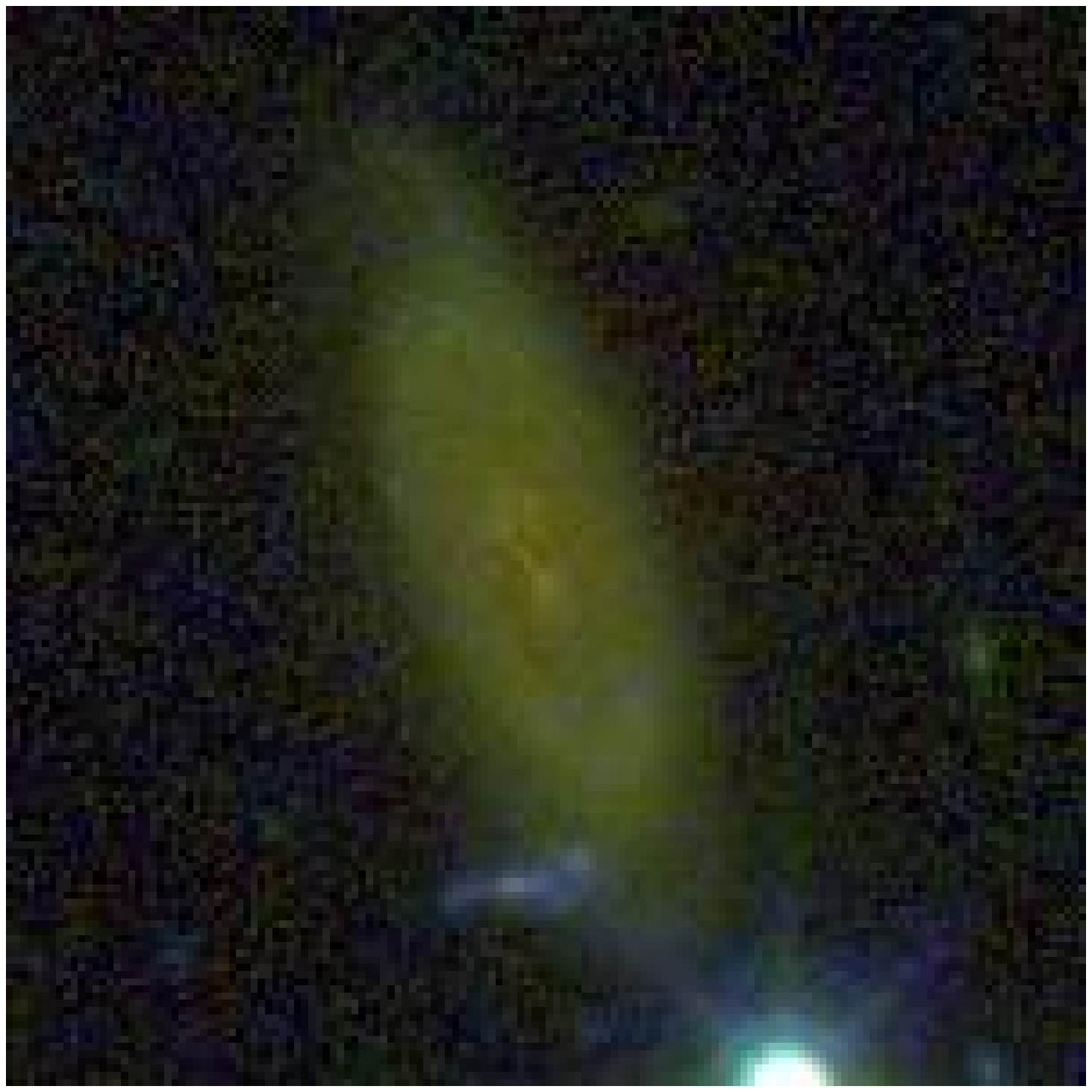} \hfill \includegraphics[width=0.23\textwidth]{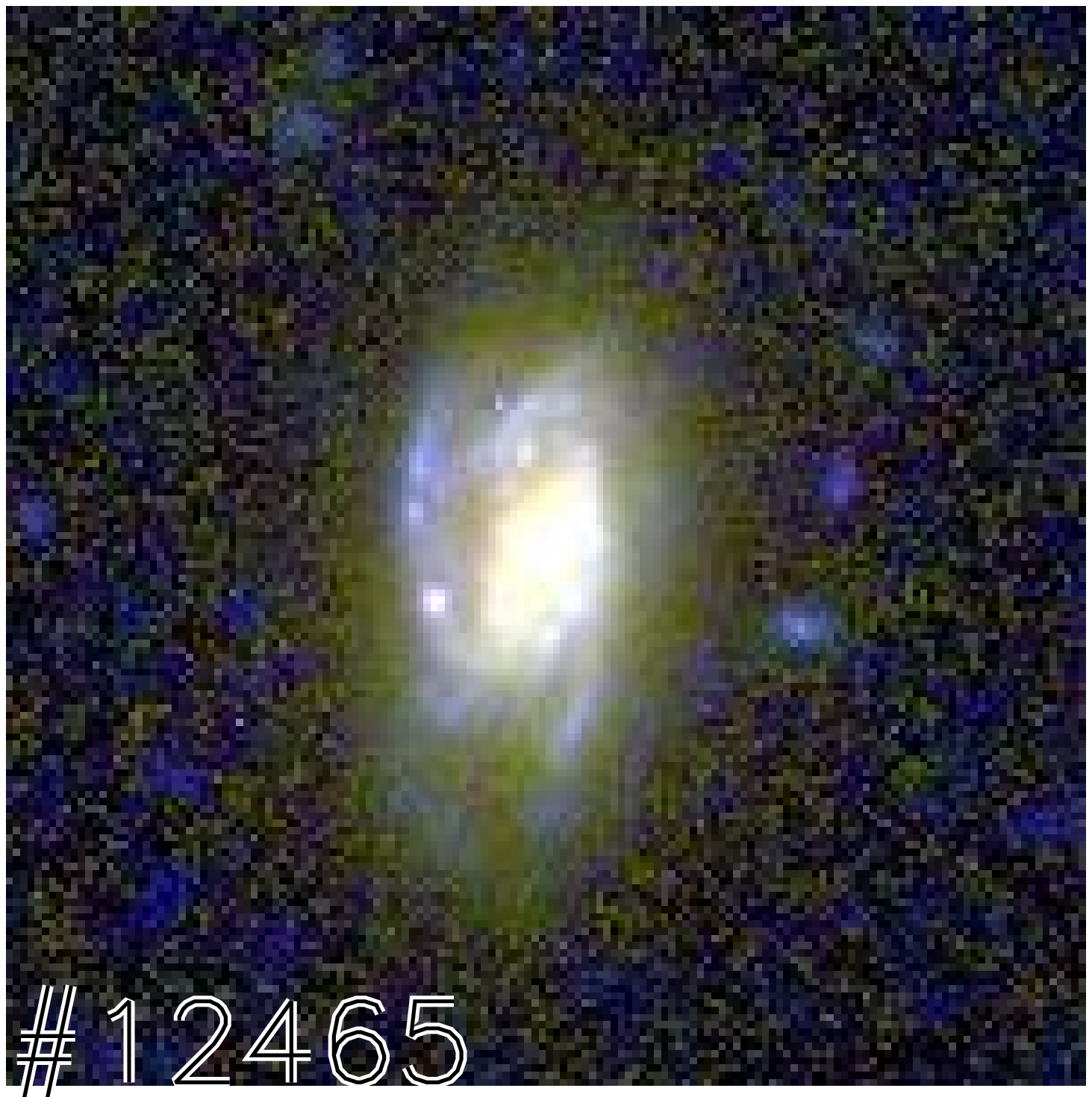}\includegraphics[width=0.23\textwidth]{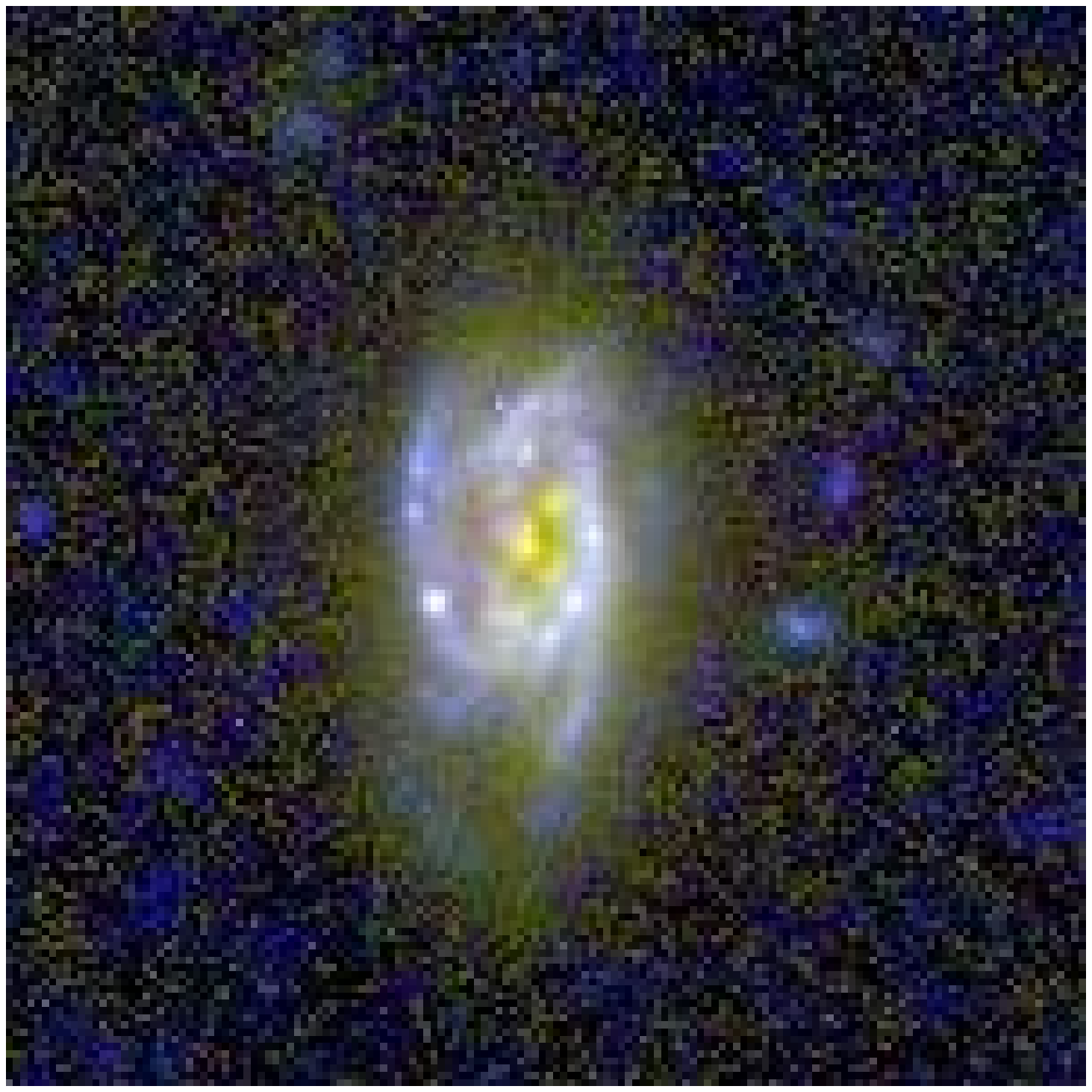}\\
\vfill

\includegraphics[width=0.23\textwidth]{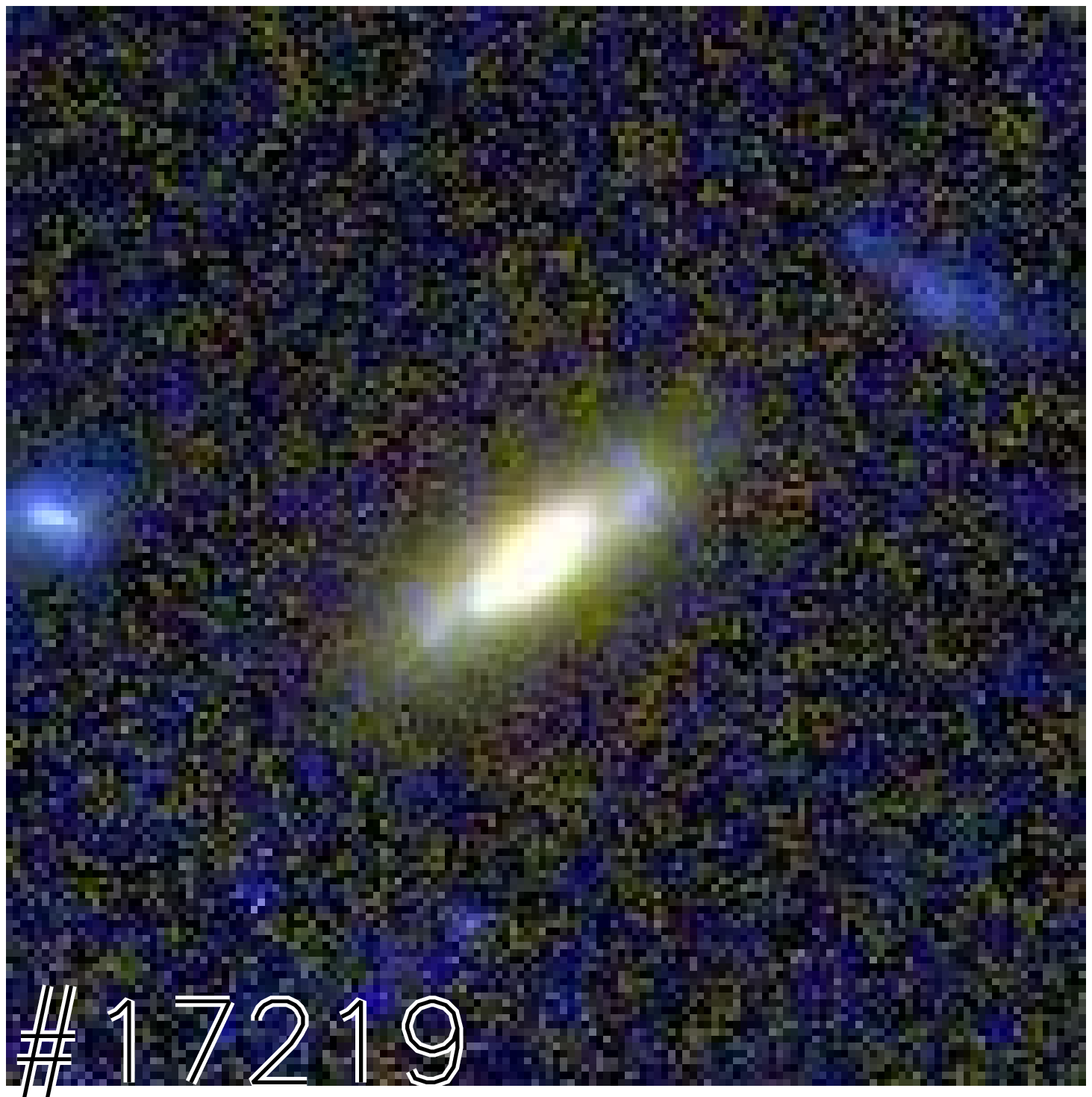}\includegraphics[width=0.23\textwidth]{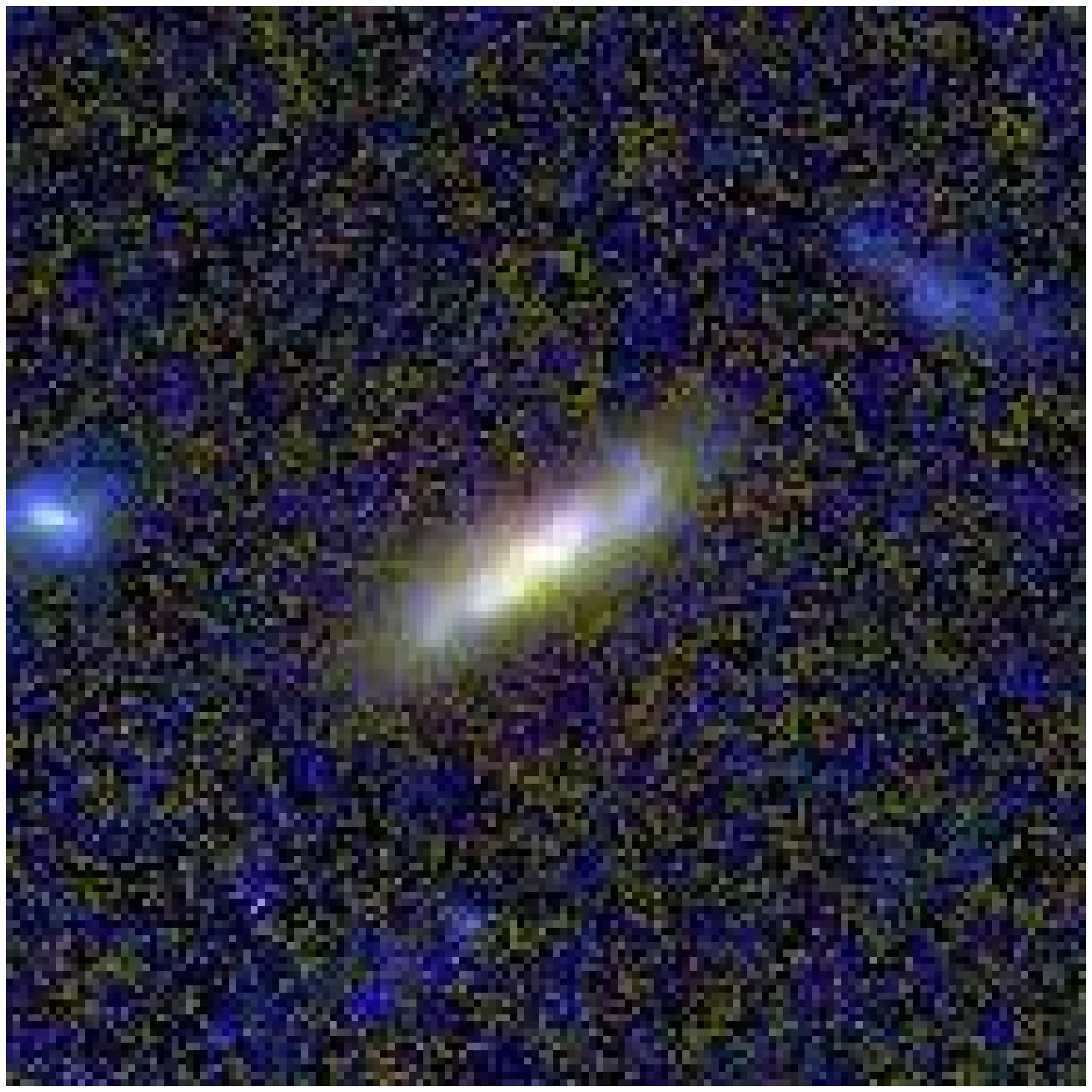} \hfill \includegraphics[width=0.23\textwidth]{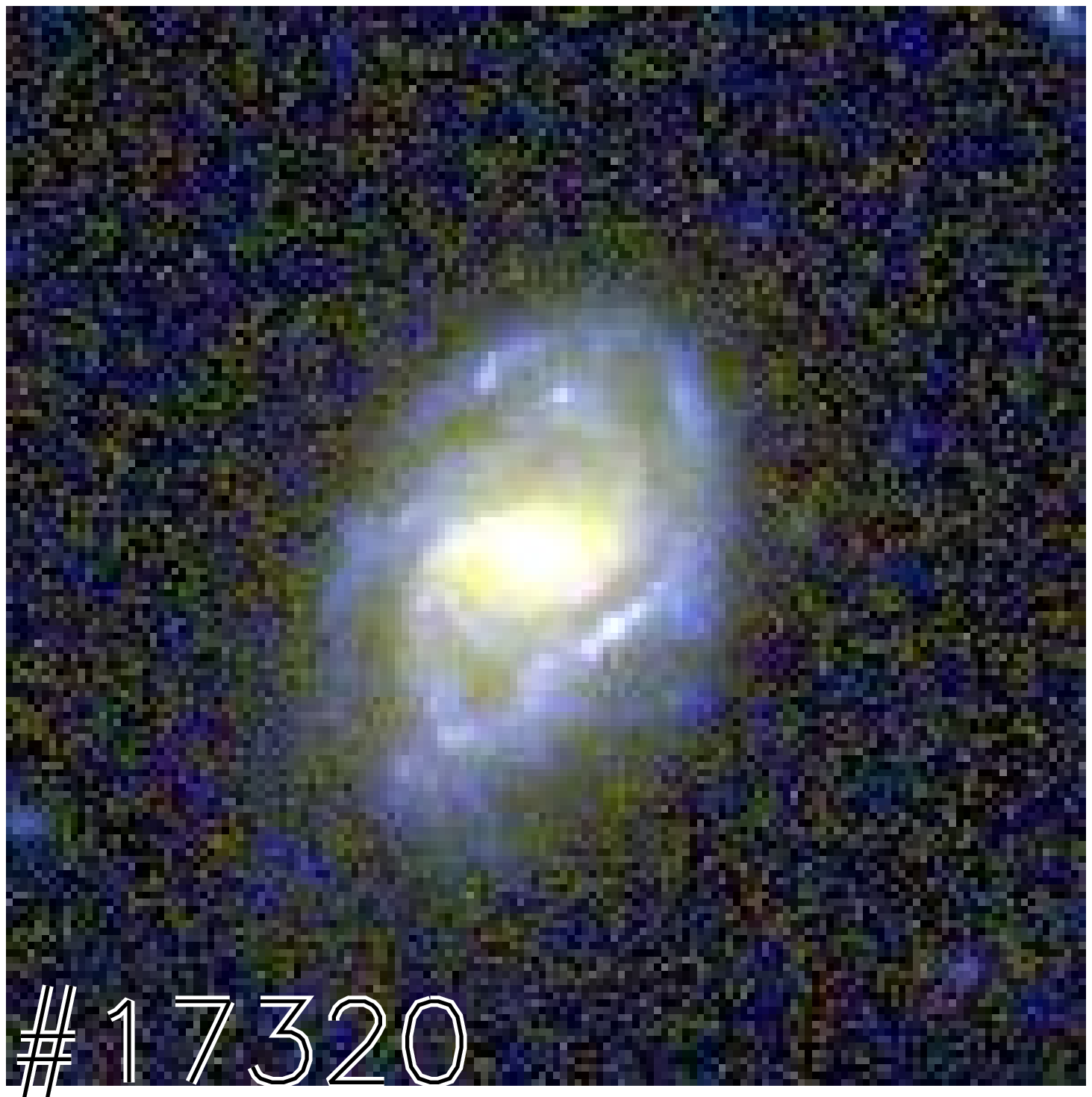}\includegraphics[width=0.23\textwidth]{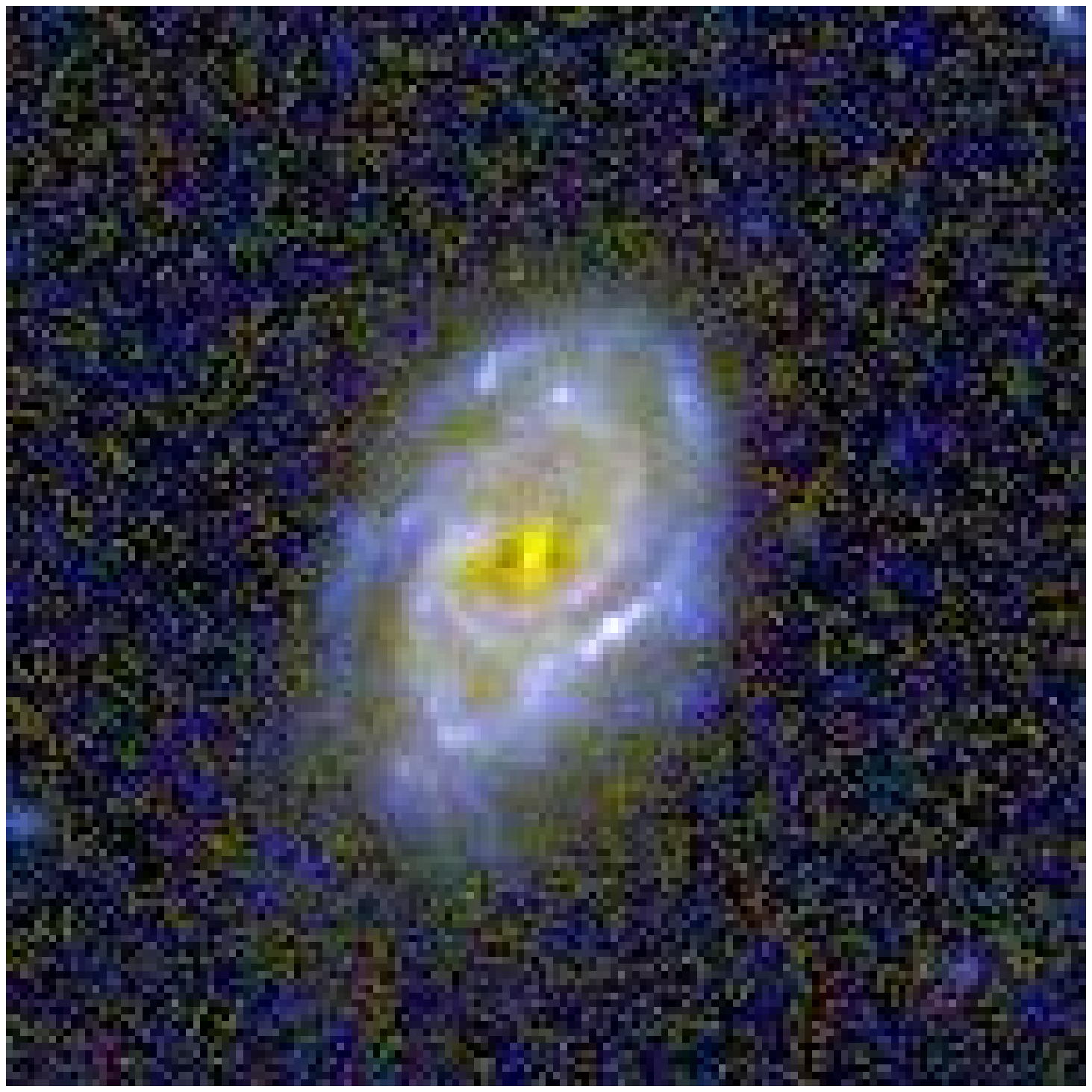}\\

\caption{Colour (red=F160W, green=F850LP, blue=F435W) 9$\times$9 arcs$^2$ postage stamps of our galaxies. For each object, identified by the ID number, the original (left) and bulge-subtracted (right) images are shown. The residuals after the bulge-subtraction are very good, allowing us to reveal the presence of bars (e.g., in \#5138, \#8099, and \#17320), spiral arms (e.g., in \#2202, \#4267), and clumpy structures (e.g., in \#12465) in the central regions of the discs.}
    \label{fig:bd}
\end{figure*}

In Figure~\ref{fig:sizes} we show the position occupied separately by the entire galaxies, bulges, and discs in the stellar mass {\it vs.} effective semi-major axis plane. 
The bulge and disc  sizes are comparable with those of quiescent and star-forming galaxies at the same mass and redshift \citep[][]{2014ApJ...788...28V}, respectively, in agreement with recent results for MS galaxies \citep{2018MNRAS.478.5410D}. Then, although these objects show reduced sSFR, their global sizes (grey filled circles in Figure~\ref{fig:sizes}) are consistent with those of massive star-forming galaxies. 
The RGB colour postage stamps (R=F160W, G=F850LP, B=F435W) of the original and bulge-subtracted galaxy images are displayed in Figure~\ref{fig:bd}.   
We also run {\it GALFIT3} in the $U$/KPNO Mosaic \citep[][]{2004AJ....127..180C} and GALEX near-UV ($NUV$, GALEX GR6 data release\footnote{http://galex.stsci.edu/GR6/}) images. 
In the ground-based $U$-band images (FWHM$\sim 1\farcs$), we used the results obtained from the B/D decomposition in the bluest HST band (F435W) as a prior information to derive the total magnitude. Hence, by leaving the bulge and disc magnitudes as the only free parameters, we fixed the remaining parameters to the known values. The bulge flux in the $U$-band contributes less that 10\% to the total flux in almost all the objects, and we verified that the results for the derived total magnitudes do not change within the errors if we completely exclude the bulge component from the profile fitting. 
Hence, we assumed that the whole near-UV flux comes from the disc component, and used PSF profiles to fit the galaxies in the GALEX/$NUV$ images, since the large FWHM ($\sim 5\farcs$) of such data prevents from recovering morphological information. 
From Figure~\ref{fig:bd} one can note that the discs of our galaxies show complex structures, such as clumpy spiral arms, or bars. For this reason we verified that the total magnitudes derived by the {\it GALFIT3} smooth models in each HST filter were in agreement with those directly measured on the bulge-subtracted images (within a Kron radius), and found that they generally agree within $\sim$ 10\%, that we adopted as flux uncertainties in the SED fitting analysis. In the few cases of larger differences between the two derived magnitudes we adopted wider error bars enclosing both measured values.

\begin{center}
\begin{table}
\caption{Structural parameters from the Bulge-Disc decomposition in the F160W (H) band.}
\begin{tabular}{|l|l|l|l|l|l|l|l|l|l|}
\hline
\hline
  \multicolumn{1}{|c|}{ID} &
  \multicolumn{1}{|c|}{B/T$_{\rm flux}$} &
  \multicolumn{1}{|c|}{B/T$_{\rm mass}$} &
  \multicolumn{1}{|c|}{mag} &
  \multicolumn{1}{|c|}{\re} &
  \multicolumn{1}{|c|}{n} &
  \multicolumn{1}{|c|}{q} &
  \multicolumn{1}{|c|}{PA}\\
\hline
  \multirow{2}{*}{2202}      & \multirow{2}{*}{0.66} &\multirow{2}{*}{0.87}&20.91 & 5.85 & 0.67 & 0.85 & 49.15\\
            &      & &20.20 & 4.73 & 5.35 & 0.75 & -59.63\\ 
  \hline
    \multirow{2}{*}{4267}     &\multirow{2}{*}{0.58} &\multirow{2}{*}{0.86}& 19.62 & 10.47 & 0.69 & 0.38 & -53.82\\
            &     & &19.26 & 4.22 & 2.85 & 0.36 & -55.27\\

 \hline
    \multirow{2}{*}{4751}     &\multirow{2}{*}{0.53} &\multirow{2}{*}{0.66}& 21.16 & 3.56 & 0.50 & 0.74 & 10.83\\
            &     & &21.02 & 1.44 & 2.79 & 0.60 & -47.65\\
\hline
    \multirow{2}{*}{5138}    &\multirow{2}{*}{0.48} &\multirow{2}{*}{0.49}& 20.11 & 10.27 & 0.50 & 0.74 & 7.74\\
            &     &  &20.19 & 4.51 & 5.15 & 0.48 & -69.61\\
\hline
   \multirow{2}{*}{8099}     & \multirow{2}{*}{0.62} &\multirow{2}{*}{0.81}&20.17 & 8.15 & 0.43 & 0.74 & 44.60\\
            &      & &19.65 & 3.44 & 4.22 & 0.59 & -52.85\\
\hline
   \multirow{2}{*}{9514}     & \multirow{2}{*}{0.67} &\multirow{2}{*}{0.78}&20.54 & 4.63 & 0.60 & 0.38 & -22.59\\
            &      & &19.75 & 2.66 & 6.21 & 0.51 & -25.07\\ 
\hline
  \multirow{2}{*}{11900}     & \multirow{2}{*}{0.33} &\multirow{2}{*}{0.38}&19.37 & 12.72 & 0.51 & 0.37 & 17.37\\
            &      & &20.12 & 1.65 & 2.49 & 0.54 & 20.76\\ 
\hline
    \multirow{2}{*}{12465}   & \multirow{2}{*}{0.38} &\multirow{2}{*}{0.47}&18.88 & 8.06 & 0.52 & 0.59 & -1.15\\
            &      & &19.42 & 4.79 & 2.82 & 0.40 & -22.09\\ 
\hline
   \multirow{2}{*}{17219}    & \multirow{2}{*}{0.48} &\multirow{2}{*}{0.66}&20.37 & 5.31 & 0.69 & 0.28 & -57.03\\
            &      & &20.44 & 2.33 & 4.25 & 0.46 & -43.56\\ 
\hline
  \multirow{2}{*}{17320}     & \multirow{2}{*}{0.40} &\multirow{2}{*}{0.56}&19.20 & 9.16 & 0.43 & 0.60 & -23.38\\
            &      & &19.66 & 3.01 & 3.96 & 0.50 & -81.02\\ 
\hline

\end{tabular}
For each galaxy the {\it galfitm} output parameters derived from the disc (first line) and bulge (second line) surface brightness fitting in the F160W band are shown. From left to right : ID = Identification number, B/T$_{\rm flux}$ = bulge-to-total F160W flux ratio, B/T$_{\rm mass}$ = bulge-to-total mass ratio, mag= F160W AB magnitude, \re = effective semi-major axis in kpc (i.e. the semi-major axis including half of the total light), n = S\'ersic index, q = minor to major axis ratio (b/a), PA = position angle in degrees.
\label{tab:sers_param}
\end{table}
\end{center}

\newpage


\section{Optical/near-IR SED fitting analysis}\label{sec:seds}
The SED of bulges and discs were derived by fitting with the {\it HyperZmass} code \citep[i.e., a modified version of HyperZ,\footnote{http://webast.ast.obs-mip.fr/hyperz/}][]{2000A&A...363..476B,2007A&A...474..443P} the magnitudes from GALEX/$NUV$ to H-band, obtained from the analysis described in the previous section. 
We adopted \citet[][hereafter, BC03]{2003MNRAS.344.1000B} stellar population models, with ages ranging from 0.05~Gyr to the age of the Universe at the observed redshifts, solar metallicity, and exponentially declining SFH (i.e., SFR $\propto$ exp(-t / $\tau$), with $\tau=$0.1, 0.3, 0.5, 1, 1.5, 2, 3, 5, 10, 15, 30, and $\infty$ = constant SFR). 
For the discs both the \citet{2000ApJ...533..682C}, and the SMC \citep{1984A&A...132..389P,1985A&A...149..330B} extinction laws were allowed in the fitting procedure, with the reddening parameter varying in the range $A_V=0-2.5$. Based on recent results from \citet{2017A&A...599A..95G} showing that a residual fraction of dust (and gas) is still present in high-$z$ Early-Type Galaxies (ETGs), we decided against assuming zero reddening for the bulges. We adopted the \citet{2000ApJ...533..682C} attenuation curve, and obtained relatively low reddening ($A_V<0.6$) for all the bulges. Similar reddening parameters are obtained assuming other extinction laws. 
In this work the SED fitting results serve two main purposes, i.e., estimating (i) the bulge and disc stellar masses (and B/T mass ratios), and (ii) the age and SFH of the discs, while the age and SFH of the bulges are better constrained from the TKRS spectra (Section \ref{sec:b_spec}). As further commented in the following, we could not use the spectroscopic information for the discs, since the stellar continuum in the outer galaxy regions (less contaminated by the bulge) is too faint to be detected. 

The stellar mass is the most robustly recovered parameter in the SED fitting analysis \citep[e.g.,][]{2007A&A...474..443P}, hence accomplishing the first goal may not be so critical when a large grid of stellar population synthesis models, covering a wide range of parameters, is used. However, while this is particularly true for the masses of bulges (i.e. old stellar populations), for composite stellar populations including very young stars ($10^6-10^8$ yr), such as the discs, there is the possibility of underestimating the stellar masses. This is because, if present, the oldest stellar populations could be outshined by the youngest ones, even if more massive, as shown by \citet{2010MNRAS.407..830M}. 
Since this could also affect the derivation of ages and SFHs for the discs, we checked against the existence of a possibly missed stellar mass contribution. We compared the IRAC observed fluxes\footnote{The photometry was taken from the catalogue by L18. We excluded from the comparison the IRAC 8.0 $\mu$m band, since it could include PAH emissions at the observed redshifts.} (not included in the B/D SED fits), sampling the part of the SED directly sensitive to the galaxy total stellar mass, with those derived from the bulge+disc SED model. This is shown in Figure~\ref{fig:seds}, where the bulge, disc, and total SEDs are shown in red, cyan, and grey, respectively. The agreement between the total SED models (grey dashed curve) and the photometric data-points (grey filled circles) at $2\lesssim\lambda_{\rm rest}\lesssim4\;\;\mu$m (on average within 10\%), indicates that our bulge and disc mass estimates are robust within the uncertainties.  
In Appendix \ref{app:med_sed}, we also derived an upper limit for the fraction of possibly missed mass by performing a similar analysis on the composite SED of our 10 objects, and found that it is $\leq 7\%$ of the total stellar mass.

\subsection{$UVJ$ diagram for bulges and discs}\label{sec:uvj}
In Figure~\ref{fig:uvj} we show the position occupied by our sample galaxies, and by bulges and discs after the B/D decomposition, in the rest-frame $(U-V)$ vs. $(V-J)$ (hereafter, $UVJ$) diagram, widely used in the literature to separate star-forming from quiescent galaxies \citep{2005ApJ...624L..81L, 2007ApJ...655...51W,2009ApJ...691.1879W}. The rest-frame $UVJ$ colours of bulges and discs were derived from their best-fit SEDs presented in the previous section convolved with he Bessel $U$, and $V$ filters and the $J$ 2MASS filter, following \citealt{2009ApJ...691.1879W}.
The colours of the entire galaxies  of our sample (grey filled circles) and those of the parent sample of MIPS-VLA detected objects at $0.45<z<1$ from the L18 catalogue (small grey open circles) are taken from the 3D-HST catalogue \citep{2014ApJS..214...24S}. Our bending galaxies are located around the quiescent/star-forming boundary in the $UVJ$ diagram, as expected for galaxies with reduced sSFR, considered in transition between the star-forming and quiescent phase \citep[e.g.][]{2010ApJ...713..738W,2011ApJ...735...53P,2018ApJ...858..100F,2018ApJ...860...60L}, i.e. they are `green valley' objects. However, when analysing  them separately,  bulges (red filled circles) and discs (cyan filled circles) lie in the quiescent, and star-forming $UVJ$ regions, respectively. This confirms that such objects hold a certain amount of star-formation (as deduced from their FIR detection), although it seems to be confined only  to the blue discs. On the other hand, their bulge show typical colours of fully quiescent, old galaxies. The only exception is represented by object \#11900, showing a redder disc, enclosed in the quiescent galaxy region of the diagram, although very close to the boundary with the star-forming region. We note that this is also visible by eye in the colour images of Figure~\ref{fig:bd} where, unlike the others, the disc  of this galaxy appears to be very faint in the blue F435W/HST filter. The fact that object \#11900 is also the most distant from the MS ridge-line (being $\sim$ 10 times below it), and shows the lowest S/N$_{\rm FIR}$ in our sample, may allow us to suspect it as a truly quiescent disc  galaxy, the old-stars heated cirrus being responsible for the FIR emission. Yet, it has one of the strongest  [OII]$\lambda\lambda$3727 emission lines (see the online version of Figure~\ref{fig:single_spectra}), suggesting that a certain amount of gas and star-formation activity is still present in this object. 

\subsection{The disc and bulge stellar ages from SED fitting}\label{sec:age_d}
To derive age and SFH for a star-forming composite stellar population one has to deal with the degeneracy and model-dependence of parameters that could produce large uncertainties.  However, as indicated by the  rest-frame $UVJ$ colours of bulges and discs, we can safely assume that the bulk of the SFR of the galaxies in our sample takes place in the discs. Hence, to reduce the parameter degeneracy, for the discs the optical/near-IR SED fitting results (cf., Section~\ref{sec:seds}) were constrained to reproduce the known value of the present SFR from IR+UV, as from Table ~\ref{tab:ir_fluxes}. In practice, all the SED fitting solutions returning values of SFRs\footnote{For each different combination of parameters (e.g., $\tau$, A$_V$, formation time, extinction law, metallicity) the {\it HyperZmass} SED fitting procedure determines the template flux in each photometric band used in the fit, and the associated $M_*$ and SFR values. The relative $\chi^2$ is derived from the comparison of the template fluxes with the observed ones as $\chi^2$ = $\Sigma_{\rm i}^{N_{\rm filt}} \left[ \frac{F_{\rm obs,i}-n F_{\rm temp,i} }{\sigma_{\rm i}} \right]^2 $, where $n$ is a normalisation constant. The best-fit solution automatically returned by the software is the one that minimises the $\chi^2$.} outside the range log(SFR(IR+UV) $\pm$0.15 were discarded (where 0.15 dex is the typical uncertainty on the SFR(IR), as derived in L18). 
Since our sample includes, by construction, galaxies with reduced sSFR(IR+UV), imposing such a limit produces the main effect of excluding solutions with high SFR, high dust extinction and very young age ($0.5-1$ Gyr). This indirectly constrains the reddening parameter within a narrower range (approximately $0.5\lesssim A_V \lesssim 1$). Moreover, given the current low sSFR, this imposes a higher sSFR in the past in order to build sufficient stellar mass in the disc. No constraints were instead adopted for the bulge SED fitting. 

For each bulge and disc, the age was computed as {\it the time $T_{50}$ since 50\% of the stars were formed}, hereafter `half-mass age' \citep[cf., ][]{2016ApJ...824...45P}. We adopted this definition instead of {\it the time since the beginning of the star formation} (T$_{\rm beg}$), because it has a milder dependence on the model SFH. We found best-fit ages of T$_{50}\sim 1-1.5$ Gyr for most of the discs, but for three of them (i.e., \#11900, \#12465, and \#17219), reaching ages of T$_{50}\sim 2-3$ Gyr. The bulge best-fit half-mass ages are much older (i.e., T$_{50}\sim 5.2 $ Gyr on average) in agreement with the results from the spectral analysis presented in the next section, but for object \#5138, although the results are still consistent within the large T$_{50}$ error bars derived from the broad-band SEDs (see Figure~\ref{fig:sedvsspec}). 

When the declining $\tau$-models are used to parametrize the SFH, an even more interesting quantity to consider is the ratio between the time since the beginning of the star formation and the star formation timescale, i.e. T$_{\rm beg}$/$\tau$, which gives a hint on the current stage of the star formation activity in the system. In fact, T$_{\rm beg}$ / $\tau \lesssim 1$ indicates that the system is still close to the peak of its star formation activity, as observed for most of our discs. 
Then, for larger T$_{\rm beg}$/$\tau$, the star formation enters the declining phase, e.g. with a SFR $\sim 35\%$ to 5\% of the peak value moving from T$_{\rm beg}$/$\tau=1$ to T$_{\rm beg}$/$\tau=3$. A T$_{\rm beg}$/$\tau\gg 1$ characterizes, instead, systems with very low, or no residual star-formation, such as our bulges. 
In Figure~\ref{fig:contours} we show the 99\%, 90\% and 68\% confidence levels of the $\chi^2$ statistics for T$_{\rm beg}$/$\tau$ {\it vs} T$_{50}$, derived for our discs for the case of ``two interesting parameters'', following \citet[][]{1976ApJ...210..642A}. It can be seen that, although in most cases the best fit solution (red filled circle) suggests $\rm {T_{\rm beg}}$/$\tau$=1-2 (e.g. declyining SFHs),  plausible solutions (i.e. within the yellow 68\% confidence levels) can be also found for $\rm {T_{\rm beg}}$/$\tau<1$ (e.g., for $\sim$ constant SFHs).
In Tables~\ref{tab:seds}, and \ref{tab:seds_B} we show the SED best-fit parameters for discs and bulges, respectively, together with the relative 68\% confidence intervals.\footnote{The 68\% confidence intervals are derived by considering all the solution for which $\chi^2\leq$min($\chi^2$)+2.3, corresponding to the case of ``two interesting parameters'' in \citet[][]{1976ApJ...210..642A}. For the discs these are the solutions enclosed in the yellow regions of Figure ~\ref{fig:contours}, then used in Section~\ref{sec:sfhs} to derive the SFHs.}

It is worth to note that, fitting the disc SEDs with stellar population models including a different treatment of the thermally pulsing asymptotic giant branch (TP-AGB) phase (e.g., \citealt{2005MNRAS.362..799M}, MA05), would lead to derive even younger ages, and consequently lower stellar masses. In fact, the composite stellar populations of the galaxy discs include from very young to intermediate age components, among which the contribution of TP-AGB stars is expected to be maximum (0.2-2 Gyr, cf.~\citealt{2006ApJ...652...85M}).  
The use of this kind of models for bulges, whose very red SEDs are interpreted by our fits as due to very old stellar populations (T$_{50}\geq 4$~Gyrs), may also result in lowering their average stellar ages and masses, if part of the near-IR light is attributed to younger stars ($1-2$ Gyr) in the TP-AGB phase. However, we directly verified on the bulge composite spectrum that this contribution is not relevant, reducing the mass-weighted age of just a $\sim 15\%$ (see Section~\ref{sec:stack}).

\begin{figure*}
\includegraphics[width=0.35\textwidth]{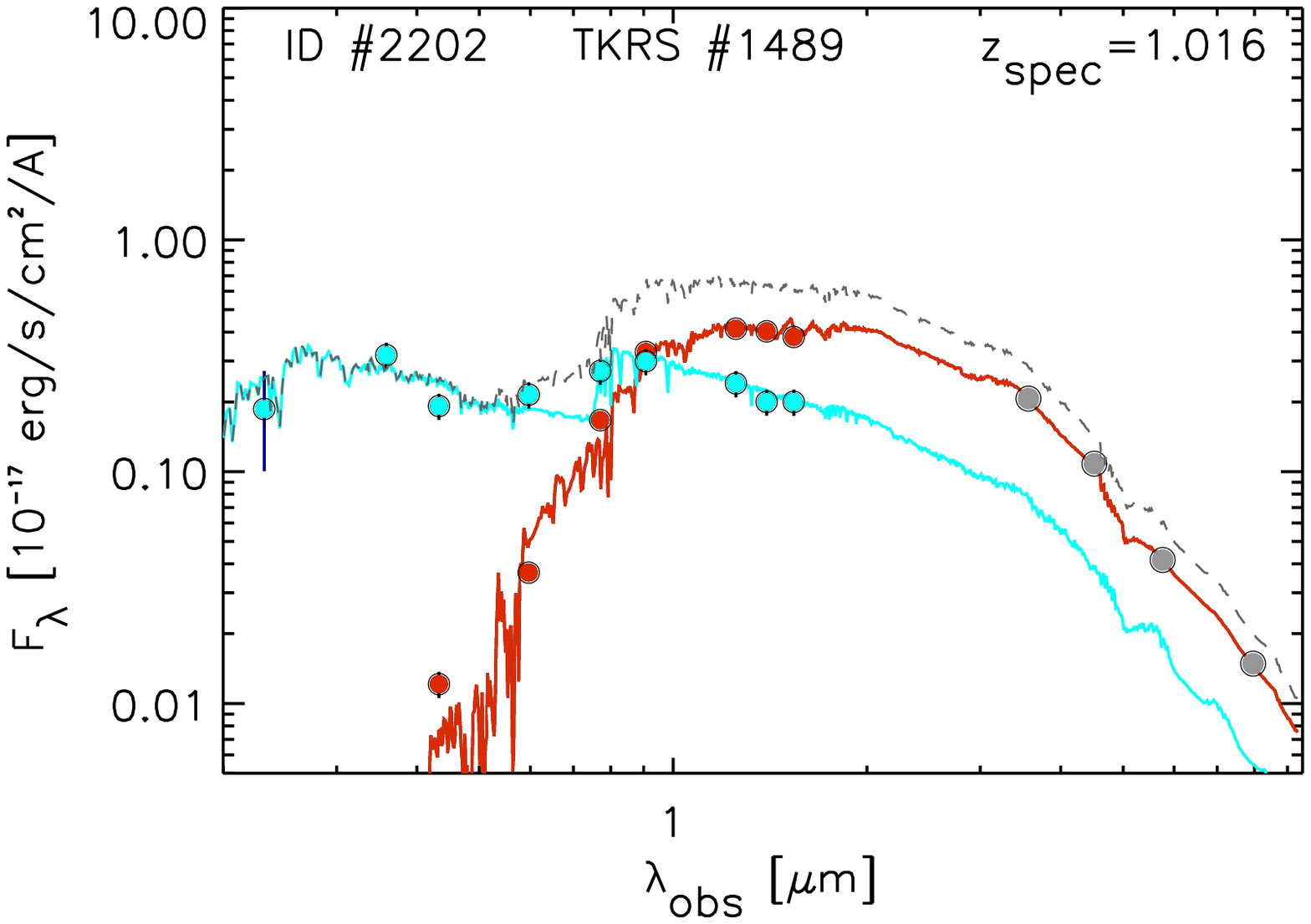}\includegraphics[width=0.35\textwidth]{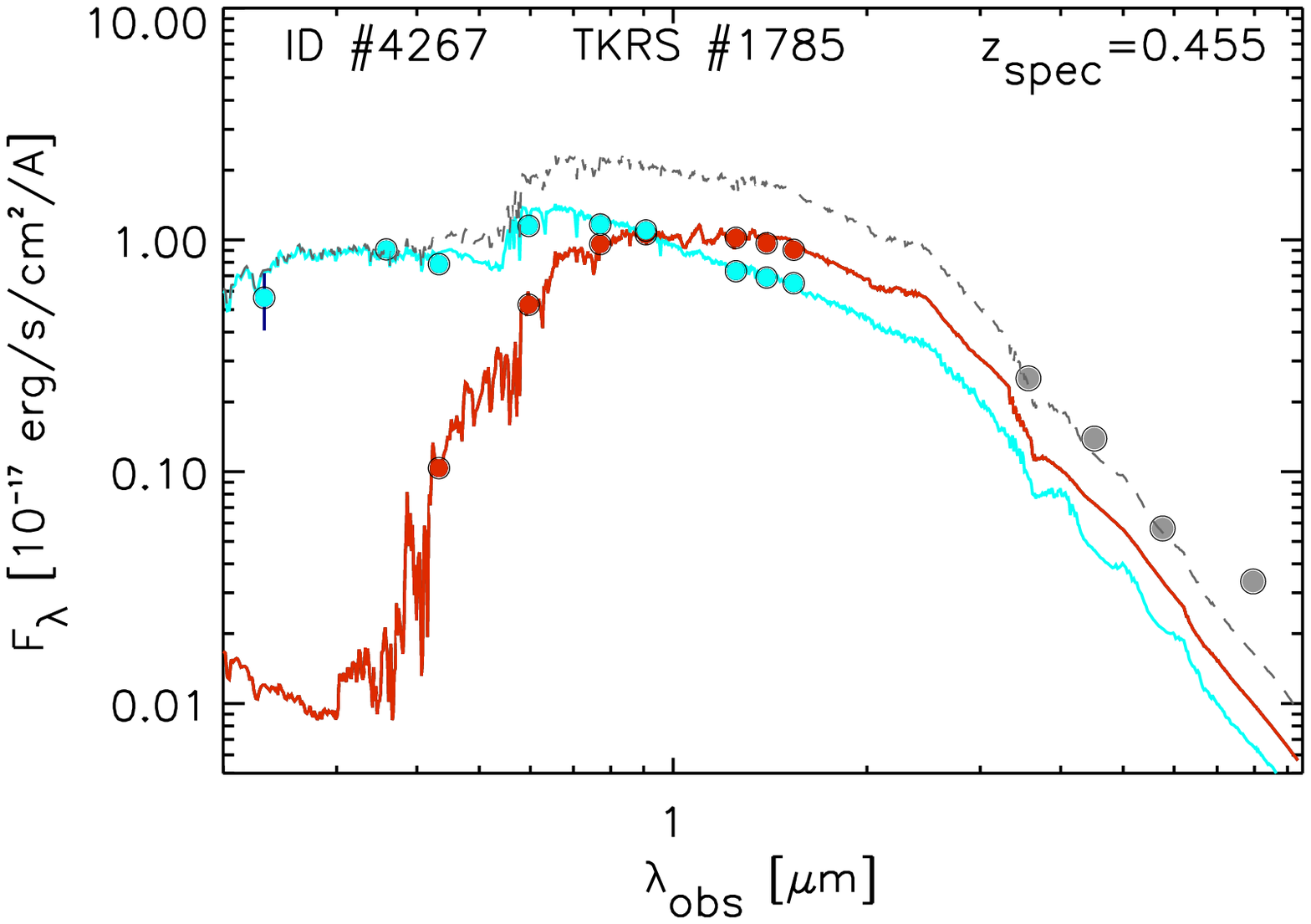}\\
        \includegraphics[width=0.35\textwidth]{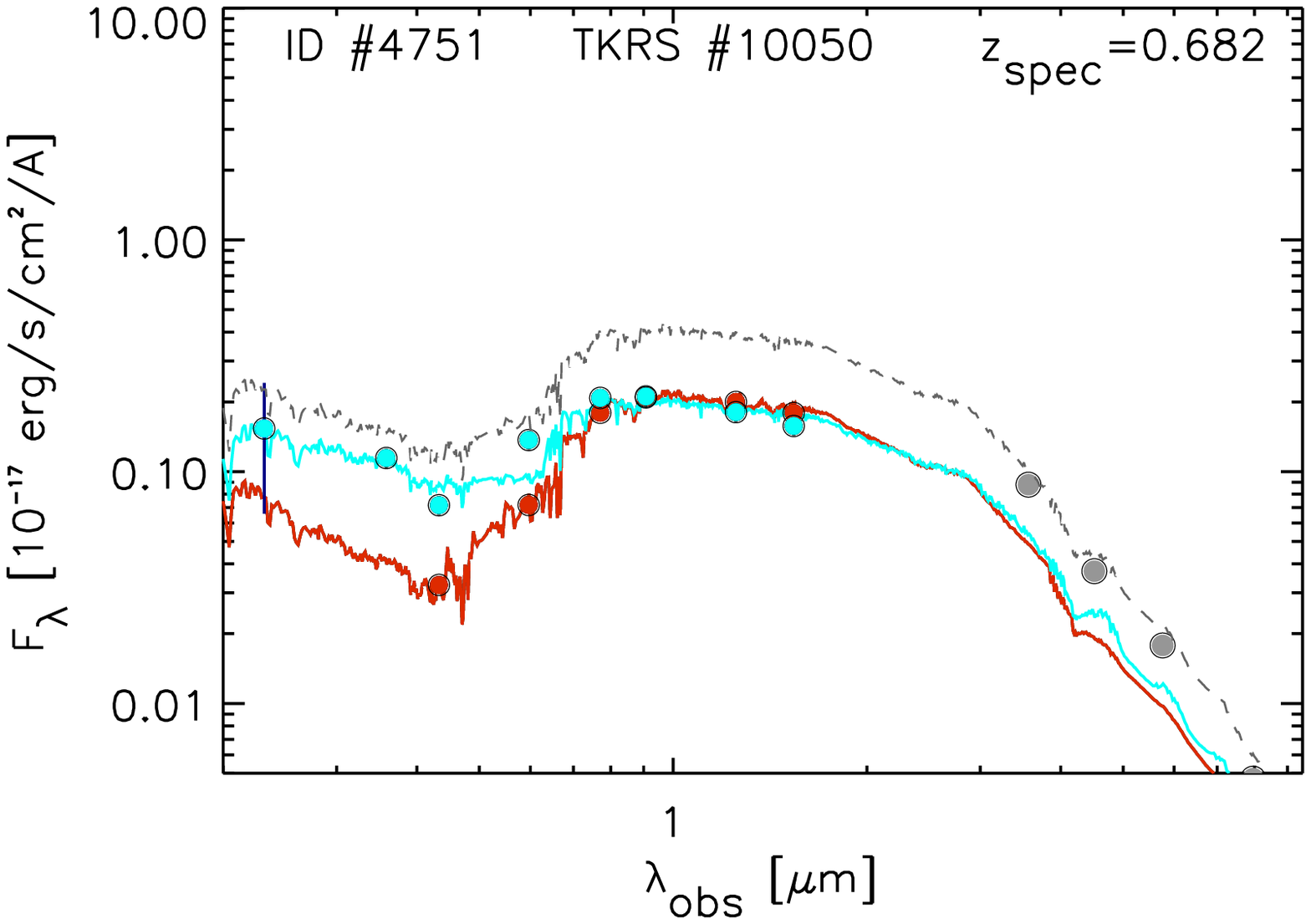}\includegraphics[width=0.35\textwidth]{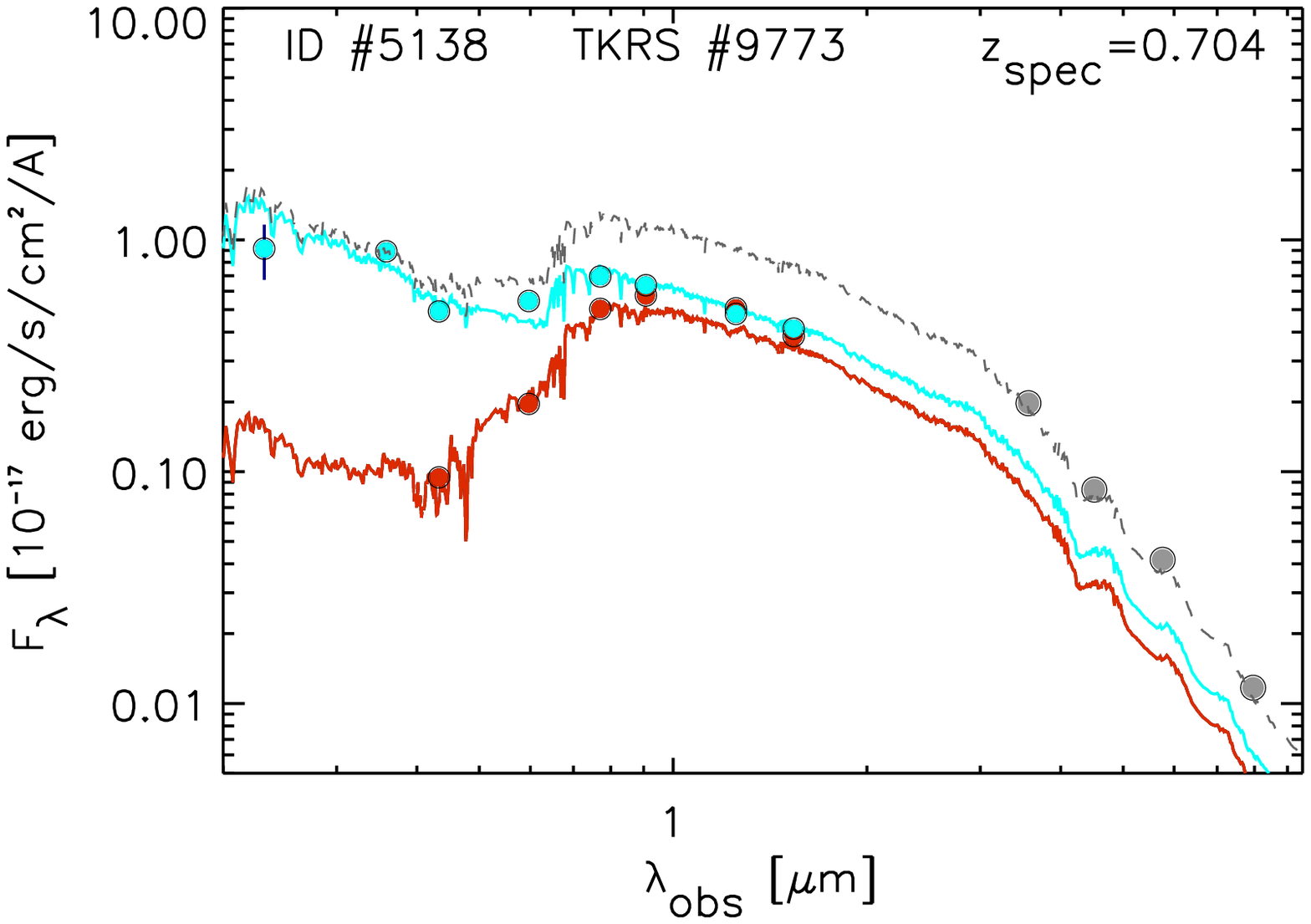}\\
        \includegraphics[width=0.35\textwidth]{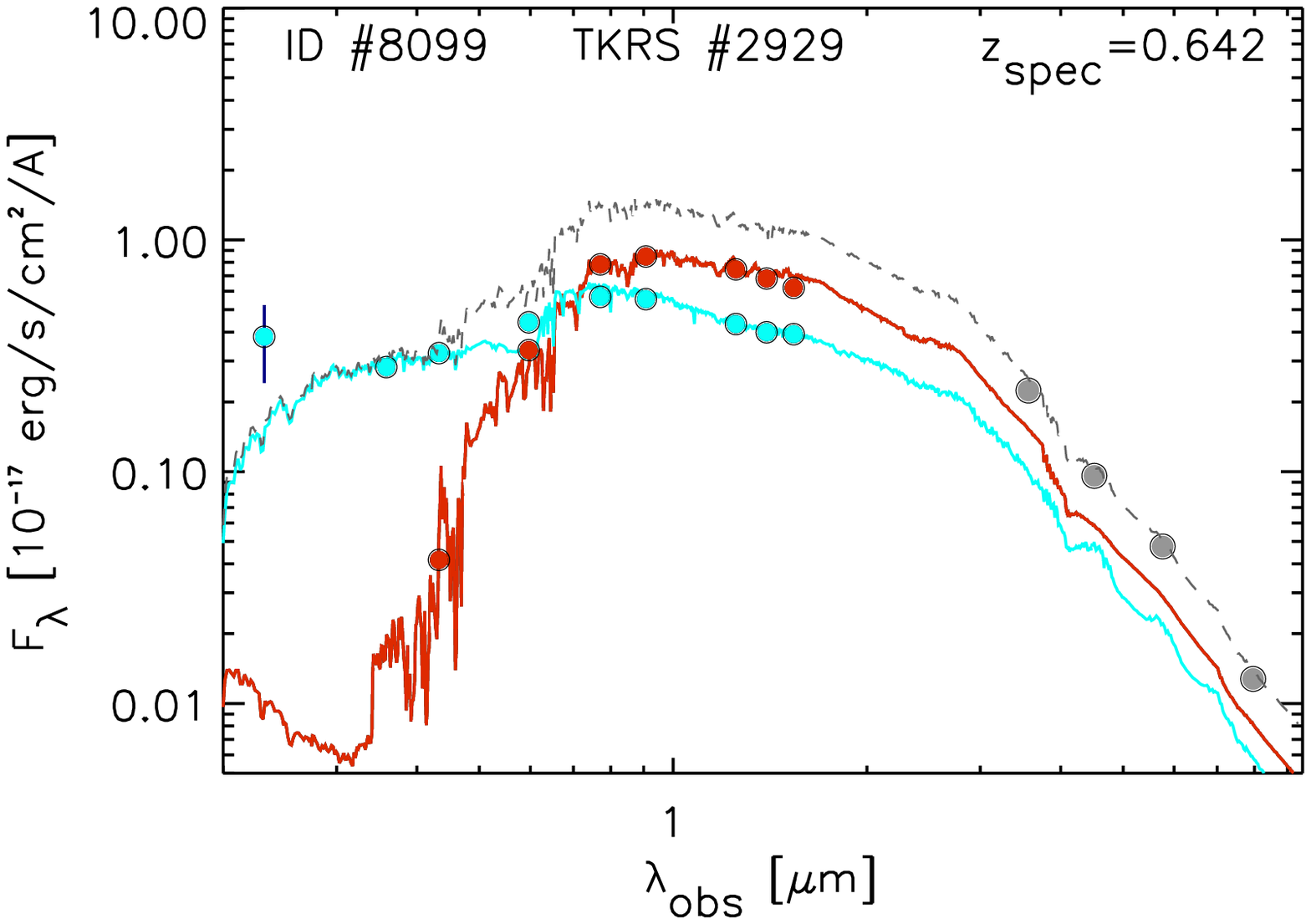}\includegraphics[width=0.35\textwidth]{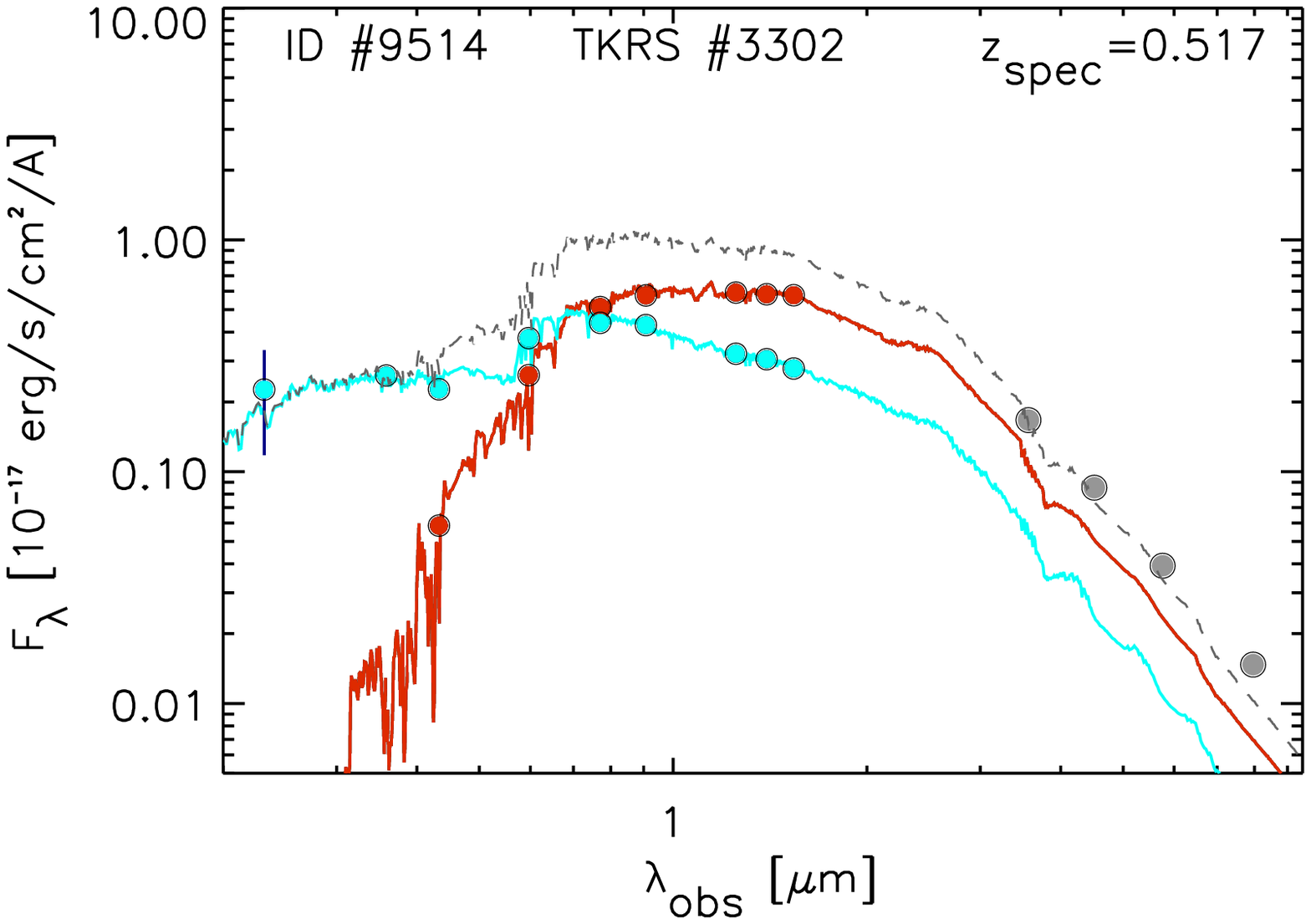}
        \includegraphics[width=0.35\textwidth]{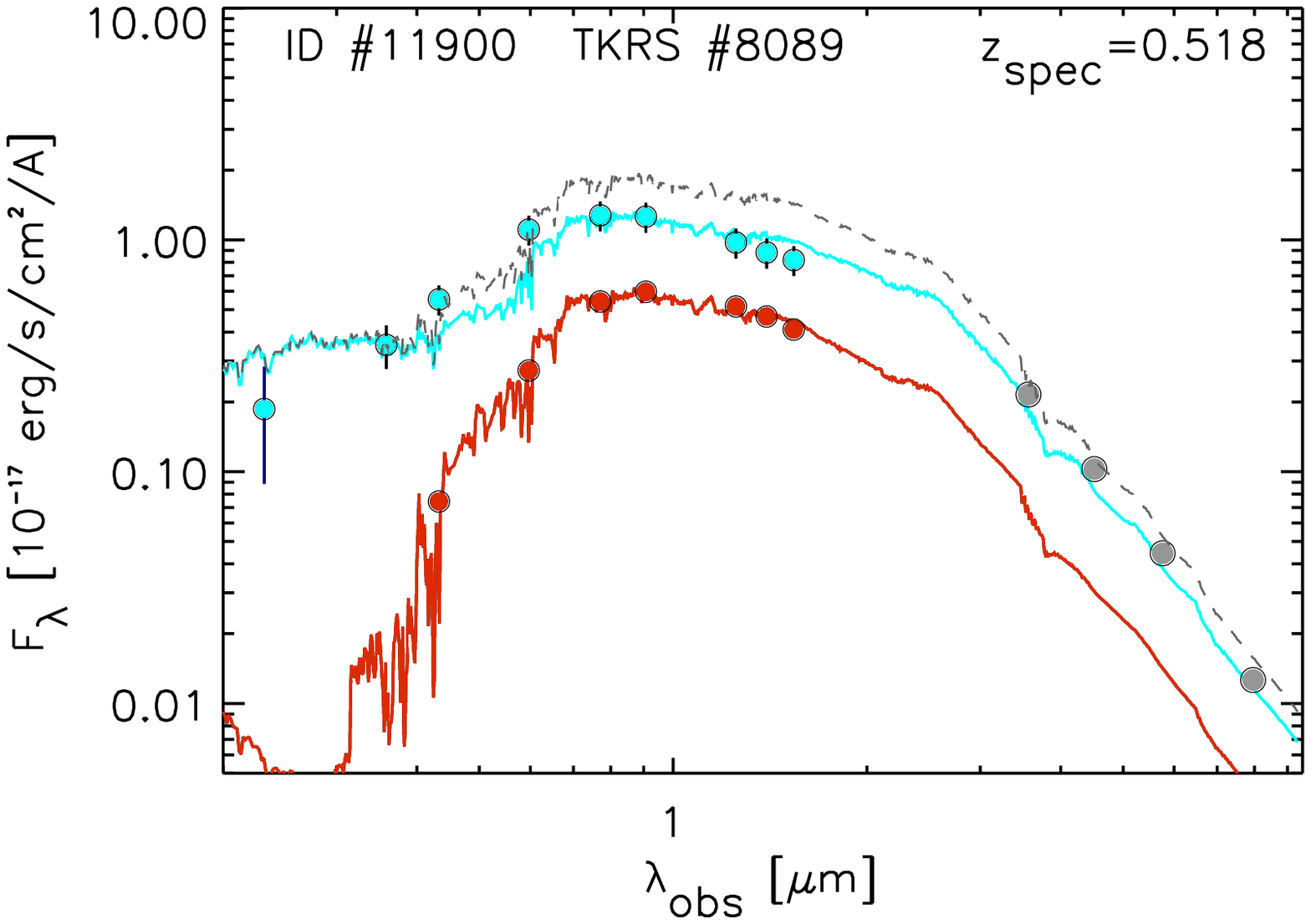}\includegraphics[width=0.35\textwidth]{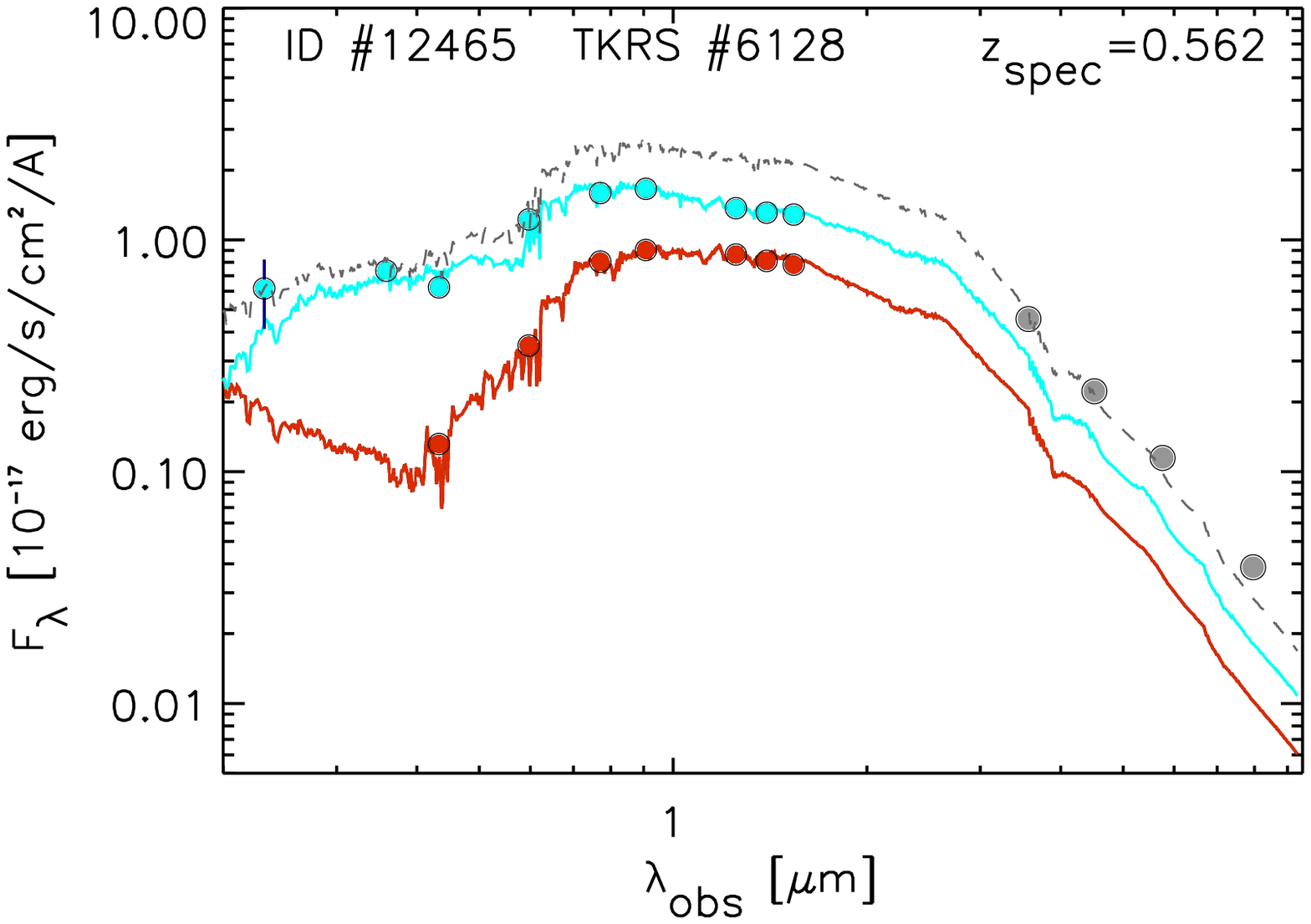}\\
        \includegraphics[width=0.35\textwidth]{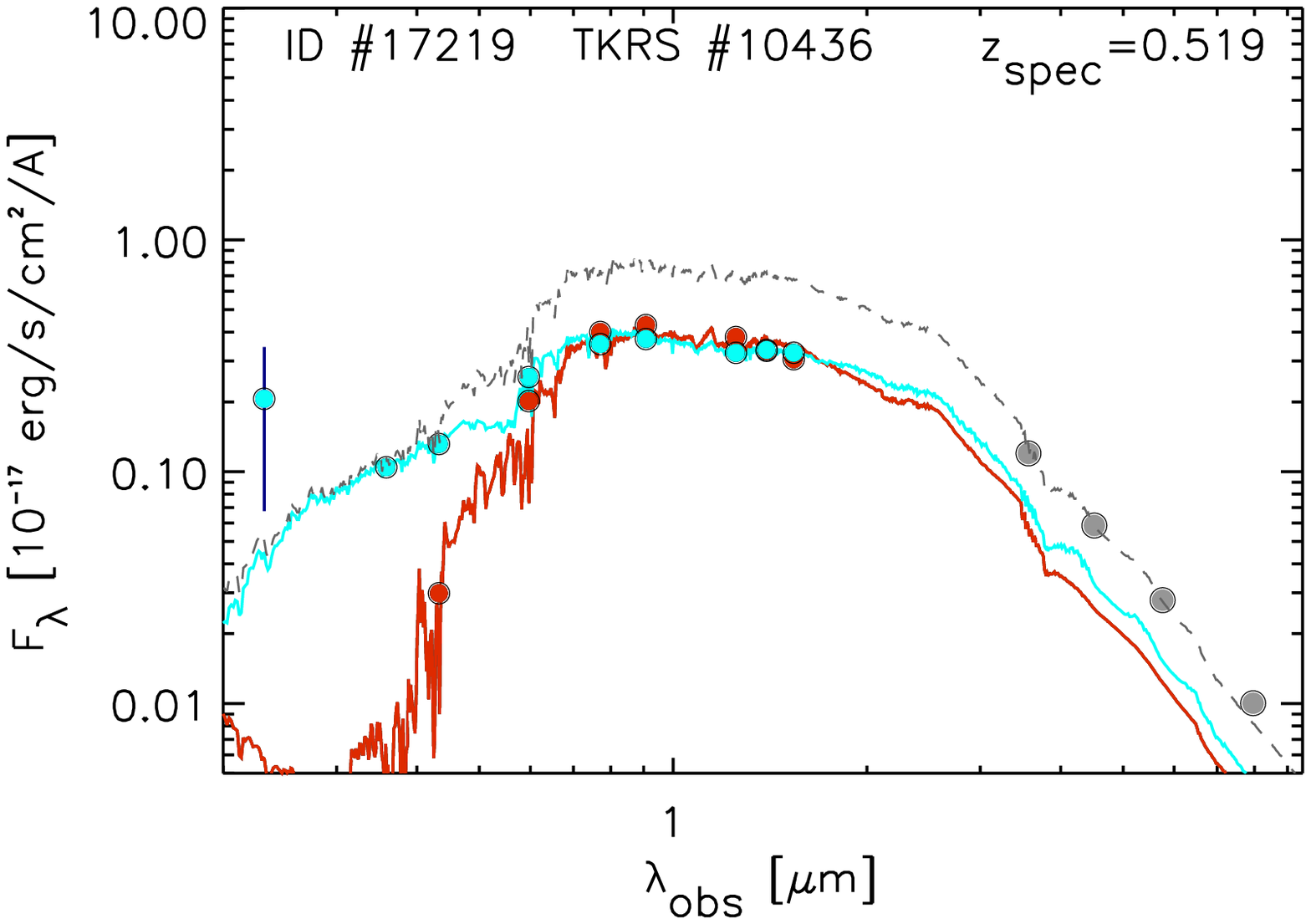}\includegraphics[width=0.35\textwidth]{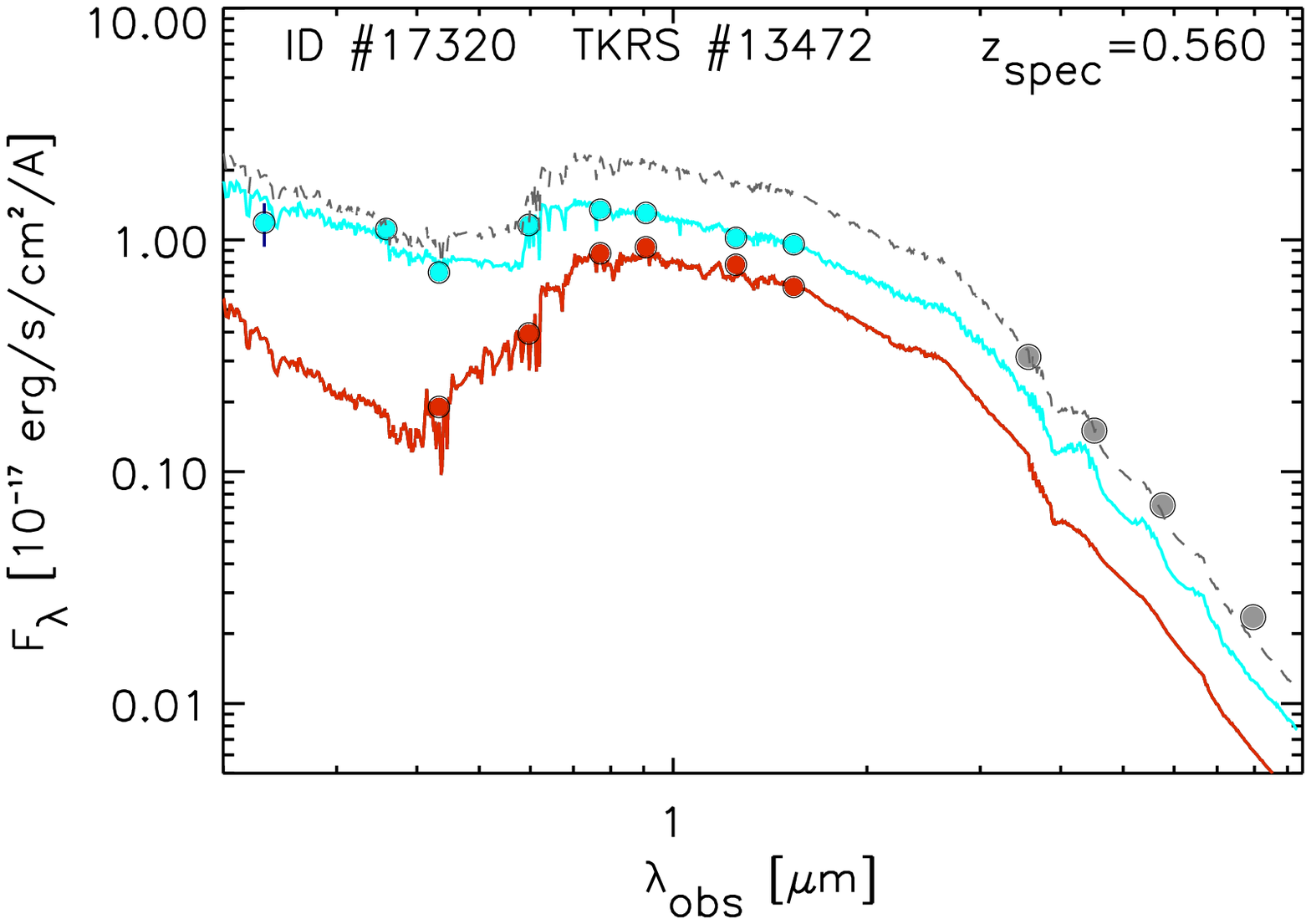}\\

    \caption{Bulge (red) and disc  (cyan) SEDs obtained by fitting the multi-band photometry derived from the B/D decomposition. The $2<\lambda_{rest}<4$ fluxes derived from the sum of the bulge and disc  best-fit models (grey dashed SEDs) are in good agreement (within 10\%) with the observed IRAC photometric data-points (grey filled circles) from \citet{2015ApJ...807..141P}.}
\label{fig:seds}
\end{figure*}

\begin{figure}
\includegraphics[width=\columnwidth]{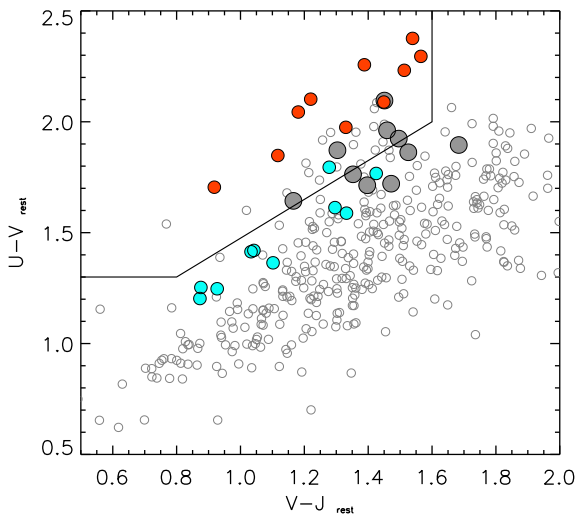}
  \caption{The rest-frame U-V vs. V-J colours of bulges (red filled circles) and discs (cyan filled circles) in our sample, derived from the SED fitting analysis described in Section~\ref{sec:seds} are compared with those of the entire galaxies (grey filled circles), and of the parent sample of MIPS/VLA-detected galaxies at $0.45<z<1$ (L18, small grey open circles), taken from the 3D-HST catalogue~\citep{2014ApJS..214...24S}. It appears that such galaxies, with reduced sSFR w.r.t. the MS, lie over or close to the border-line between the quiescent and star-forming regions in the UVJ diagram, while after the decomposition, bulges and discs show typical colours of quiescent and star-forming objects, respectively (with only one exception, see details in Section~\ref{sec:uvj}).} 
  \label{fig:uvj}
\end{figure}

\begin{figure*}
\includegraphics[width=\textwidth]{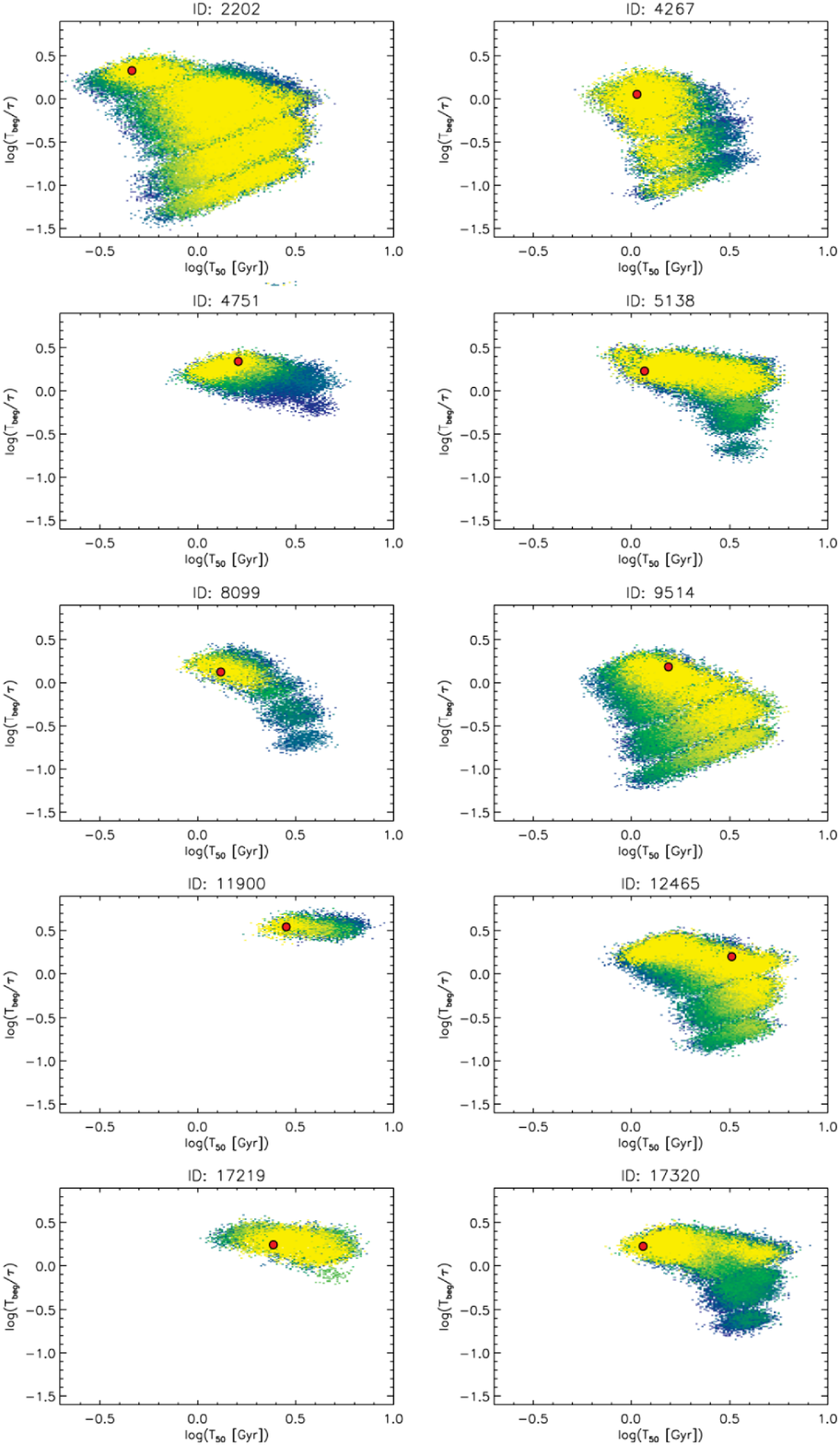}\\
\caption{Confidence levels at 99\% (blue), 90\% (green), and 68\% (yellow) for T$_{\rm beg}$/$\tau$ {\it vs} T$_{50}$, in logarithmic scale. The red filled circle shows the best fit solution. The coloured areas were obtained by smoothing the originally discrete data distribution along the x and y axes, using a Gaussian filter with  $\sigma$ similar to the average distance among the data points ($\sigma \sim 0.05$).}\label{fig:contours}
\end{figure*}


\begin{figure}
 \includegraphics[width=\columnwidth]{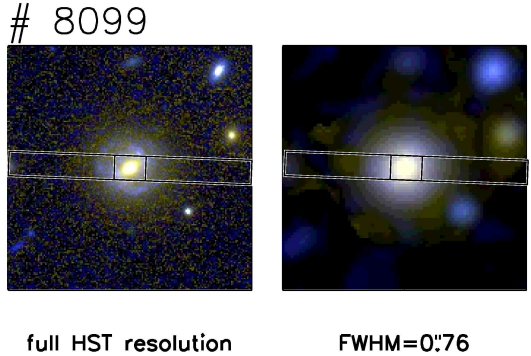}    
\caption{HST colour image in full resolution (left) and smoothed to the TKRS spatial resolution (right). The slit position and the smaller window from which the bulge spectrum was extracted is also shown.}
    \label{fig:ori_smooth}
\end{figure}


\begin{table*}
\caption{SED fitting best-fit parameters, and the relative 68\% confidence intervals, for the discs.}
\begin{tabular}{|r|r|r|r|r|r|r|r|r|r|r|r|r|}
\hline
\hline
ID   & $A_V$ & $A_V$(68\%) &$T_{50}$ &$T_{50}$(68\%) & T$_{\rm beg}$ &T$_{\rm beg}$(68\%)& $\tau$&$\tau$(68\%)& $\frac{T_{\rm beg}}{\tau}$ & $\frac{T_{\rm beg}}{\tau}$(68\%) & log$\frac{M_*}{M_{\odot}}$ & log$\frac{M_*}{M_{\odot}}$(68\%) \\
     & [mag] &[mag]             &[Gyr]   &[Gyr]               &                     &                                &                         &                                      \\
\hline
  2202 &  1.05 &  0.45$-$1.15 & 0.46 &  0.38$-$2.96 &  0.64 &  0.55$-$5.25 &  0.3 &  0.3$-\infty~$&  2.14 &  0.00$-$2.69 &  10.46 &  10.34$-$10.74 \\
  4267 &  0.65 &  0.50$-$0.75 & 1.07 &  0.75$-$1.68 &  1.70 &  1.18$-$3.31 &  1.5 &  1.0$-$30~    &  1.13 &  0.09$-$1.80 &  10.14 &  10.10$-$10.24 \\
  4751 &  1.10 &  1.10$-$1.40 & 1.61 &  1.08$-$1.70 &  2.20 &  1.59$-$2.35 &  1.0 &  1.0$-$1.0    &  2.20 &  1.61$-$2.30 &  10.21 &  10.15$-$10.27 \\
  5138 &  0.60 &  0.15$-$0.60 & 1.17 &  0.97$-$4.43 &  1.70 &  1.28$-$6.75 &  1.0 &  0.5$-$5.0    &  1.70 &  1.10$-$2.56 &  10.43 &  10.32$-$10.70 \\
  8099 &  0.80 &  0.75$-$0.80 & 1.31 &  1.08$-$1.96 &  2.00 &  1.60$-$3.11 &  1.5 &  1.0$-$3.0    &  1.33 &  1.00$-$1.70 &  10.37 &  10.33$-$10.46 \\
  9514 &  0.70 &  0.70$-$0.85 & 1.55 &  1.04$-$3.82 &  2.30 &  1.54$-$6.87 &  1.5 &  1.0$-$15~    &  1.53 &  0.38$-$1.80 &  10.02 &  9.94$-$10.22  \\
 11900 &  0.55 &  0.50$-$0.60 & 2.83 &  2.58$-$3.09 &  3.50 &  3.24$-$3.76 &  1.0 &  1.0$-$1.0    &  3.50 &  3.25$-$3.75 &  10.72 &  10.65$-$10.75 \\
 12465 &  0.80 &  0.75$-$1.40 & 3.23 &  1.09$-$4.83 &  4.75 &  1.61$-$7.75 &  3.0 &  1.0$-$15~    &  1.58 &  0.52$-$2.40 &  10.97 &  10.77$-$11.09 \\
 17219 &  1.05 &  0.85$-$1.10 & 2.43 &  1.94$-$5.03 &  3.50 &  2.65$-$7.49 &  2.0 &  1.5$-$5.0    &  1.75 &  1.20$-$2.50 &  10.26 &  10.22$-$10.43 \\
 17320 &  0.95 &  0.40$-$0.95 & 1.15 &  1.09$-$4.72 &  1.68 &  1.60$-$7.12 &  1.0 &  1.0$-$5.0    &  1.68 &  1.25$-$2.30 &  10.60 &  10.55$-$10.82 \\

\hline
\end{tabular}
\label{tab:seds}
\end{table*}

\begin{table*}
\caption{SED fitting best-fit parameters, and the relative 68\% confidence intervals, for the bulges.}
\begin{tabular}{|r|r|r|r|r|r|r|r|r|r|r|r|r|}
\hline
\hline
ID   & $A_V$ & $A_V$(68\%) &$T_{50}$ &$T_{50}$(68\%) & T$_{\rm beg}$ &T$_{\rm beg}$(68\%)& $\tau$&$\tau$(68\%)& $\frac{T_{\rm beg}}{\tau}$ & $\frac{T_{\rm beg}}{\tau}$(68\%) & log$\frac{M_*}{M_{\odot}}$ & log$\frac{M_*}{M_{\odot}}$(68\%)\\
     & [mag] &[mag]             &[Gyr]   &[Gyr]               &                     &                                &                         &                                      \\
\hline
2202 &  0.60 &  0.50$-$0.60 &  4.15 &  2.80$-$4.89 &  4.50 &  2.87$-$5.24 &  0.5 &  0.1$-$0.5 &   9.00 &  8.00$-$42.50  &  11.34 &  11.25$-$11.42 \\
4267 &  0.35 &  0.35$-$0.40 &  8.43 &  6.34$-$8.42 &  8.50 &  6.41$-$8.49 &  0.1 &  0.1$-$0.5 &  85.00 &  16.50$-$85.00 &  10.92 &  10.90$-$10.92 \\
4751 &  0.35 &  0.35$-$0.40 &  4.56 &  4.32$-$4.78 &  5.25 &  5.01$-$5.47 &  1.0 &  1.0$-$1.0 &   5.25 &  5.25$-$5.25   &  10.45 &  10.45$-$10.47 \\
5138 &  0.30 &  0.00$-$0.50 &  1.14 &  1.08$-$4.83 &  1.28 &  1.26$-$5.52 &  0.2 &  0.2$-$0.5 &   6.39 &  5.36$-$6.67   &  10.33 &  10.28$-$10.55 \\
8099 &  0.00 &  0.00$-$0.05 &  6.04 &  4.64$-$6.89 &  6.25 &  4.71$-$7.24 &  0.3 &  0.1$-$0.5 &  20.83 &  12.00$-$67.50 &  11.00 &  10.96$-$11.04 \\
9514 &  0.60 &  0.40$-$0.60 &  3.90 &  2.88$-$7.29 &  4.25 &  2.95$-$7.99 &  0.5 &  0.1$-$1.0 &   8.50 &  7.00$-$35.00  &  10.65 &  10.56$-$10.78 \\
11900 & 0.05 &  0.00$-$0.15 &  4.65 &  2.44$-$7.25 &  5.00 &  2.52$-$7.95 &  0.5 &  0.1$-$1.0 &  10.00 &  7.50$-$52.50  &  10.50 &  10.38$-$10.60 \\
12465 & 0.45 &  0.45$-$0.60 &  6.71 &  4.51$-$6.70 &  7.75 &  5.20$-$7.74 &  1.5 &  1.0$-$1.5 &   5.17 &  5.17$-$5.50   &  10.99 &  10.92$-$11.01 \\
17219 & 0.15 &  0.10$-$0.30 &  7.93 &  4.65$-$7.92 &  8.00 &  4.72$-$7.99 &  0.1 &  0.1$-$0.5 &  80.00 &  13.50$-$80.00 &  10.56 &  10.52$-$10.59 \\
17320 & 0.00 &  0.00$-$0.30 &  4.81 &  1.95$-$6.70 &  5.50 &  2.30$-$7.74 &  1.0 &  0.5$-$1.5 &   5.50 &  5.00$-$5.50   &  10.70 &  10.54$-$10.80 \\
\hline
\end{tabular}
\label{tab:seds_B}
\end{table*}

\section{TKRS spectroscopy: the bulge spectra}\label{sec:b_spec}
To reliably reconstruct the SFHs of our sample galaxies, the ideal would be to derive the stellar ages of both the bulge and the disc components from rest-frame optical spectroscopy, by fitting separately their stellar continua. However, since our galaxies are characterised by massive compact bulges and fainter diffuse discs (cf. Figure~\ref{fig:sizes}), even in the relatively deep TKRS spectra, in the outer part of the galaxies (where the bulge component is negligible) only the brightest emission lines can be detected (if any), with no underlying continuum. 
Moreover, in some cases the situation is worsened by the fact that the slit orientation does not coincide with the source position angle (e.g., when a close neighbour was included in the slit to maximise the number of targets in TKRS).
For these reasons, we could exploit the TKRS spectra to derive the bulge stellar ages, while no further constrains can be given on the age of the discs. Hence, we only used the central parts of the spectra, where the stellar continuum is dominated by the bulge component. 

We extracted the 1D spectrum from smaller windows in the slits, with lengths of 1$\farcs$5-2$\farcs$, chosen as a compromise to include most of the bulge-light (50\% on average) with the lowest possible contamination by the disc component (30\%, on average, cf. Appendix \ref{app:decont}), and fixed widths of $1\farcs$0, as defined in the TKRS slitmask design. As an example, in Figure \ref{fig:ori_smooth} the slit position and the used extraction window for object \#8099 are shown both on the full-resolution HST colour image, and on the same image degraded to the spatial resolution of the ground-based DEIMOS.

\subsection{Individual spectra}\label{sec:bulge_fit}
We fitted the bulge spectra using the {\it SIMPLEFIT} code, a modified version of the IDL procedure used in \citet{2004ApJ...613..898T}, which allows  fitting together a linear combination of stellar population templates, at given kinematics  (by masking the [OII]$\lambda\lambda3727$, [OIII]$\lambda\lambda5007$, and the Balmer emission line regions). When present, nebular emission lines can also be fitted, after subtracting the best-fit continuum model. 
The code adopts the basic assumption that the galaxy SFH can be approximated as a sum of discrete bursts. Hence, to fit each spectrum, we used a set of single stellar population (SSP) models of 19 different ages, chosen to be equally spaced by $\sim$0.1 dex (i.e., 0.08, 0.1, 0.15, 0.2, 0.25, 0.3, 0.4, 0.5, 0.6, 0.8, 1.0, 1.25, 1.5, 2.0, 2.5, 3.0, 4.0, 5.0, 6.5 Gyr), from the MILES stellar libraries \citep{2010MNRAS.404.1639V}.
 
\begin{figure}
\includegraphics[width=\columnwidth]{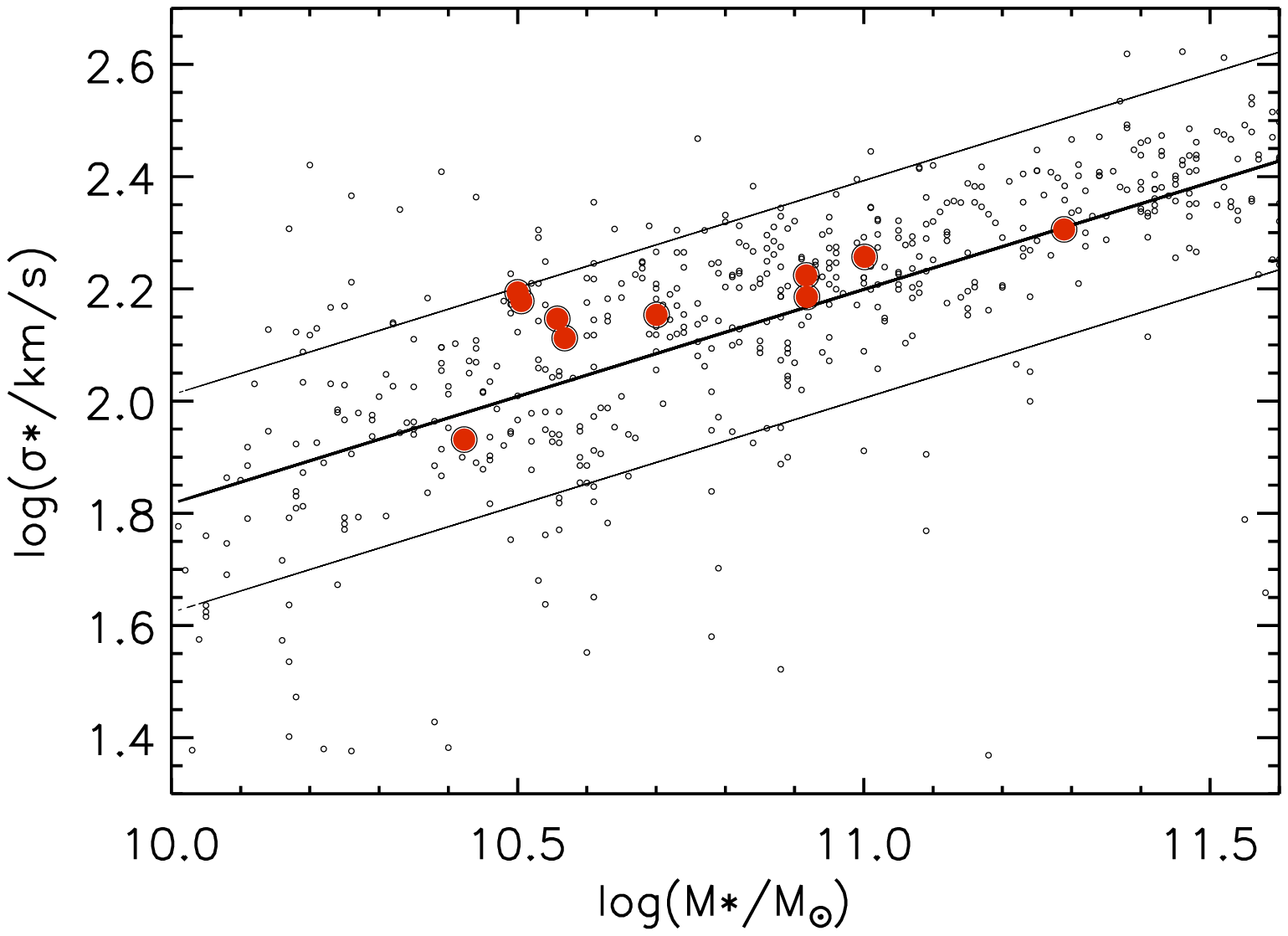}
\caption{Comparison of the empirically derived bulge velocity-dispersions with those measured in the local Universe by \citet{2013MNRAS.431.1383T} for ETGs of similar stellar masses. The black solid lines show the linear fit to the $M_*-\sigma_*$ in the considered mass range and its standard deviation.}
\label{fig:sigma_mstar}
\end{figure}

The resulting best-fit model is constructed by {\it SIMPLEFIT} as a linear combination of the input SSPs, with dust extinction modeled as an additional free parameter, using the attenuation curve of \citet[][]{2000ApJ...539..718C}.  
In particular, the code returns the fractional contributions of each SSP to the best-fit model 5500 \AA\ rest-frame luminosity ($L_{5500}$), from which we derived the light-weighted age, then converted into mass-weighted age of the bulges (through the corresponding $V$-band mass-to-light ratio,  M/L$_V$,  of the MILES models).   
Since {\it SIMPLEFIT} requires the velocity dispersion ($\sigma_*$) as input parameter in the fitting procedure, and the S/N of our spectra does not allow a solid measurement of this quantity, we estimated the $\sigma_*$ of the bulges from the virial relation, $\sigma_*= \sqrt{\frac{M_{\rm dyn}G}{5\re}}$ \footnote{ This corresponds to an estimate of the velocity dispersion measured within a radius \re (i.e, $\sigma_e$), which is a good proxy for our bulge spectra, extracted from rectangular apertures including on average $\sim$50\% of the bulge total light (see Appendix ~\ref{app:decont})}, following \citet{2006MNRAS.366.1126C}. To this purpose we used the circularised bulge effective radii (\re$_{, \rm circ}$= \re$\sqrt{\rm q}$) derived in Section~\ref{sec:bd}, and the bulge stellar masses from the SED fitting as a proxy of the dynamical masses ($M_{\rm dyn}$), assuming that the baryonic matter dominates the mass budget of the bulges as it does in ETGs inside the effective radius \citep{2007MNRAS.382..657T,2013MNRAS.432.1709C}. In Figure \ref{fig:sigma_mstar} we show that the bulge velocity dispersions estimated with this method are consistent with those derived from the kinematical analysis of similarly massive ETGs in the local Universe \citep[][i.e., 100$-$200 km/s]{2013MNRAS.431.1383T}. As a further check, we fitted the spectra with a grid of velocity dispersions, in the range $\sigma_*$= 100$-$200 km/s (with steps of 10 km/s), to evaluate the impact of $\sigma_*$ in the age determination, and found that the derived age is virtually insensitive to the chosen value of $\sigma_*$ (e.g., varying within the errors derived from simulations, as shown in Appendix~\ref{app:simul}).  

\begin{figure*}
\includegraphics[width=\textwidth]{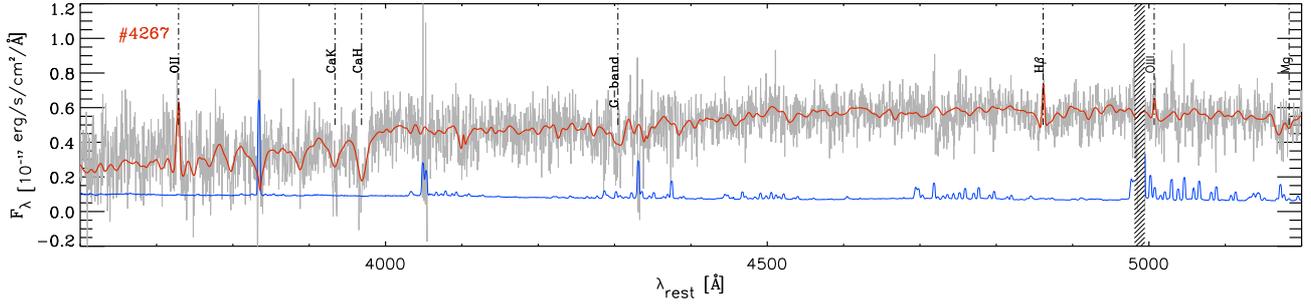}
\caption{The best-fit obtained with {\it SIMPLEFIT} for one of the individual bulge spectra using MILES/BAsTI models and single [Z/H]=0.06 metallicity. The spectrum and the best-fit continuum model are shown in grey and red, respectively, while the 1$\sigma$ uncertainties are shown in blue. The shaded area marks the largest masked region in the fit. The spectra and best-fits of the whole sample can be found as online material.}
\label{fig:single_spectra}
\end{figure*}

\begin{figure}
\includegraphics[width=\columnwidth]{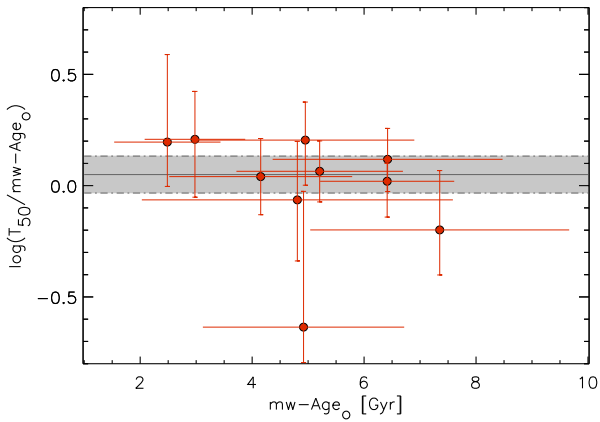}
\caption{Comparison of the bulge mass-weighted ages from individual `original' spectra (mw-Age$_{\rm o}$), derived as described in Section~\ref{sec:bulge_fit} and shown in Table ~\ref{tab:single_spec}, and the best fit ages from the best fit SEDs, T$_{50}$ (Section~\ref{sec:age_d}, and Table~\ref{tab:seds_B}). The grey shaded strip shows the mean logarithmic T$_{50}$/mw-Age$_{\rm o}$ ratio (solid black line) with its errors (standard deviation of the mean, dot-dashed black lines).}
\label{fig:sedvsspec}
\end{figure}

According to the Mass-Metallicity ($M_*-Z$) relation derived by \citet{2010MNRAS.404.1775T} for local ETGs, near solar metallicity ($Z_{\odot}$) should be appropriate for bulges with log($M_*$/\msun)$>10.3$, assuming that they behave like ETGs. Note that both our spectra and the SDSS ones used by Thomas et al. sample the central part of bulges/galaxies essentially within one effective radius, so we do not need to apply relative corrections to account for possible metallicity gradients. The bulge of our own Galaxy, with a mass of  $2\times 10^{10}\,\msun$ \citep{2016A&A...587L...6V}, exhibits a wide range of metallicities, from $\sim$ 1/10 to $\sim 3$ times solar, with an average close to solar \citep{2017A&A...599A..12Z} and is dominated by old ($\gsim 10$ Gyr) stellar populations \citep{2018ApJ...863...16R}. Thus, considering the average S/N of our spectra, to limit the degeneracies we fitted all ten bulges with a single metallicity, using the MILES templates with [Z/H]=0.06. This slightly super-solar metallicity corresponds on average to the highest allowed value for our bulge masses, considering the typical $M_*-Z$ relation scatter, so being the most conservative choice to account for age-metallicity degeneracy. Similar results are obtained by fixing the metallicity to the value predicted by the $M_*-Z$ for each bulge.
In Figure~\ref{fig:single_spectra} we show the spectrum and the best-fit model for object \#4267 (average S/N per pixel of $\sim$ 5.5), while the individual spectra and best-fits of the whole sample are available as online material. The light- and mass-weighted ages derived from the individual bulge spectra are given in the Appendix (Table~\ref{tab:single_spec}). 
From this analysis it appears that almost all the bulges of our bending galaxies are very old, with ages approaching the age of the Universe at the time of the observation. 
We should note that  several of the best fits also included a very young stellar population (0.08 Gyr), contributing $\lesssim$20\% to $L_{5500}$ (and $\lesssim 2\%$ to the stellar mass), that we can ascribe to the contamination from the central disc light falling into the window used to extract the spectra, which may be further enhanced by the effect of seeing smearing. This is justified by the results from the B/D decomposition, and SED fitting analysis, described in Sections \ref{sec:bd}, and \ref{sec:seds}, showing that at high HST resolution the bulges of these galaxies have colours consistent with fully quiescent objects (see Figure~\ref{fig:uvj}). 
We also tried  the exercise of subtracting the disc components from the TKRS spectra, using the information from the bulge and disc best-fit SEDs, as detailed in Appendix~\ref{app:decont}. The results obtained from the disc-decontaminated spectral fits showed that the contribution from the youngest SSP is substantially reduced and ages are increased by $\sim 1$ Gyr. 
However,  the procedure of disc subtraction is just based on the broad-band best-fit SED (the only one available for the disc), and further lowers the already modest S/N of our spectra.  Therefore, we prefer to stick primarily on the `original' (not `decontaminated') spectra for our analysis.
In Figure~\ref{fig:sedvsspec} we compare the mass weighted ages, derived from the `original' spectra, (mw-Age$_{\rm o}$) with the bulge half-mass ages from SED fitting, T$_{50}$ (Section~\ref{sec:age_d}). This comparison is meaningful, since the half-mass ages and mass-weighted ages are very close quantities\footnote{We also verified on our data that the bulge mass-weighted ages derived from the best fit SEDs are slightly lower, but very similar to the half-mass ages reported in Figure~\ref{fig:sedvsspec} and Table \ref{tab:seds_B} (log(T$_{50}$/mw-Age$_{\rm SED}$)$\sim (0.016\pm 0.001)$ Gyrs, on average). The lw-Age$_{\rm SED}$ and mw-Age$_{\rm SED}$ are almost equivalent for most of the bulges, as a consequence of the very short $\tau$ of the best-fit SEDs.} \citep[as shown e.g. by][]{2017A&A...602A..35T}. The bulge ages from SEDs and spectra are in good agreement within the errors, with an average ratio of log(T$_{50}$/mw-Age$_{\rm o}$)=$0.05\pm 0.08$ (grey strip in the figure). The error bars on the mw-Age$_{\rm o}$ (from simulations, see Appendix~\ref{app:simul}), are quite large, due to the average S/N$\lesssim 5$ of the individual TKRS spectra. Errors on T$_{50}$ are the confidence intervals from the SED fitting, reported in Table~\ref{tab:seds_B}.

\begin{figure*}
\includegraphics[width=\textwidth]{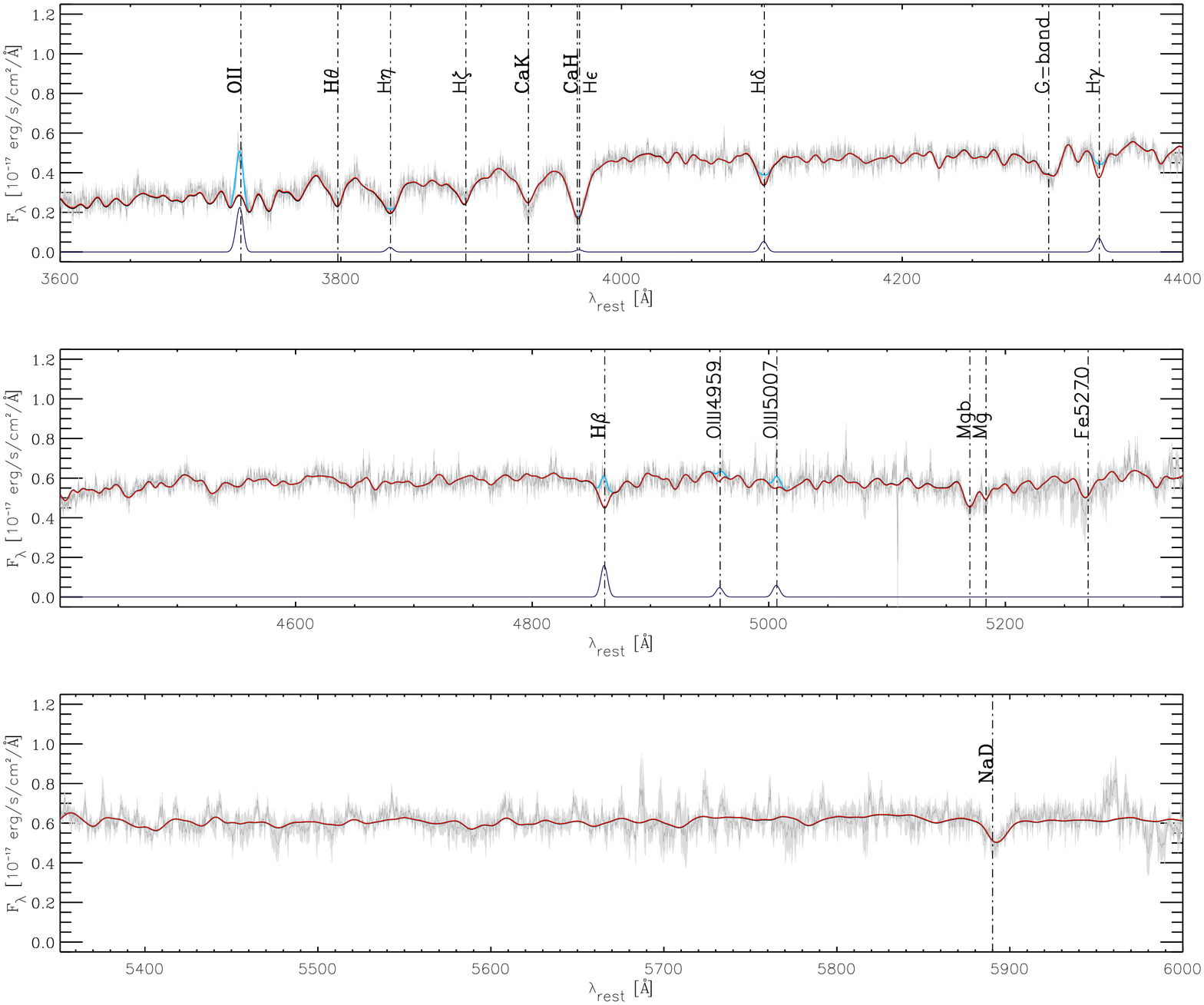}
\caption{The thin dark-grey line and the light-grey shaded area show the composite spectrum and its 1$\sigma$ error. The continuum best fit models obtained with single ([Z/H]=0.06) and double ([Z/H]=0.0, 0.22) metallicities, are shown by the black and the red lines, respectively, though the red line completely overlaps the black one, which remains unseen. The fitted emission lines are marked in cyan on the spectrum. Emission line fitting is also reported in the bottom part of the panel. The most prominent spectral feature are labelled.}\label{fig:stack_spec}
\end{figure*}

\subsection{Bulge stellar ages from the composite spectrum}
\label{sec:stack}
As mentioned above, our bulge spectra have relatively low S/N  that does not allow us to retrieve robust measurements of galaxy velocity dispersions ($\sigma_*$). Moreover, as we verified through simulations (see Appendix \ref{app:simul}), such typical S/N could lead to quite large uncertainties also in the derived light-weighted and mass-weighted ages and SFHs \citep[see also][]{2009ApJ...704L..34C,2016A&A...592A..19C,2018MNRAS.478.2633G}.
Hence, to achieve a higher S/N we built a composite bulge spectrum by stacking the individual spectra, as detailed below.

All the spectra were brought back to the rest-frame wavelength, resampled to 0.3~\AA, and normalised to the same flux at $4300 <\lambda_{\rm rest} <4700~\AA$. The same procedure was applied to the noise spectrum in quadrature. Before the stacking, all the resampled spectra (and relative noise) were cut in wavelengths in the range 3400-6000 \AA, to exclude regions sampled by less than 6 objects. 
The spectra were coadded using a weighted mean, with weights equal to the inverse square of the noise at each wavelength.
Following \citet{2012ApJ...755...26O}, we  adopted the {\it jackknife} method to derive the final stacked spectrum $f_{\rm jn}$ and its 1$\sigma_{\rm jn}$ noise,
respectively from their Eq. (2) and (1). 
This procedure gives a more realistic noise estimate, e.g. accounting for spectral regions contaminated by atmospheric absorptions in individual spectra. 

This final composite spectrum reaches an average S/N$\sim 17$ per pixel (S/N$\sim 14$ in the 4000~\AA\ break region), enabling us to accurately estimate the central velocity dispersion. To this purpose, we used the penalised pixel fitting (PPXF) software \citep{2004PASP..116..138C,2017MNRAS.466..798C} and derived a velocity dispersion of $\sigma_*=$ 164 km/s, which is comparable to those expected for similarly massive bulges (see Section~\ref{sec:bulge_fit}, and Figure~\ref{fig:sigma_mstar}).
Then, we fitted the spectrum with SIMPLEFIT following the same procedure described in Section~\ref{sec:bulge_fit}, but by fixing the $\sigma_*$ to the measured value.   
As for the individual spectra (Section~\ref{sec:bulge_fit}), we used MILES models with a single slightly super-solar metallicity. 
However, since in the composite spectrum some metallicity-sensitive features are now well visible (e.g., Mgb 5170\AA, Fe 5270\AA, and NaD 5890\AA), we also explored to which extent the age-metallicity degeneracy could affect our results, by leaving the metallicity as a free parameter and fitting with templates of both solar ([Z/H]=0.0), and super-solar ([Z/H]=0.22) metallicity. 
The uncertainties on the best-fit parameters (mass-weighted age, mass-weighted metallicity, A$_V$, and SFH), were estimated through Monte Carlo simulations, by artificially adding Gaussian noise based on the error spectra at each wavelength pixel (i.e., by assuming that the errors are normally distributed). Then the 68th percentile intervals are derived from 1,000 realisations.

The composite spectrum and the best-fit results are shown in Figure \ref{fig:stack_spec}. It is worth noting that the SIMPLEFIT best-fit models from the single- and the double-metallicity fits (black and red curves, respectively) completely overlap each other, including for the metallicity-sensitive features, showing that a single metallicity [Z/H]=0.06 is well adequate to reproduce the spectrum. 
The data are fitted very well, including the Balmer lines that were masked during the fitting procedure, with the exception of their cores clearly affected by (weak) emission. The emission lines were then fitted on the residuals from the continuum model, and their best fit is shown in Figure \ref{fig:stack_spec} by the black line in the bottom part of the panels.
In Section~\ref{sec:sfhs} we try to derive (set constraints on) the SFH of the bulge and disc separately, using this composite spectrum of the bulges. 

Since the PPXF code allows one to estimate the $\sigma_*$ and fit the stellar population continuum at the same time, we have the opportunity to test the stellar population fitting results (mass-weighted age, mass-weighted metallicity, and SFH) against different fitting tools. We used the same templates, and two options for metallicity, as with the SIMPLEFIT code. A detailed description of the PPXF stellar population fitting procedure is given in Appendix~\ref{app:ppxf}. 
The PPXF best-fit models, derived with the two choices of metallicity, also overlap each other, and with the SIMPLEFIT best-fit curves, although they are not shown in figure \ref{fig:stack_spec} for simplicity.
The mass-weighted ages, and metallicities derived with the two different fitting codes and options for metallicity are reported in Table~\ref{tab:stack_spec}. The resulting mass-weighted age ranges from 5.7 to 6.3 Gyr, with SIMPLEFIT and PPXF giving fully consistent results. With both the codes, when leaving the metallicity as a free parameter in the fits, a slightly super-solar [Z/H] is recovered, in good agreement with the [Z/H]=0.06 adopted for the fixed-metallicity fit. We also verified that the results (SFH, mass-weighted age, mass-weighted metallicity) remain stable using a larger range of metallicities, e.g., including in the template library used in the fit those with the highest available metallicity among the MILES models ([Z/H]=0.40), and/or sub-solar metallicity templates.
As mentioned in Section~\ref{sec:age_d}, we checked that the use of TP-AGB heavy models in the spectral fitting does not substantially affect the results on the bulge mass-weighted ages and SFHs. We fitted the composite spectrum with the \citet[][]{2011MNRAS.418.2785M} high resolution models (with solar metallicity, and using the SIMPLEFIT software), and found that a population of $\sim 2$~Gyr stars contributes to half of the light at 5500~\AA, corresponding to $\sim 30 \%$ of the total stellar mass. However, the remaining $\sim 70\%$ of the mass is still accounted for by the oldest stellar population ($\sim 6.5$ Gyr). This would lead to derive mass-weighted ages only $\sim 15\%$ younger for our bulges, that would not affect the results of the paper.

\begin{table*}
\begin{center}
\begin{small}
\begin{tabular}{c|c|c|c|c|c}
\hline
Template &Fitting Code &Mass-weighted&Mass-weighted & $A_V$  \\
Metallicities&            & Age [Gyr]   & [Z/H][dex]      &  [mag]\\
\hline
\multirow{3}{*}{[Z/H] = [0.06]}           &SIMPLEFIT & $6.04^{+0.30}_{-0.30}$ &0.06&$0.71\pm 0.02$\\
                                &PPXF       & $6.32^{+0.12}_{-0.28}$ &0.06&$ 0.52\pm 0.02$\\
\hline
\multirow{3}{*}{[Z/H] = [0.0, 0.22]}       &SIMPLEFIT &$5.77^{+0.19}_{-0.42}$ & 0.04$\pm$0.03& 0.70$\pm0.02$\\
                                &PPXF       &$5.73^{+0.48}_{-0.69}$  &$0.05\pm0.03$&$ 0.52\pm0.02$ \\
\hline
\end{tabular}
\caption{Bulge best-fit parameters derived from the spectral fitting with SIMPLEFIT and PPXF described in Section~\ref{sec:stack}. The columns from left to right show: 1) metallicities of the templates used in the fit; 2) Used fitting code; 3) Mass-weighted age; 4) Mass-weighted metallicity 5) Reddening parameter. The reported uncertainties are drawn from the simulations described in Sections~\ref{sec:stack}, and Appendix ~\ref{app:ppxf}, for SIMPLEFIT and PPXF, respectively, as the 16th and 84th percentiles of the parameter distributions.}\label{tab:stack_spec}
\end{small}
\end{center}
\end{table*}

\begin{center}
\begin{figure*}
\includegraphics[width=0.5\textwidth]{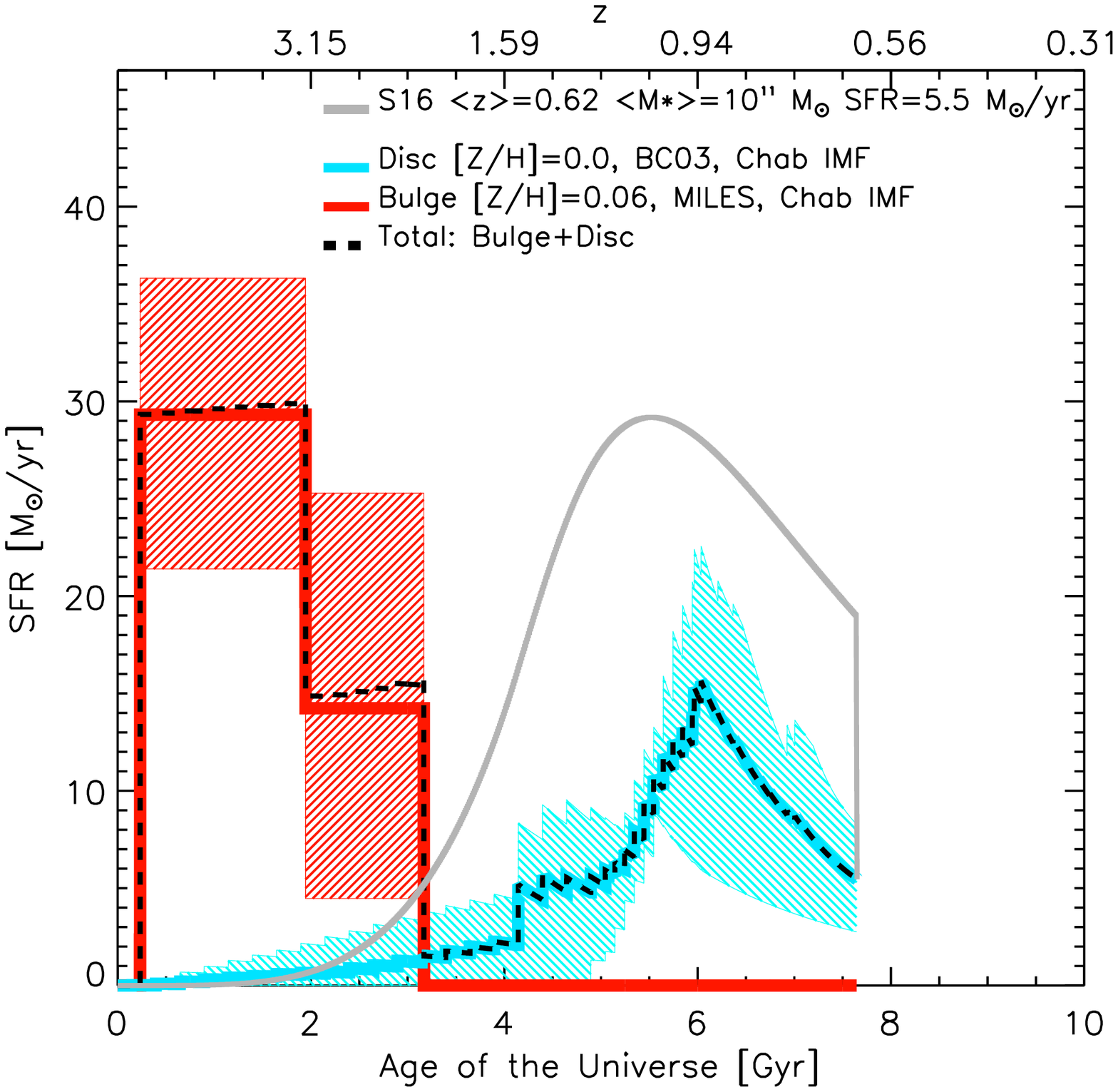}\includegraphics[width=0.5\textwidth]{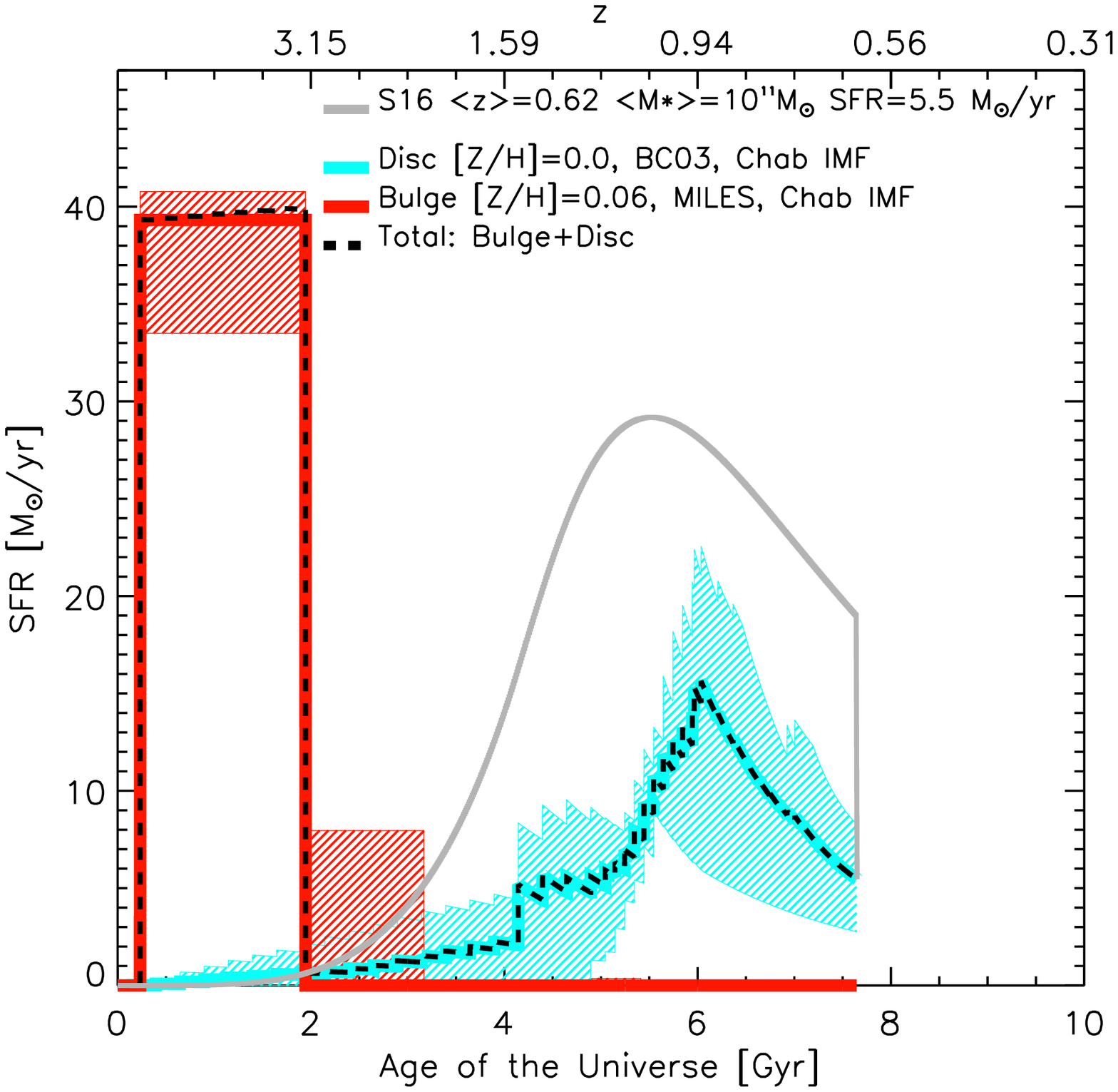}\\
\caption{Average SFH of our sample galaxies at $\langle z\rangle=0.62$, obtained as the sum (black dashed line) of the bulge (red) and disc (cyan) SFHs, from the composite spectrum, and the broad-band SED fitting, respectively. The results derived by fitting the bulge spectrum with the SIMPLEFIT and PPXF routines, and single metallicity (Z/H=0.06), are shown respectively in the left, and right panels. The coloured shaded areas are the confidence intervals (16th and 84th percentiles) for the bulge and disc SFHs. In both panels the stellar mass, and the bulge to total mass ratio have been renormalized to $M_*= 10^{11} \msun$, and B/T$_m\sim$0.67, and SFR $\sim 5.5 \msun/yr$ (average properties of our sample). The grey curve shows the SFH of a galaxy with the same $M_*$ and SFR that would have evolved along the MS throughout its life, according to the parametrization of S16. The last vertical part of the grey curve accounts for the fact that the galaxy $\sim 2.5$ times lie below the MS at the time of observation. }\label{fig:sfh}
\end{figure*}
\end{center}

\begin{figure*}
\includegraphics[width=0.5\textwidth]{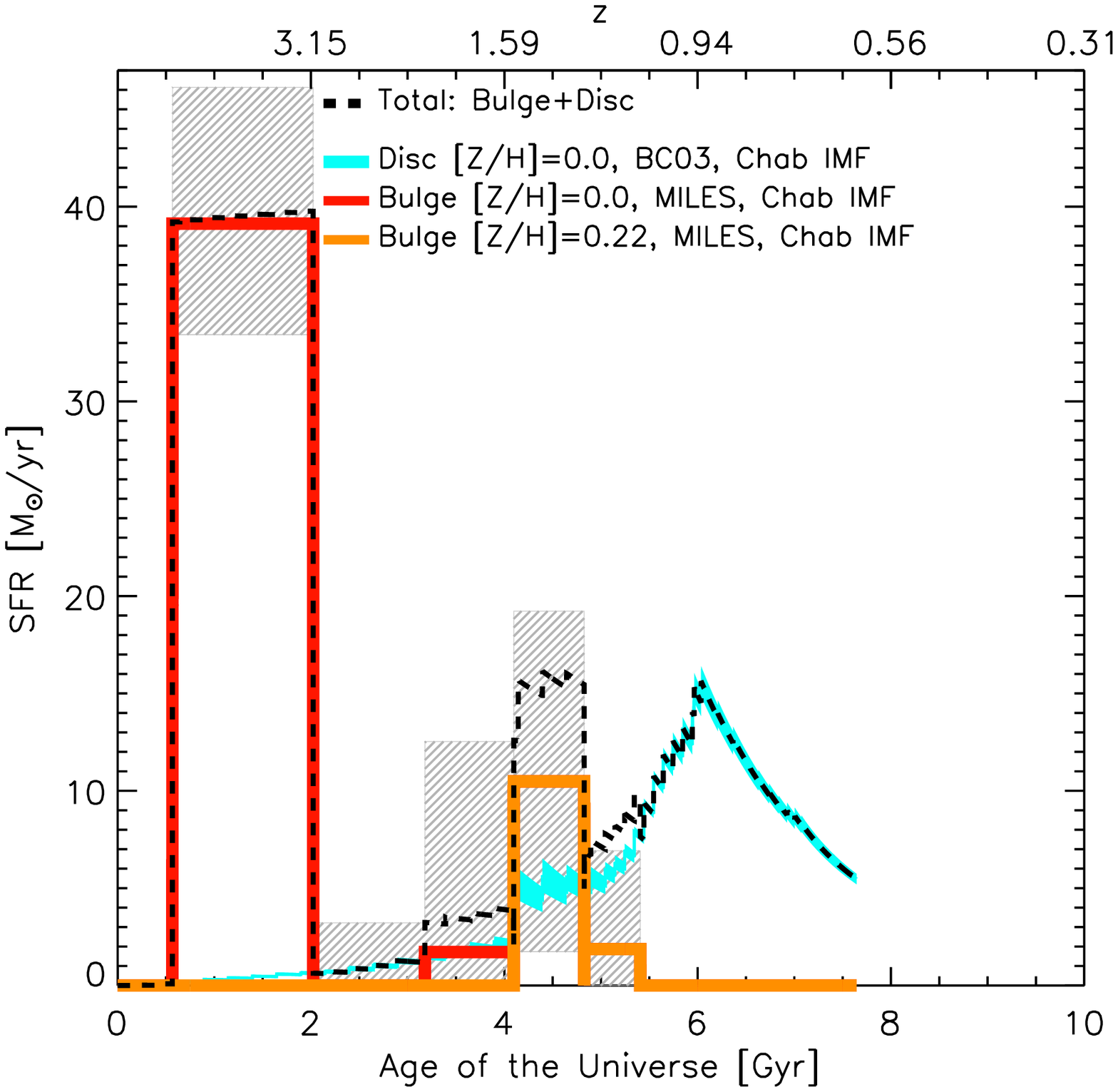}\includegraphics[width=0.5\textwidth]{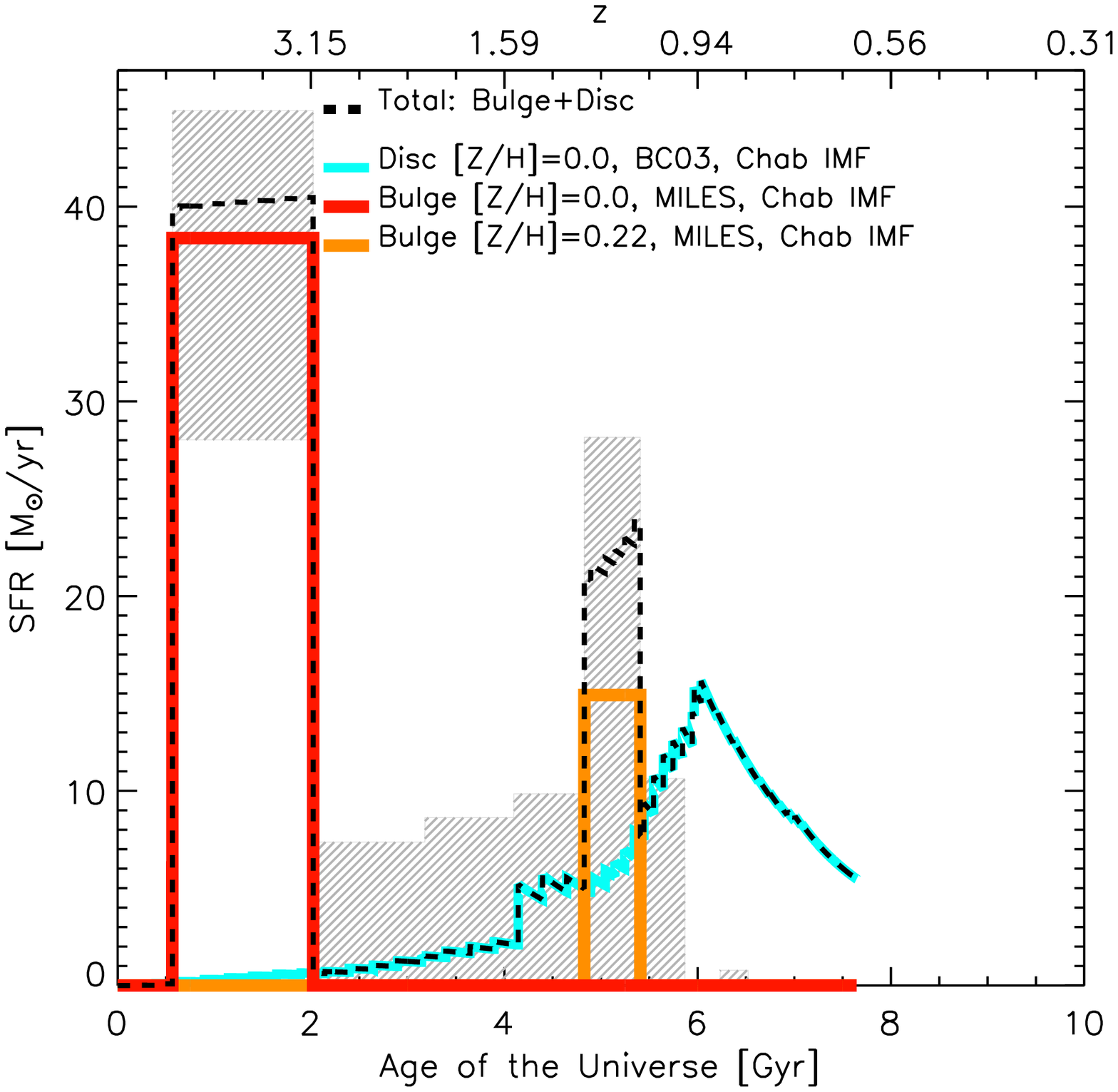}\\
\caption{As in Figure ~\ref{fig:sfh}, the average SFH of our sample is shown (black dashed line), where the bulge SFH is the result of the spectral fitting allowing a combination of both solar ([Z/H]=0.0), and super-solar ([Z/H]=0.22) metallicities. The mass fractions relative to the two populations are shown in red, and orange, respectively. The left and right panels present the results from SIMPLEFIT, and PPXF routines. Only the confidence intervals (16th and 84th percentiles) for the bulge SFH are shown here, for simplicity (grey shaded areas).}\label{fig:sfh2}
\end{figure*}

\section{Star Formation Histories}\label{sec:sfhs}
We reconstructed the average SFH of our sample galaxies over the period from cosmic time $t=0$ to $t\simeq 7.5$ Gyr, corresponding to $z=0.62$ (the average redshift of the sample). This average SFH is obtained as the sum of the bulge and disc average SFHs, inferred as detailed in the following. 

\noindent {\bf Bulge.} 
The average bulge SFH was derived from the best fit of the stacked spectrum, after applying the {\it jackknife} method mentioned in Section ~\ref{sec:stack}. 
The SFH of the bulge composite stellar population (CSP) is here given as the fraction of stellar mass formed in different bursts (SSPs) as a function of cosmic time. 
Hence, the procedure adopted by both {\it SIMPLEFIT}, and PPXF to fit the bulge spectrum with a linear combination of 19 different SSPs (see Sections ~\ref{sec:bulge_fit}, and ~\ref{sec:stack}), directly provides its SFH, giving the fraction (or `weights') of the bulge stellar mass formed in each of the 18 time bins.

\noindent {\bf Disc.}
The average disc SFH of the sample was derived based on the SED fitting results for the individual objects. As detailed in Section~\ref{sec:seds}, we used declining $\tau$-models in the SED fitting analysis, that are suitable to estimate the galaxy $M_*$, mean age, and dust extinction. However, the characteristic shape, with the SFR peaking at the onset of the star formation, unlikely represents the true SFH of  star-forming systems, not even of those with reduced sSFR.
To overcome this problem, we used the probability distribution derived from the SED-fit $\chi^2$ test to reconstruct the average SFH of the discs, as briefly described in the following. 
For each disc, we averaged out all the SFHs from the {\it HyperZmass} solutions included in the yellow regions of Figure~\ref{fig:contours}, i.e., within the 68\% confidence levels, and retrieving a SFR consistent within the errors with the SFR(IR+UV). This is illustrated in more details in Appendix~\ref{app:disc_sfh}, and Figure~\ref{fig:discsfh}. The average disc SFH was then obtained by combining the SFHs thus obtained of the 10 discs.
We note that  for each object, both the current SFR and the disc $M_*$ (the latter well constrained by the photometry) span a relatively small range of values within the 68\% confidence intervals (cf. Table~\ref{tab:seds}), while the only parameters that can substantially vary are the star formation timescale, and the time since the beginning of star formation ($\tau$, and T$_{\rm beg}$, respectively). 

The results are reported in Figure~\ref{fig:sfh}, where the disc SFH is shown in cyan, and the bulge SFH, obtained with SIMPLEFIT (left panel), or PPXF (right panel), with a fixed [Z/H]=0.06 metallicity is shown in red. 
The SFH confidence intervals (16th and 84th percentiles) for the bulge (red shaded area) were drawn from Monte-Carlo simulations (cf., Section~\ref{sec:stack}), while those for the disc (cyan shaded area) were derived from the distribution of the 10 individual SFHs. 
It is worth to note that the shape of the SFH of our average disc is similar to a delayed-$\tau$ model, rising up to a peak value, and then declining. However, the error bars in Figure \ref{fig:sfh} show that a flat SFH at late times is also an acceptable result. This is related to the fact that, as mentioned in Section \ref{sec:age_d}, for the individual discs only in very few cases solutions with T$_{\rm beg}\lesssim \tau$ can be significantly rejected (cf., Figure \ref{fig:contours}, and Table \ref{tab:seds}).

In each panel, the dashed black line marks the total SFH, derived by summing the disc and the bulge components, normalised to the case of an average galaxy in our sample, at $z=0.62$, with total (B+D) $M_*=10^{11}\msun$, B/T$_{\rm mass}$=0.67, and SFR=5.5~$\msun$/yr. The total SFH thus obtained is compared with that inferred by integrating the SFR of our average galaxy back in time, by assuming that it would have spend all its life on the bending main sequence of S15, according to the parametrization of S16 (grey curve in Figure~\ref{fig:sfh}). The results and implications of this comparison are discussed in the next section.

The impact on the global SFH of the use of different metallicities in the bulge spectral fit (cf., Section~\ref{sec:stack}) is also shown in Figure~\ref{fig:sfh2}. As in the previous figure, the results obtained with both SIMPLEFIT and PPXF are reported.

From Table~\ref{tab:stack_spec}, and Figure~\ref{fig:sfh} we note that our results are robust against the used fitting code since a good agreement (within the errors) is found between the {\it SIMPLEFIT} and the PPXF best-fit mass-weighted ages, metallicities, and SFHs. 
Regardless of the used code, the mass-weighted age obtained by leaving the metallicity as a free parameter are slightly younger than that derived with fixed metallicity ([Z/H]=0.06). This is due to the appearance at later time of a stellar population with higher metallicity (shown in orange in Figure~\ref{fig:sfh2}), compared to the [Z/H]=0.0 of the dominant older population. However, the super-solar metallicity burst is responsible for the formation of a negligible mass fraction, and we do not exclude that it is just the product of age-metallicity degeneracy in the fit. This further confirms that the choice of metallicity does not affect the results of this work. 

\section{Discussion}\label{sec:disc}
In this section we discuss the main implications of the results of this work for our understanding of the nature of the bending of the MS, which in turn has important implications for tracing the evolution and mass assembly history of galaxies in general. 

In the literature, the most popular interpretation of the MS bending is that it is a sort of transition phase, in which star-forming galaxies, once reached a certain critical mass ($M_0$, corresponding to the knee in the stellar mass function of star-forming galaxies), start to quench their star-formation, and leave the MS to evolve, along the green valley, towards the red sequence \citep{2015ApJ...801...80L,2016ApJ...817..118T,2016MNRAS.457.2790T}. 

In addition, the debate is open not only on which is the most efficient quenching mechanism (e.g., among AGN feedback, halo quenching, morphological quenching, etc.), but also on which process is able to switch off the star-formation in relatively long timescales (slow quenching), so to allow us to detect the star-forming-to-quiescent transition phase, i.e. the MS bending itself (cf., S16). 
Since the most massive star-forming galaxies are generally bulge-dominated, among the slow quenching processes, many authors favoured the ``morphological'' (or gravitational) quenching \citep{2009ApJ...707..250M,2014ApJ...796....7G}. In this scenario, the buildup of a central mass concentration (bulge) at later epochs would stabilise the gas disc  against fragmentation, thus preventing the formation of clumps, and gradually quenching the star-formation \citep{2015Sci...348..314T, 2014ApJ...788...11L, 2015ApJ...811L..12W}. 
Another proposed slow quenching mechanism, is the starvation (or strangulation) \citep{1980ApJ...237..692L,2000ApJ...540..113B,2006MNRAS.368....2D,2015Natur.521..192P}, in which the supply of cold gas to the galaxy is halted for some reason (e.g., via shock-heating of the circum-galactic medium once the host halo exceeds a critical mass, \citealt{2003MNRAS.345..349B}, or due to radio mode AGN feedback, \citealt{2006MNRAS.365...11C}), but the star-formation can continue until the remaining available gas is exhausted. 

A further possibility, is that the bending of the MS is not an evolutionary phase, undergone by the MS galaxies as a whole, but it is instead the result of averaging out a mixed population of systems, with different evolutionary paths and star formation histories. In this view, galaxies with reduced sSFR at the high-mass end of the MS could be either in the act of quenching (downturn), or in a rejuvenation process (upturn) following the accretion of new gas, e.g., through minor, or major mergers \citep{2017MNRAS.470.3946S}. 
One argument against this interpretation would be that at the high-mass end of the MS, statistically, one would expect that most of the galaxies (close to their critical mass, $M_0$) are turning down, rather than rising. 
On the contrary, alternative fast quenching mechanisms (such as quasar mode AGN feedback, \citealt{2004ApJ...600..580G}) may dominate, preventing us from revealing the majority of the galaxies on the way to be quenched.
Moreover, the progressive flattening of the high-mass end slope of the MS with decreasing redshift, observed by many authors, may support the hypothesis that many rejuvenated galaxies populate the MS bending, when considering both that the number density of quiescent galaxies (available to be ``rejuvenated'') strongly increases at $2<z<1$, and that minor mergers are suggested to drive the bulk of the morphological transformation in spheroid progenitors at $z < 1$ \citep{2018MNRAS.480.2266M}. 
As shown in the following, the results presented in this work provide new clues to shed light on these issues, although other questions remain open to debate. 
 
\subsection{The MS bending is not due to the {\it late growth} of bulges}
In Figure~\ref{fig:sfh} the SFH of our sample is compared with that of a galaxy with the same $z$, $M_*$, and SFR that would have evolved along the MS of S15 throughout its life, undergoing a slow downfall (grey solid line), according to the prescription of S16 \citep[see also][]{2017A&A...608A..41C}. Since a galaxy with similar properties lies below the ridge of the S15 MS at $z\sim 0.62$, when assuming a similar SFH one has to account for a further drop in SFR, caused by some event. This is crudely represented in Figure~\ref{fig:sfh} by the grey vertical line, since a more sophisticated parametrization is beyond the scope of this paper. It appears from this figure that the average SFH of the ``bending galaxies'' in our sample (black dashed line) is remarkably different from that predicted for a galaxy evolving along a bending MS of S15.  
The main difference between the two SFHs is that while for the grey curve almost all the galaxy stellar mass was assembled within the last 3 Gyrs, in our galaxies $>$50\% of the total stellar mass (i.e., the bulge) formed very early, at $z>2.5-3$. 
 This result remains robust irrespective of the use of different fitting codes, and of the assumed metallicity.
Clearly, these galaxies experienced a very fast growth of their bulge at early times, such as in the quasi-exponential linear main sequence integrations \citep[e.g., ][]{2009MNRAS.398L..58R,2010ApJ...721..193P,2014ApJS..214...15S}, followed by quenching, while the disc begun forming stars late, and this process is still ongoing at the time of observation. 
It is worth noting that such SFH, with the bulk of the stellar mass formed early, and a young stellar population appearing only recently, is very similar to the spatially resolved SFH found for M31 \citep{2017ApJ...846..145W,2018arXiv181102589T}. 

The evidence that the bulge stellar mass of such galaxies, located in the bending part of the MS, was already formed at $z>2.5-3$, put an important constraint on the nature of the MS bending itself, since it rules out the possibility that it is somehow linked to the {\it growth of bulges at later epochs}.
In fact, if this was the case, one would expect to find relatively newly formed bulges ($<3$ Gyr) in galaxies at the MS high-mass end, and mostly in those with the lowest sSFR, while this is not observed in our sample. We note that similar conclusions against a direct connection between the bulge formation in massive galaxies and the MS bending are supported by other studies, starting from different approaches \citep{2011MNRAS.411..993F,2018MNRAS.478...41S,2018MNRAS.tmp.3056P}.

\begin{figure}
\centering
\includegraphics[width=0.5\textwidth]{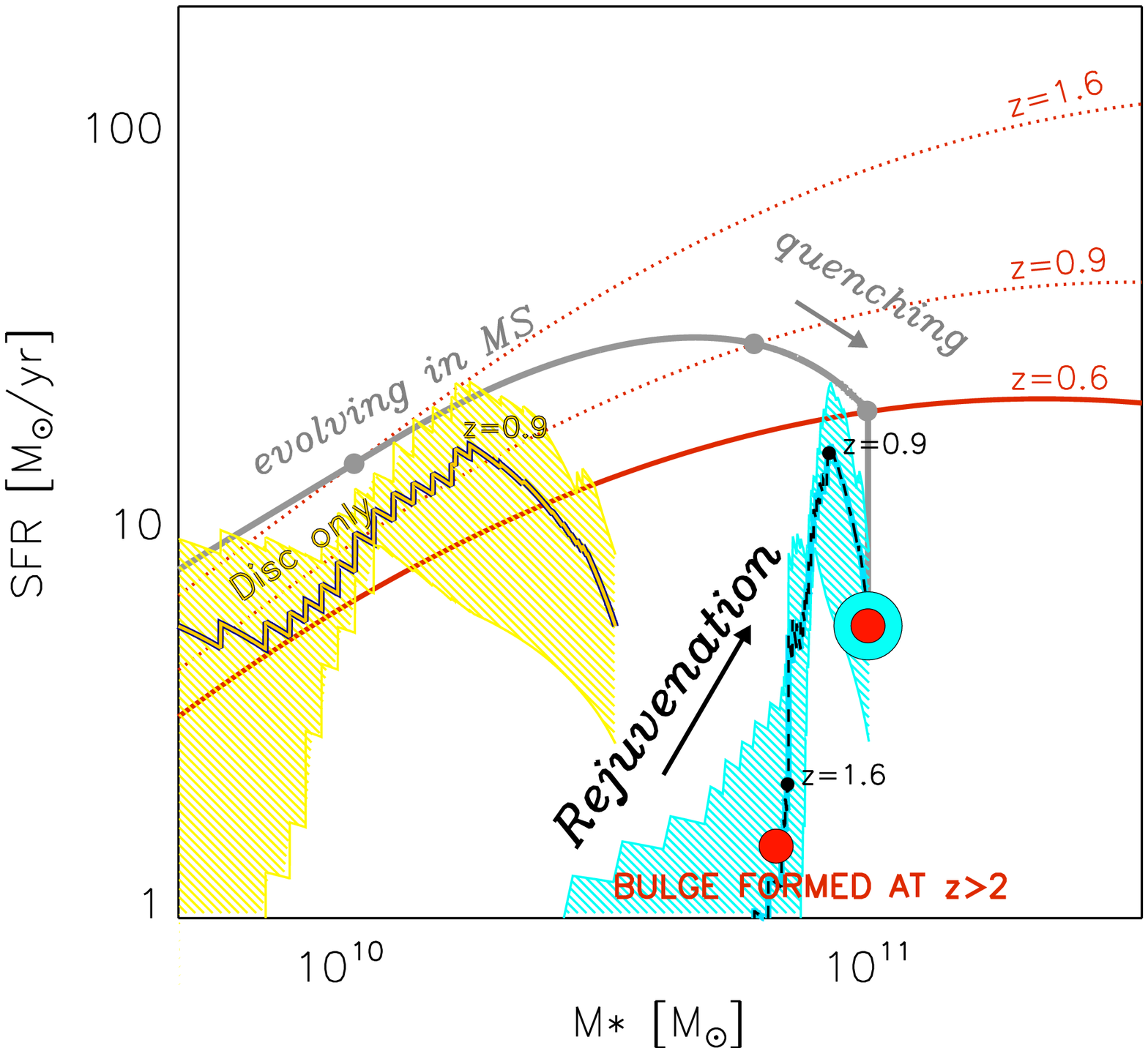}\\
\includegraphics[width=0.5\textwidth]{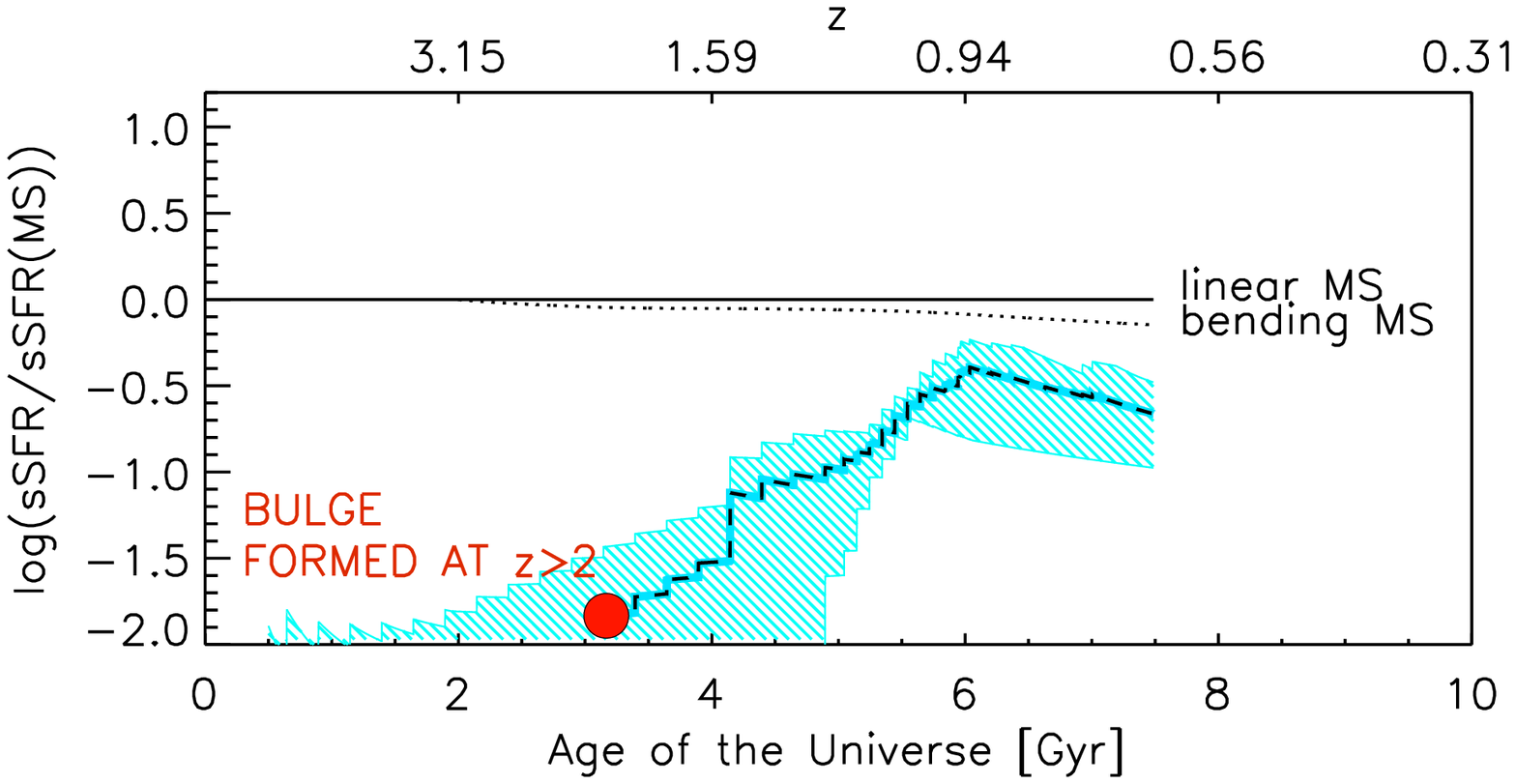}\\
\includegraphics[width=0.5\textwidth]{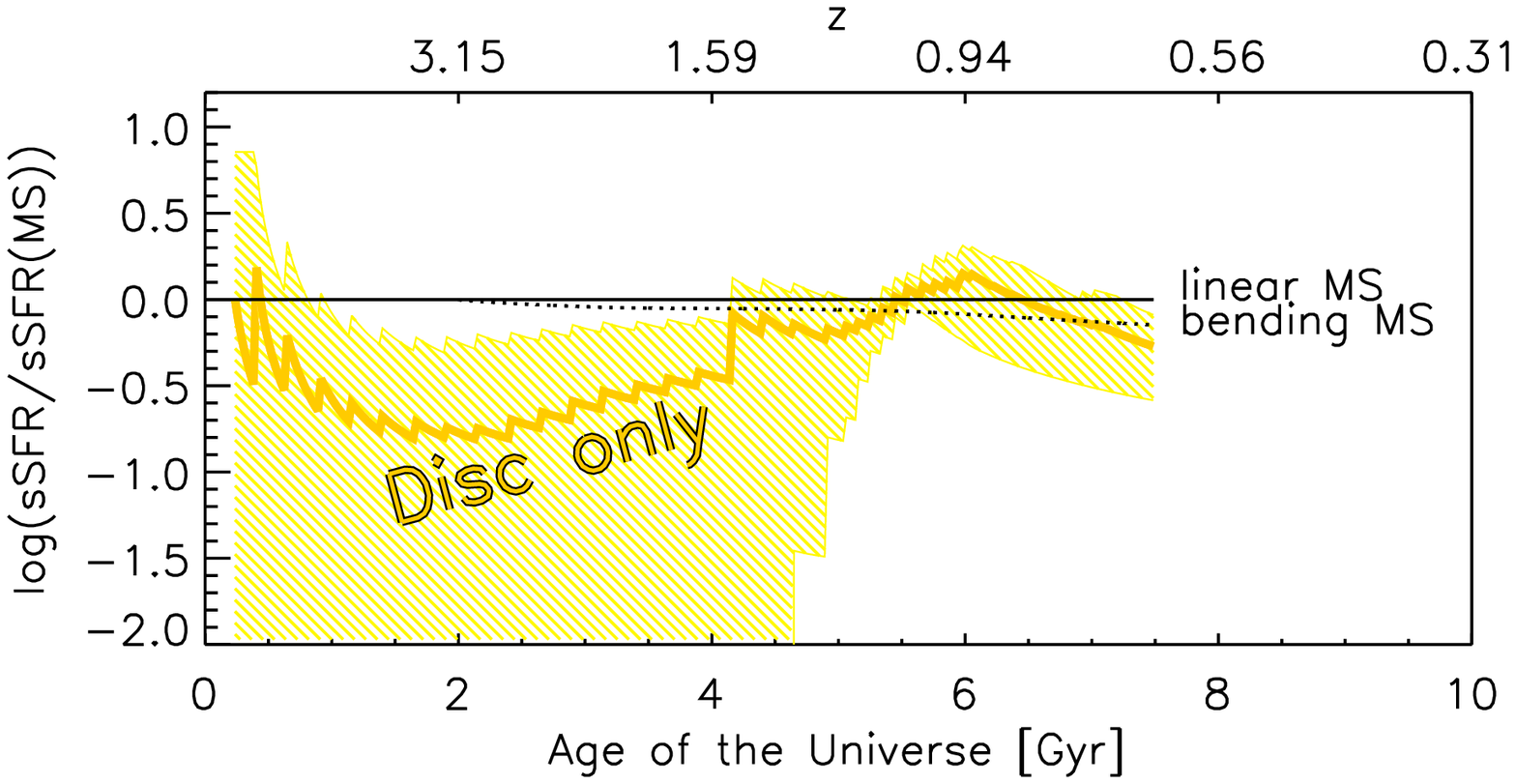}\\
\caption{{\bf Top Panel:} Evolutionary track of our average galaxy on the $M_*-$SFR plane, compared with that of a galaxy with similar $z$, $M_*$, and SFR, evolving along the bending MS of S15 (grey curve). 
The black dashed curve, and the cyan shaded area, show the mass growth history of our average galaxy, with the relative uncertainties. The mass growth history of the bulge is not drawn, since based on our SFHs we cannot establish whether it formed in MS or in SB mode (see Section~\ref{sec:red_harring}). The red filled circle shows the locus occupied by the galaxy once the bulge stellar mass has been fully formed (e.g., $\sim 100$ times below the MS). 
For comparison, we also show the pathway that would be followed by the ``disc only'', e.g. considered as an independent system (yellow curve, and shaded area). The MS of S15 at the average redshift of the sample, $z\sim 0.6$ (solid red line), and at $z\sim 1.6$, and $z\sim 0.9$ (dotted red lines) are shown as reference. Redshift steps (z=0.9, 1.6) are also labelled on all the curves. 
{\bf Middle and Bottom panels:} the distance from the linear MS of \citet{2014ApJ...793...19S} of our average galaxy, and of our average ``disc only'', are shown as a function of the cosmic time (the symbols and colour codes are as in the top panel). The bending to linear MS ratio is also shown (dotted line). Our average galaxy (middle panel) drops 100 times below the MS after the bulge is formed, and then approaches it again after $\sim$ 1 Gyr, rejuvenated. On the contrary, when the bulge mass is not considered, the average disc (bottom panel) more closely follows the MS.}.\label{fig:evol_on_MS} 
\end{figure}

\subsection{Rejuvenated galaxies in the MS bending}\label{sec:red_harring}
Taken at face value, the SFH shown in Figure \ref{fig:sfh},  with a deep minimum in SFR at t=2-3 Gyr, and a relatively wide gap in time ($\gtrsim$ 1 Gyr) between the bulk of the bulge and disc formation, suggests that our sample galaxies have entered the green valley, and the MS bending, as a result of a ``rejuvenation'' episode.
We use here the term ``rejuvenation'' to indicate a process by which a galaxy, after having passed a substantial amount of time well below the star forming main sequence, having little or no star formation activity, witnesses a subsequent rise in its SFR that brings it close to, or within the MS at later times.

This is better seen in the cartoons shown in Figure \ref{fig:evol_on_MS}. In the top panel we report the evolutionary track of our average total galaxy on the $M_*-$SFR plane, compared with that of a galaxy with similar $M_*$ and SFR evolving on the S15 bending MS throughout all its life. The SFH shown in Figure \ref{fig:sfh} tells us that the bulge mass ($\rm M_*\sim 6.7 \times 10^{10} \rm M_{\odot}$) was fully assembled at $z>2$, although we do not have detailed information about the {\it mode} of star formation (i.e., {\it main sequence} or {\it starburst}) controlling the bulge growth at early times. On the other hand, from a statistical point of view, we could speculate that most of the Early-Type galaxies observed in the local Universe, (accounting for $\sim 70\%$ of the total stellar mass density, see ~\citealt{2006ARA&A..44..141R}), should have formed their stars on the MS, since the starburst galaxies only contribute to $\sim $15\% to the total SFR density \citep{2011ApJ...739L..40R}, and assume the same for our average bulge. However, for simplicity, we only show in Figure \ref{fig:evol_on_MS} the average pathway (black dashed curve, with cyan shaded confidence intervals) followed by the disc on the $M_*-$SFR diagram since when the bulge was fully formed, and already below MS, e.g. at $z\sim 2$ (red filled circle).
We point out that our derived SFH, and evolutionary path on the $M_*-$SFR plane, do not intend to constrain how the MS bending globally evolves with cosmic time, but only to trace back how the selected galaxies, with reduced sSFR, have reached the MS bending at $z\sim 0.6$. 

In the middle panel of Figure \ref{fig:evol_on_MS} we report the distance of our average galaxy from the linear MS of \citet{2014ApJ...793...19S}, $\Delta$MS=log(sSFR/sSFR(MS)), as a function of the cosmic time. This shows that our objects do not evolve horizontally along the bending MS (e.g., as the grey curve in the top panel). Conversely, they drop $\sim$ 100 times below the MS after a major episode of star formation, responsible for the assembly of the bulges, and then rise again towards the MS strip, rejuvenated by the accretion of new gas, in the outer parts, which originates the young star-forming discs. This result remains robust within the (cyan shaded) SFH uncertainties, showing that even when considering the 1$\sigma$ upper bound, with the highest allowed sSFR, our average galaxy remains well below the MS ($\Delta$MS$<-1$) at early times.

It is also interesting to examine the evolutionary path on the $M_*-$SFR diagram, and on the $\Delta$MS {\it vs} cosmic time diagram, followed by the discs of our galaxies, in case they are considered as independent systems (e.g., without the bulges). 
This is shown by the yellow track (and shaded confidence intervals) in the top, and bottom panels of Figure~\ref{fig:evol_on_MS}, obtained by normalising the SFH of our average disc to the $M_{*,\rm disc}$, rather than to the total stellar mass, $M_{*,\rm disc}+M_{*,\rm bulge}$. It is evident that, if taken independently, the discs of our objects behave like typical star-forming galaxies, forming the bulk of their stellar mass along the MS. 

Against the proposed rejuvenation scenario, one may object that the disc stellar population, and consequently the derived SFH, could be biased towards young ages, since the youngest stars may outshine the older ones, when present. In this case our galaxies would include an early-formed bulge, and an equally old disc, which keeps accreting gas for longer, until the entire galaxy is fully quenched \citep{2016ARA&A..54..597C}.
However, this would mean that in our analysis we are missing a large amount of the disc mass, formed in the earliest epochs. But, as already remarked in Section ~\ref{sec:seds}, this is at odds with the consistency between the IRAC photometry (sensitive to the galaxy stellar mass) and the total (bulge+disc) best-fit SED model, shown in Figure~\ref{fig:seds}, and in Figure~\ref{fig:med_sed} of the Appendix. Since, based on this last comparisons, the mass eventually missed in the disc should not exceed $\sim 7\%$ of the galaxy $M_*$ (cf., Appendix~\ref{app:med_sed} and Figure~\ref{fig:sfh_extram}), we do not expect this effect to be of great importance.

\subsection{Residual {\it vs} new gas in high-z quiescent and `green valley' galaxies} 
It has now been proven, both indirectly through stacking of far-IR and sub-mm data, and directly through H$\alpha$, [OII], or CO observations, that a certain amount of gas ($M_{\rm gas}$/$M_*$ up to $5-10\%$) is present also in high-redshift ($0.7\lesssim z\lesssim 2$) quiescent galaxies \citep{2017A&A...599A..95G,2018NatAs...2..239G,2017ApJ...841L...6B,2018ApJ...860..103S}. 
However, while some studies suggested an ``external'' origin of the gas, others supported the idea that it is left over from the formation epoch of the bulk of the galaxy. 
\citet{2017A&A...599A..95G} opened the possibility that the younger stellar age, and the extended [OII] emission, found in the outer regions of their quiescent galaxies at $z\sim 1.5$, is due to widespread rejuvenation episodes. For a sample of quiescent galaxies at $z=1-2$, \citet{2017ApJ...841L...6B} also suggested that the ionised gas should have been accreted at later time, based on the measured low gas-phase metallicity. An opposite conclusion was instead drawn by \citet{2018ApJ...860..103S} for three out of their four CO-detected galaxies, based on the alignment between the stellar rotation curves from optical spectra, and the projected components of the CO velocity gradients. Recently, some studies of the SFH of quiescent galaxies at intermediate redshifts have also shown that a non negligible fraction ($\sim 16\%$) of them has experienced rejuvenation episodes in the past \citep{2019arXiv190311082C,2019ApJ...877...48C}. 

We expect that new insight on the origin of the gas discs of our ``bending galaxies'' may arise from the analysis of the emission line properties, and gas-phase metallicity of our sample. 
We leave this analysis for a future work, since an object-by-object optimised extraction of the spectra (and in larger windows, instead of just in the central part of the slit), would be more appropriate for this purpose. 
Moreover, an accurate estimate of the gas-phase metallicity from a composite spectrum would require a dedicated analysis, e.g., through simulations, to quantify possible biases related to the stacking procedure (e.g., due to galaxies for which alternatively the [OIII], or H$\beta$ lines fall in the oxygen atmospheric absorptions A-band).  

If extended to a larger sample of galaxies, the combination of B/D decomposed SFHs with the analysis of emission lines properties, end eventually kinematics, would potentially help to quantify the fraction of rejuvenated, and quenching galaxies in the green valley, and so to understand whether the MS bending is mostly caused by slow-quenching, rejuvenation, or both. 
    
\begin{figure}
\includegraphics[width=0.5\textwidth]{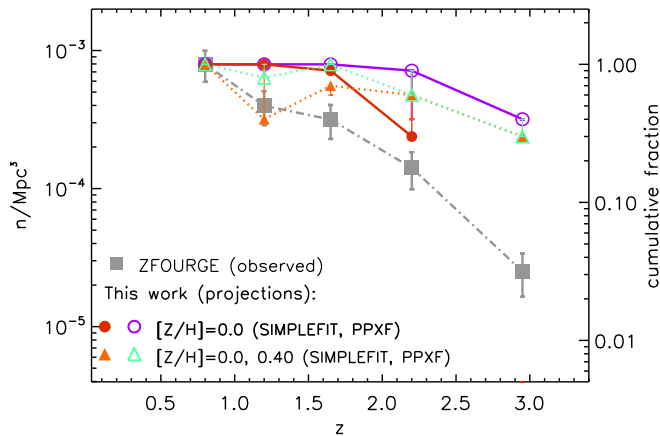}
\caption{Number density evolution projected from the SFH of the bulges in our sample normalised to the value observed for ETGs at $z\sim 0.8$ by \citet[][i.e., ZFOURGE survey]{2014ApJ...783L..14S}. The ZFOURGE ETG number density evolution, in the redshift range of $0.8<z<3$, is also shown for comparison (grey filled squares). The filled, and open coloured symbols are the projection from bulge SFHs derived with the SIMPLEFIT, and the PPXF fitting tools, respectively. Then, red, and purple circles are relative to the fits with the single metallicity of [Z/H]=0.06, while orange, and green triangles to those with both solar and supersolar metallicities ([Z/H]=0.0, 0.22, cf., Section \ref{sec:sfhs}, Figures \ref{fig:sfh}-\ref{fig:sfh2}, and Table~\ref{tab:stack_spec}). The rightmost y axis shows the cumulative relative fractions of bulges observed at z=0.6, that would have been detected as quiescent-UVJ at higher redshifts.}\label{fig:num_dens}
\end{figure}

\subsection{Implications for the number density evolution of Early-Type Galaxies}\label{sec:num_dens}
The finding that the majority of the bulges in our galaxies below MS have ages approaching the age of the Universe at the time they are observed, raises a more general question related to the mass assembly history of ETGs in general. In fact, the inferred old ages, implying that our bulges were already formed at the earliest epochs, provide interesting clues when compared to the rapid evolution of the number density of ETGs with redshift, that increases by a factor of $\sim 100$ from $z\sim 4$ to $z=0$, as reported by many works \citep[e.g.][]{2008ApJ...676..781W,2009A&A...500..705M,2009A&A...501...15F,2011ApJ...739...24B,2013ApJ...777...18M,2014ApJ...783L..14S}. 
To evaluate this issue in details, we used the bulge SFH and its uncertainties, derived in Section~\ref{sec:sfhs}, to trace back the fraction of our bulges that would have been detected as quiescent systems at higher redshifts, as described in the following.

We used the GALAXEV code \citep{2011ascl.soft04005B} and MILES models to predict the evolution of the UV, and VJ colours with redshift for a CSP with the same SFH as our average bulge. For each object in our sample, we assumed the SFH derived from the composite spectrum, rescaled to its own formation epoch, corresponding to the difference between the time of observation and the upper bound of the first (highest) age bin ($t_{\rm form}=t_{\rm obs}-t_{\rm bin1,up}$). This provides a constraint on the epoch in which each bulge stopped to form stars, entering the UVJ diagram. 
We repeated this exercise for all the bulge SFHs shown in Figures~\ref{fig:sfh}, and ~\ref{fig:sfh2}, e.g., derived by fitting the spectrum with SIMPLEFIT or PPXF, and single (slightly super-solar), or double (solar+supersolar) metallicity\footnote{We note that for this test we replaced the templates with [Z/H]=0.06, and [Z/H]=0.22, with those with the closest metallicity, available in the GALAXEV library i.e., [Z/H]=0.0 and [Z/H]=0.40, respectively.}. 

Since estimates of bulge number-density evolution with cosmic time are not available in the literature, we compared our predictions with the observed number densities of similarly massive ($M_*>10.^{10.6}~M_{\odot}$) ETGs at $0.8<z<3$ by \citet[][ZFOURGE sample, grey squares in Figure~\ref{fig:num_dens}]{2014ApJ...783L..14S,2016ApJ...830...51S}. The number density derived for our bulges in the comoving volume included in the GOODS-N field for the studied redshift interval, $z=0.45-1$, is $0.64\times10^{-4} Mpc^{-3}$.
We remark that our sample does not include pure ellipticals, but bulges, and this exercise is entirely dependent on the assumption that the SFHs of bulges should be similar to those of ellipticals. This assumption is motivated, on the one hand, by the well documented analogy of stellar population, structure, and kinematics of bulges and low-luminosity ETGs in the local Universe (see \citealt{2016ARA&A..54..597C}, and references therein). On the other hand, by the full consistency of the colours, sizes, and spectra of the analysed bulges with those of similarly massive ETGs at the same redshift, shown in Sections~\ref{sec:bd}, and~\ref{sec:seds}. 
The main aim of this comparison is understanding whether the ``evolutionary rate'' of the number density deduced from the SFHs of our bulges may be consistent, or not, with that observed for ETGs. 
To this purpose in Figure~\ref{fig:num_dens}, the number density of our bulges, and its evolution back in cosmic time, are normalised to match the ZFOURGE value at $z\sim 0.8$. 
In other words, we compare the cumulative fraction of our bulges, and of ETGs at $z\sim 0.8$ (as reported on the right y axis), that are found to be in place as quiescent UVJ, at each redshift.  The red and purple circles (and solid-line tracks) show the results from the single metallicity fits, while the orange and green triangles (and dotted-line tracks) are from the double metallicity fits, derived with SIMPLEFIT and PPXF, respectively.

It appears from the figure that the results are quite sensitive to the used SFH. For instance, even the two SFHs derived from the PPXF and SIMPLEFIT fits with a single ($\sim$ solar) metallicity, which produce very similar results in terms of mass-weighted age (Table~\ref{tab:stack_spec}), give rise to quite different projected number densities at $z>2$. This is related to the cosmic epoch in which the SFR fully drops to zero (age of the Universe, $t_{\rm U}=2$ and 3 Gyrs, respectively), since even a small burst of star formation is capable to keep the object UVJ colours blue. 
The purple track, implying that 100\% of the ETGs observed at $z\sim0.8$ were already in place at $z\gtrsim 2.2$, and $\sim$ 40\% at $z\sim 3$, is strongly in contrast with the observed rapid evolution of the ETG space density. The red track also overpredicts (by a factor 2.5, and 1.5, respectively) the cumulative fraction of ETGs found at $z\sim1.5$, and $z\sim 2.2$, but then tells us that none of our bulges would have been detected as quiescent at $z>2.5$.  
The orange and green tracks, derived from the two-metallicity fits show a downturn at $1\lesssim z\lesssim1.5$, caused by the second smaller burst of star formation at $t_{\rm U}=4-5$ Gyr, which moves the system out from the quiescent region of the UVJ diagram for relatively short time.  
However, we can ignore this effect for two reasons. First, as mentioned in Section~\ref{sec:sfhs}, these minor bursts, forming stars with younger age and super-solar metallicity, are possibly just the consequence of age-metallicity degeneracy in the fit. Second, we can argue that their impact on our predicted cumulative fraction of quiescent galaxies {\it vs} redshift, is artificially enhanced, since we are assuming the same SFH (from the composite spectrum) for all the objects (most of which are at similar redshift).

Hence, our bulge SFHs would lead to predict a somewhat larger fraction of ETGs at $z=1.5-3$, with respect to what observed. Although this result needs to be validated based on a complete sample of ETGs at intermediate redshift, we note that if some age differences are present between bulges and pure elliptical or lenticular galaxies, one would expect the latter to be older. This might further increase the number density of ETGs predicted at the earliest epochs. 
We note that our results are consistent with the finding from \citet{2016A&A...592A..19C}, based on a sample of local ($0.02<z<0.3$) ETGs with $M_*<10^{10.75}$~M$_{\odot}$. In fact, the SFHs of their galaxies also suggested that a large fraction of those systems (or at least of their stars) should be already in place at $z>2$, or even at $z>3$, depending on the used template libraries in the spectral fitting. 
This further supports the idea that SFHs of bulges and ETGs with similar masses are comparable. Moreover, recently, an increasing number of quiescent galaxies have being spectroscopically identified at $z\geq 3$ \citep[][D'Eugenio et al. in preparation]{2012ApJ...759L..44G, 2017Natur.544...71G,2018A&A...618A..85S}. In particular, \citet{2018A&A...618A..85S} using the Keck-MOSFIRE near-IR spectrograph, confirmed the redshift and the old ages (with half of the stars formed at $z>5.5$) of a dozen of quiescent galaxy candidates at $3\leq z\leq 4$. Interestingly, they showed that, although the UVJ criterion is still efficient to select a clean sample of quiescent galaxies at $z\geq 3$, it could lead to miss a fraction of this population. This would imply an underestimation of the space density of quiescent galaxies at high redshift for samples just based on the UVJ colour selection. On the other hand, the high sSFR of very high-redshift galaxies appears to demand quenching being active also at very early times to avoid an overgrowth of cosmic stellar mass \citep{2016MNRAS.460L..45R}.

This result is potentially very compelling since it would imply that we are missing a certain amount of stellar mass at $z>1.5$. Although only conjectural, it is possible that part of the ETG stellar mass is undetected at high-$z$, for a combination of different reasons. In fact, on the one hand, some quiescent bulges could be hidden within luminous star-forming discs. This would entail an underestimate of the global stellar mass of the systems, since their colours (and M/L ratios) would be dominated by the blue component. This effect may have a growing importance at $z>1.5-2$, where the discs have very high sSFR, and galaxies are observed (and resolved) in bluer rest-frame by the current facilities. In the next future, the high-sensitivity and resolution of the James Web Space Telescope (JWST) near-IR camera will help to shed light on this issue, enabling the search for old faint bulges possibly hidden in galaxy discs at high-z. 
On the other hand, a certain fraction of the stellar mass could have been already formed at the earliest epochs, but not yet fully assembled into massive ETGs, according to the paradigm of the hierarchical mass assembly \citep[][and many others]{1993MNRAS.264..201K,1996MNRAS.283L.117K,1996MNRAS.283.1361B,2000MNRAS.319..168C,2006MNRAS.370..645B,2006MNRAS.366..499D}. 
We further stress that shedding light on this issue, e.g., with similar tests performed on complete ETG samples, is crucial to better understand the mass assembly history of massive galaxies. 

\section{Summary and Conclusions}\label{sec:concl}

The main aim of this paper is investigating the nature of the bending of the star-forming galaxy MS at the high-mass end, by testing some scenario proposed to explain its origin. In particular, we compared the SFHs of massive galaxies with reduced sSFR (that seemingly cause the $M_*-$SFR relation to bend) with those predicted for objects evolving on the MS throughout their life, and undergoing a ``slow downfall'' of the SFR towards the quiescence, possibly induced by the growth of a central bulge (e.g., through ``morphological-quenching''), or by other processes. 
Hence, we selected 10 massive ($M_*>2\times10^{10}$\msun) star-forming galaxies at $0.45<z<1$ in the GOODS-N field, located a factor of $2.5-10$ below the linear MS. The sample only includes objects securely star-forming (i.e., with reliable FIR detection), and for which a DEIMOS/TKRS spectrum with a S/N$> 3$ in the region of the 4000~\AA\ break was available.

To derive the galaxy SFHs we used a quite novel approach, consisting in separately reconstructing the bulge, and disc  SFHs, so to avoid the systematics produced when averaging out the light of the two (young and old) populations. The bulge SFHs were inferred from the TKRS spectra (extracted in small windows to minimise the contamination from the disc  light), while the disc  SFHs were derived from the SED fitting analysis, after a B/D decomposition. The main results are summarised in the following: 

\begin{itemize}

\item 
Based on photometric multi-colour B/D decomposition, and SED fitting analysis of the bulge and disc  components, we showed that our sample galaxies have high bulge-to-total mass ratios ($\langle$B/T$_{\rm mass}\rangle \sim$0.67), and extended star-forming discs. Their global sizes are compatible with those of similarly massive star-forming galaxies at the same redshift. Then, when taken separately, the bulge and disc components occupy the loci of quiescent and star-forming galaxies, respectively, in the $M_*-$\re\ diagram. Similarly, while the entire galaxies show colours typical of ``green'' galaxies, with reduced sSFR, being located on the quiescent/star-forming boundary in the UVJ diagram, bulges and discs clearly split up in the red/quiescent and blue/star-forming regions.   
\item 
By stacking (weight-averaging) the 10 individual spectra we obtained a composite spectrum with an average S/N $\sim 17$ (S/N $\sim 14$ in the 4000~\AA\ break region). Based on extensive simulations, we verified that such S/N is high enough to allow a robust age determination, within 10\% uncertainties, while the ages derived from individual spectra (S/N $\sim 4.5$) could suffer larger (30-50\%) uncertainties.  

\item The average mass-weighted age derived by fitting the bulge composite spectrum with the SIMPLEFIT code and a slightly supersolar metallicity ([Z/H]=0.06) is 6.04$\pm$0.30 Gyr. 
This result remains robust (well consistent within the uncertainties) against the used fitting code (i.e., SIMPLEFIT, and PPXF), and when both solar and super-solar metallicities are used in the fit. 
Also, the mass-weighted ages estimated from the individual bulge spectra (mw-Age$_o$) are almost all maximally old, and in good agreement with the half-mass ages (T$_{50}$), roughly corresponding to the mass-weighted ages, derived from the broad band SED fitting. 
Conversely, the disc half-mass ages estimated from the broad-band SED fitting are very young, ranging from T$_{50}=0.46$ to 3.23 Gyr, and being on average $\langle$ T$_{50} \rangle$=1.66 Gyr.

\item 
We found that the average bulge+disc SFH of our selected ``bending galaxies'' is not consistent with the that predicted for a galaxy with the same $z$, $M_*$, and SFR evolving along a bending MS.  
In fact, while the evolution along the MS would require that almost all the galaxy stellar mass should have been assembled within the last 3 Gyrs, in our galaxies $>$50\% of the total stellar mass  formed very early (in the bulge), at $z>3$. 
This evidence is incompatible with a scenario in which the bending of the MS is caused by the late growth of the bulges, which would have contributed to slowly extinguish the star-formation in the discs (e.g., via morphological quenching).

\item  The derived SFH, imply that our average galaxy, does not spend all its lifetime on the (bending) MS, but, after a major episode of star-formation responsible for the bulge assembly, drops $\sim$ 100 times below the MS. Then, it rises again only at later epochs (after $\gtrsim$ 1 Gyr), originating the disc. Such large gap in time between the bulk of the bulge and disc formation, suggests a scenario in which the earlier formed, old ETG is rejuvenated by the accretion of new gas, then settled into a disc. 
Although this finding only concerns the galaxies with the lowest sSFR in the MS bending at $z\sim 0.6$, analysed in this paper, in principle these extreme objects could be regarded as those that much strongly contribute to reduce the average sSFR in the MS high-mass end. 
This raises a question about the nature of the MS bending itself, which does not seem to include (only) galaxies that are following a downturn towards the full quenching of their star-formation, but (also, and perhaps mostly) rejuvenated systems turning up due to renewed star-formation activity. This effect needs to be taken into account when attempting to use sub-MS galaxy number counts to infer the quenching timescale of galaxies.

\item To evaluate whether the old ages of our bulges are consistent with the observed rapid evolution of the number density of massive ETGs, we used their derived SFH to predict the evolution of their UVJ colours with time. Then, we derived the  cumulative fraction of bulges in our sample that would have been detected as quiescent systems, in the UVJ diagram, at each redshift.  
We performed this test using the bulge SFH obtained by fitting the spectrum with the different codes, and/or template metallicities, to evaluate the results from all the solutions investigated in our analysis. The projections from the bulge SFHs lead to overpredict the number density of ETGs at $z=1-3$, i.e., from a factor 2.5 to a factor of 15, depending on the used SFH. Such a possible discrepancy needs to be carefully examined, since it could  have important consequences on the mass assembly history of bulges, and even of the entire class of ETGs. In fact, if confirmed (e.g., on a mass-complete sample of ETGs), this might be the first direct evidence that the stellar mass included in the local bulges (ETGs) was not entirely formed in situ, but hierarchically assembled during the cosmic epochs, according to the $\Lambda$CDM paradigm of galaxy formation.

\end{itemize}

\section*{Acknowledgements}
We acknowledge funding from the INAF PRIN-SKA 2017 program 1.05.01.88.04. This work is based on observations taken by the 3D-HST Treasury Program (HST-GO-12177 and HST-GO-12328) with the NASA/ESA Hubble Space Telescope, which is operated by the Association of Universities for Research in Astronomy, Inc., under NASA contract NAS5-26555. We thank Laura Morselli and Paolo Cassata for useful discussions. We also thank the referee for suggestions that helped improve the quality of the paper.
Some of the data presented herein were obtained at the W. M. Keck Observatory, which is operated as a scientific partnership among the California Institute of Technology, the University of California and the National Aeronautics and Space Administration. The Observatory was made possible by the generous financial support of the W. M. Keck Foundation. The authors wish to recognise and acknowledge the very significant cultural role and reverence that the summit of Maunakea has always had within the indigenous Hawaiian community.  We are most fortunate to have the opportunity to conduct observations from this mountain.




\bibliographystyle{mnras}
\bibliography{new_ref} 



\appendix
\section{Deriving the Composite Disc SFH}\label{app:disc_sfh}
\begin{figure*}
\includegraphics[width=\textwidth]{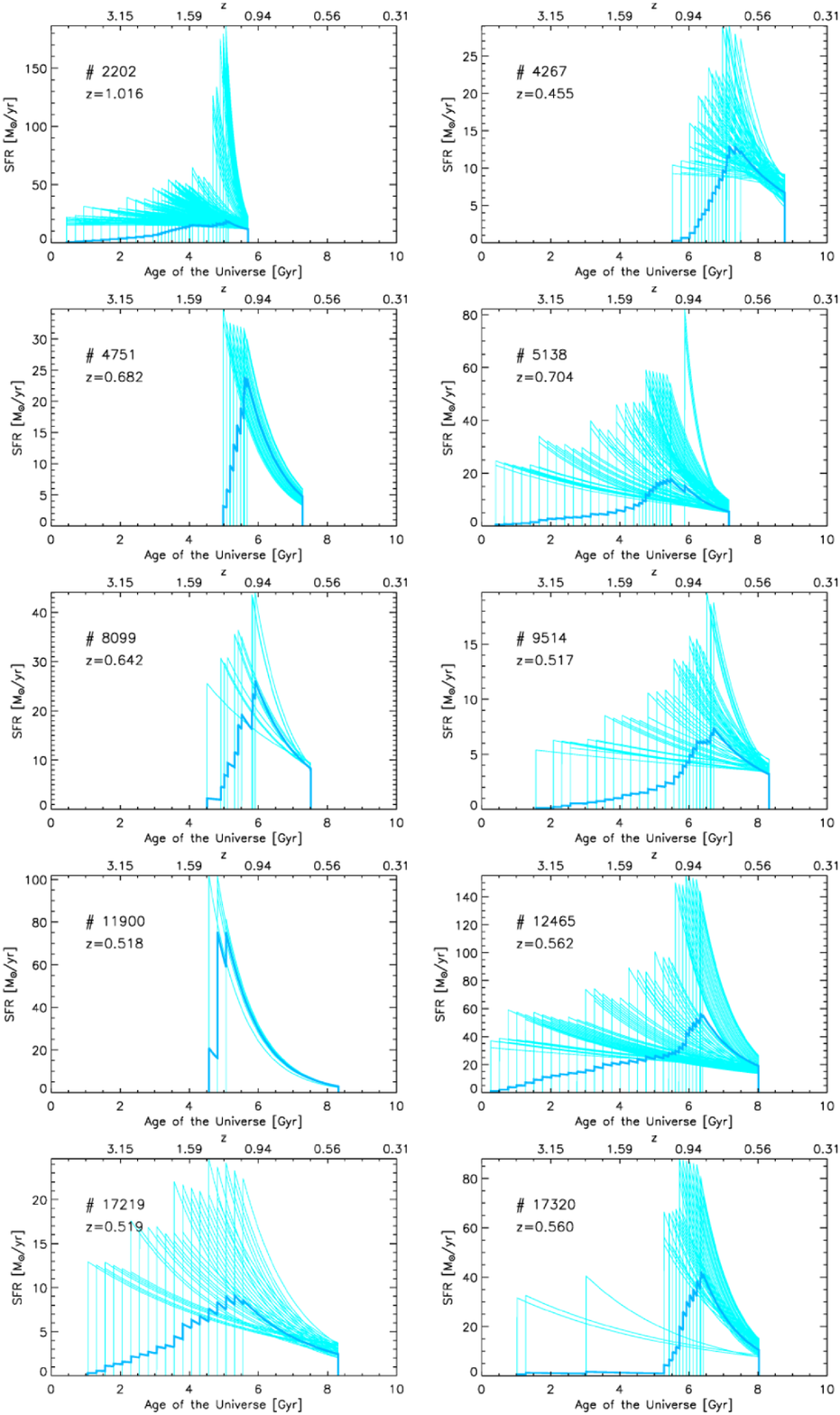}\\

    \caption{The procedure used to derive the composite SFHs of the discs from the SED fitting, described in Section~\ref{sec:sfhs}, is exemplified here. The cyan curves show all the SFHs corresponding to the SED fitting solutions included within the 68\% confidence intervals of the $\chi^2$ probability distribution, and with a SFR compatible with that derived from the IR+UV. The turquoise curve show the mean SFH for each disc, then used to derive the average disc SFH of the sample as detailed in the text.}
    \label{fig:discsfh}
\end{figure*}

In Figure~\ref{fig:discsfh} we show the SFHs derived from the SED fitting results as described in Section~\ref{sec:sfhs} for the individual discs of our sample galaxies. For each object, all the SFHs (declining $\tau$ models, shown in cyan) associated to the allowed solutions within the 68\% confidence intervals, and with a SFR compatible with that derived from the IR+UV (yellow regions in Figure~\ref{fig:contours}, see Section~\ref{sec:sfhs}) were averaged out to derive the disc SFH (turquoise curve). The average disc SFH and confidence intervals of the sample, were then obtained by combining the 10 individual disc SFHs, after having reported all the objects at the mean redshift $\langle z\rangle=0.62$. The total average SFH, shown in Figure ~\ref{fig:sfh}, was obtained by summing up the bulge and disc SFHs, normalised to the average bulge-to-total, and disc-to-total mass ratios of the sample, respectively (e.g. B/T$_{\rm mass}$=0.67, D/T$_{\rm mass}$=0.33).       

\section{Decontaminating the 1D spectra from the disc component}\label{app:decont}
 Although the bulge component dominates the TKRS spectral continuum in our galaxies, we can expect a certain degree of contamination from the disc in the central part of the spectra (where the stellar density of the disc is higher). Hence, to reliably infer the bulge stellar ages, we developed a procedure to decontaminate the TKRS spectra from the disc components, using the information from the B/D decomposition and the SED fitting analysis, as detailed in the following. For each galaxy:
\begin{enumerate}

\item The {\it galfitm} F850LP output images obtained both for the bulge model, and for the bulge-subtracted residual (the disc) were convolved with the seeing of the spectroscopic observations.

\item The bulge and disc light fractions falling into the spectral extraction windows were directly measured from the degraded images. This analysis showed that our extracted spectra can include from 36\% to 75\% of the total bulge light ($\sim 50\%$ on average), and from 15 to 45\% ($\sim 30\%$ on average) of the total disc light. 

\item These light fractions were used to rescale the bulge and disc best-fit templates derived from the SED fitting (red and blue lines in Figure \ref{fig:seds}, respectively), so to reconstruct a sort of synthetic spectrum (bulge+disc) that should ideally overlap with the real one. 
We verified that after a modest renormalization, the shapes of the synthetic spectrum matched pretty well with that of the real one. 

\item Finally, the decontaminated bulge spectrum was obtained by subtracting the rescaled synthetic disc template from the real spectrum.  

\item We stacked the decontaminated individual spectra, following the same procedure used for the original spectra, to build a composite decontaminated spectrum to be compared with the composite original spectrum.
\end{enumerate}

We fitted again both the individual, and the composite decontaminated bulge spectra with the same procedure used for the original ones (and just single [Z/H]=0.06 metallicity), and found that the contribution from the youngest (0.08 Gyr) SSP to $L_{5500}$ was substantially reduced, or completely disappeared after the disc  component was subtracted. This is visible in Table~\ref{tab:frac}, where we show the comparison between the fractional contributions (both in light at 5500~\AA, and in mass) from SSP of different ages to the best-fit model obtained from the original and decontaminated composite spectrum. After the decontamination, the light fractional contribution from the young 0.08 Gyr SSP decreases from $\sim 22\%$ to $\sim 10\%$ in the composite spectrum.   

\begin{table}
\begin{tabular}{r|rr|rr}
\hline
  \multicolumn{1}{c}{} &
  \multicolumn{2}{c}{ORIGINAL} &
  \multicolumn{2}{c}{DECONTAMINATED} \\
\hline
  \multicolumn{1}{c}{Age [Gyr]} &
  \multicolumn{1}{c}{f$_{\rm light}$ (\%)} &
  \multicolumn{1}{c}{f$_{\rm mass}$ (\%)} &
  \multicolumn{1}{c}{f$_{\rm light}$ (\%)} &
  \multicolumn{1}{c}{f$_{\rm mass}$ (\%)}\\
\hline
0.08&22.36  & 1.23  &10.00  & 0.50 \\
5.00& 9.41  & 9.90  & 0.00  & 0.00 \\
6.50&68.23  &88.87  &90.00  &99.50 \\
\hline
\end{tabular}
\caption{Light fractional contribution at 5500~\AA\ (f$_{\rm light}$), and the corresponding mass fractional contribution (f$_{\rm mass}$) to the final best-fit model of the composite spectrum, from of the used SSP templates of different ages. We show here the comparison between the results derived from the fit with a single [Z/H]=0.06 metallicity for original and decontaminated spectra, i.e., before and after the process of decontamination from the disc  light, detailed in Section \ref{app:decont}.}\label{tab:frac}
\end{table}

\begin{table}
\begin{tabular}{|r|r|r|r|r|r|r|r|r|r|r|}
\hline
  \multicolumn{1}{|c|}{ID} &
  \multicolumn{1}{c|}{lw-Age$_{\rm o}$} &
  \multicolumn{1}{c|}{mw-Age$_{\rm o}$} &
  \multicolumn{1}{c|}{lw-Age$_{\rm d}$} &
  \multicolumn{1}{c|}{mw-Age$_{\rm d}$} \\
\hline
  2202 & 3.87$\pm$ 2.24 & 4.81$\pm$  2.78 &    6.32$\pm$  4.72 & 6.48$\pm$  4.84\\
  4267 & 5.02$\pm$ 1.61 & 6.42$\pm$  2.05 &    5.48$\pm$  2.02 & 6.45$\pm$  2.38\\
  4751 &  2.97$\pm$0.95 &  4.15$\pm$ 1.32 &    4.82$\pm$  3.31 & 5.24$\pm$  3.67\\
  5138 & 4.42$\pm$ 1.62 & 5.25$\pm$  1.92 &    6.15$\pm$  3.56 & 6.47$\pm$  3.73\\
  8099 & 4.02$\pm$ 1.15 & 5.21$\pm$  1.49 &    5.28$\pm$  1.82 & 6.44$\pm$  2.22\\
  9514 & 2.28$\pm$ 0.87 & 2.96$\pm$  1.13 &    4.66$\pm$  2.97 & 4.98$\pm$  3.18\\
 11900 & 6.42$\pm$ 2.02 & 7.35$\pm$  2.31 &    5.13$\pm$  1.84 & 6.43$\pm$  2.30\\
 12465 & 4.91$\pm$ 0.92 & 6.41$\pm$  1.20 &    5.11$\pm$  1.90 & 6.43$\pm$  2.38\\
 17219 & 4.20$\pm$ 1.65 & 4.95$\pm$  1.95 &    5.82$\pm$  4.28 & 6.47$\pm$  4.75\\
 17320 & 2.74$\pm$ 0.82 & 2.98$\pm$  0.90 &    6.50$\pm$  2.56 & 6.50$\pm$  2.56\\
\hline
\end{tabular}
\caption{Light-weighted (lw) and mass-weighted (mw) ages for the individual original (`o' subscript) and decontaminated (`d' subscript) spectra, i.e., before and after the process of decontamination from the disc  light, detailed in Section \ref{app:decont}.}\label{tab:single_spec}
\end{table}

\section{Spectral fitting of simulated galaxies}\label{app:simul}
To reliably estimate the uncertainties on the best-fit parameters derived from the {\it SIMPLEFIT} spectral fitting procedure (Section~\ref{sec:b_spec}), as a function of the average S/N, we used simulated galaxy spectra as described in the following.

\begin{figure}
\includegraphics[width=0.47\textwidth]{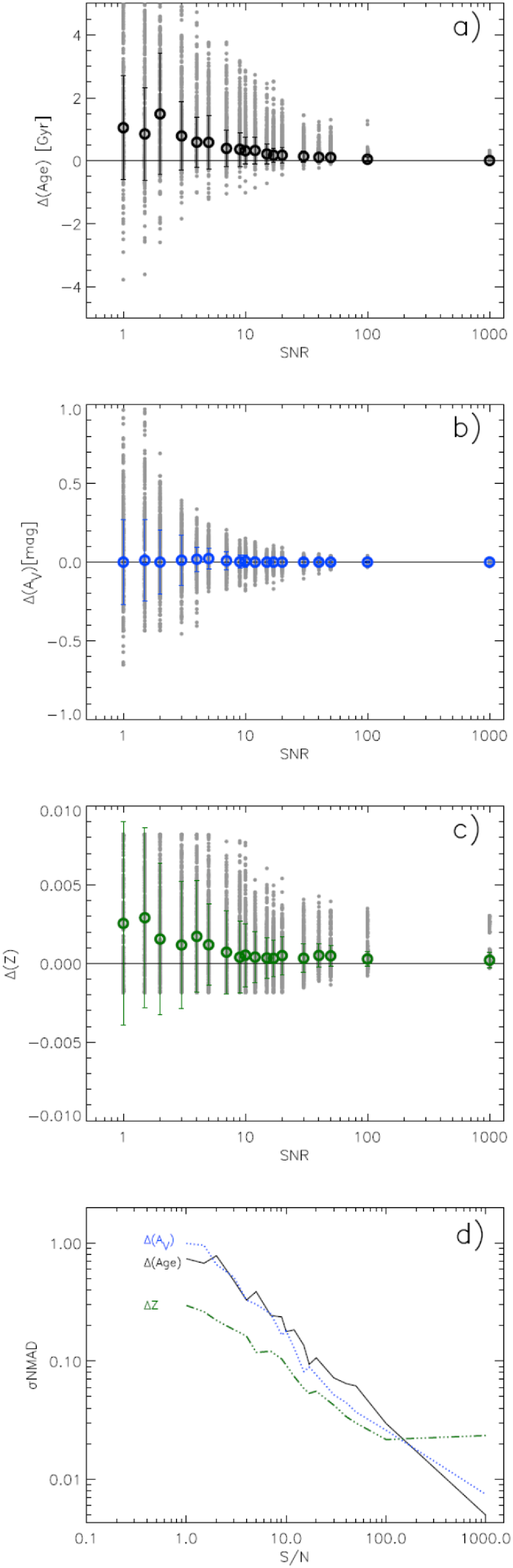}\\
\caption{Results from simulated galaxies fitted with {\it SIMPLEFIT}. In panels a), b),  c), we report $\Delta x_i =x_{OUT}-x_{IN}$ as a function of the S/N, where $x$=Age, $A_V$,  and Z, respectively. Individual mock galaxies are shown by grey dots, and median quantities by large open circles. The error bars show the median absolute deviations (MAD) at each S/N. The curves in panel d) represent the variation with the S/N of the $\sigma$NMAD, used as a S/N-dependent estimate of the uncertainties on each parameter, as labelled in the figure.}\label{fig:param_snr}
\end{figure}

Since in this work we analysed TKRS spectra extracted in the central part of the galaxies, and dominated by the light from the bulge components (as shown in the previous section), we are mostly interested in testing the fitting code against old stellar populations (age$\geq$1 Gyr). We used the MILES-BaSTI SSP models with [Z/H]=0.06 (i.e., the expected metallicity for the central regions of our bulges, cf. Section~\ref{sec:bulge_fit}) to generate a total of 6480 galaxies, with 18 different values of average S/N, in the range 1-1000 (360 objects for each S/N), 9 different ages (i.e., 1, 1.25, 1.5, 2, 2.5, 3, 4, 5, 6.5 Gyr), and 4 possible $V$-band optical depths ($\tau_V$=$A_V$/1.086=0, 0.2, 0.4, 0.6).
The MILES templates (with intrinsic resolution of 2.51 \AA) were normalised at 5500 \AA\ and convolved with a Gaussian kernel so to match the average velocity dispersion of our sample ($\sigma *$=164 km/s, as measured from the stacked spectrum, cf. Section~\ref{sec:stack}). By assuming that the noise is wavelength-independent, we added random Gaussian noise to the mock spectra, derived as the ratio between the average model flux and the S/N value for each object. This assumption is justified by the fact that, beyond the effects of the atmospheric absorptions, the noise of our real spectra is almost constant with wavelength (e.g., see Figure~\ref{fig:single_spectra}). 
We fitted our simulated spectra with {\it SIMPLEFIT}, using the same template libraries, and SSP age-grid as for the real data. 
To also check for the reliability in recovering the metallicity as a function of the S/N, we repeated the fit with the metallicity as a free parameter, as done for the real spectra.
Although simulated spectra do not include emission lines, to reproduce the real conditions, in the fitting procedure we applied the same masks used for the real data in the [OII]$\lambda\lambda$3727, [OIII]$\lambda\lambda$5007, and Balmer line regions. The results are shown in Figure~\ref{fig:param_snr}, where the difference between the output and input parameters in the fit $\Delta x_i =x_{OUT}-x_{IN}$ (where $x$=Age, $A_V$, and [Z/H]) are shown as a function of the S/N.  
In each panel the grey dots represent individual mock galaxies, and the red open circles and error bars are the median, and the standard deviation, computed at each S/N. 

To give an estimate of the uncertainties as a function of the S/N, we derived the normalised median absolute deviation, defined as $\sigma$NMAD = 1.48 $\times$ median( $\vert \Delta x_i -$ median($\Delta x_i$)$\rvert$)/$x_{IN}$ \citep{1983ured.book.....H}. Then we inferred the parameter errors as err($x$) = $\sigma$NMAD $\cdot$ $x$. The variation of the $\sigma$NMAD with the S/N for age, $A_V$, and [Z/H] is shown in the panel (d) of Figure~\ref{fig:param_snr}.

Figure~\ref{fig:param_snr} shows that all the parameters require a $\langle$S/N$\rangle \geq$ 12 to be estimated with an accuracy $\lesssim 10\%$. 
At S/N$<100$ the galaxy mass-weighted ages tend to be overestimated by a certain quantity, decreasing with increasing S/N (e.g., by 1, 0.5 and 0.25 Gyr at $\langle$ S/N$\rangle\sim$ 3, 8, and 17). However, we point out that this does not depend on the {\it SIMPLEFIT} code, since a similar bias is found using other fitting codes \citep[e.g., pPXF and STARLIGHT, see][]{2016A&A...592A..19C,2018MNRAS.478.2633G}. 
Based on these results, we concluded that, while the best-fit parameters of our individual spectra (S/N$\sim 4.5$) could be subject to quite large uncertainties, a robust estimate of mass-weighted age, $A_V$, and [Z/H] can be obtained from the stacked spectrum.   

\section{Test against the used spectral-fitting code: PPXF fit of stellar populations}\label{app:ppxf}
As mentioned in Section~\ref{sec:b_spec}, the PPXF code, beyond measuring the velocity dispersion, can fit the stellar continuum. This gives us the opportunity to test the SIMPLEFIT results on stellar population properties (stellar mass-weighted ages, SFHs, and reddening) with a different fitting tool.
As described in \citet{2017MNRAS.466..798C}, PPXF includes the option to set the so-called {\it regularisation} parameter in the fit, which suppresses noise during the SFH recovery, with the result of smoothing the mass-weights relative to the different SSP models over a wider range of ages (and allowed metallicities). This can be useful to retrieve a reliable SFH, when many degenerate solutions are equally consistent with the data. 
However, this procedure may bias the fit towards smoothed solution, when a too strong regularisation prior is applied. 
Hence, we followed here a more conservative approach, recently validated in \citet{2018MNRAS.480.1973K}, based on simulations obtained with the wild bootstrapping \citep{Wu1986}. We generated 1,000 synthetic spectra by resampling the flux for each pixel ($y_i$) from the residuals ($r_i$) of the best-fit model ($y_{i, best}$), derived from the PPXF (unregularized) run on the real spectrum. The resampling is made so that: $y_i=y_{i, best}+v_i*r_i$, where $v_i$ is a random variable following a Rademacher distribution, e.g. only allowed to assume the values +1 or -1. 
We fitted the synthetic spectra with PPXF in the {\it linear} mode \citep[i.e., by fixing the velocity moments, and the dust reddening (A$_V$) to the known best-fit values, cf.][]{2017MNRAS.466..798C}. The fiducial solution, and the relative uncertainties, are represented by the median, and 16th and 84th percentiles of the mass-weights distribution. As explained in \citet{2018MNRAS.480.1973K}, this approach allows one to estimate the uncertainties on the mass weights (i.e., on SFH, and mass-weighted age), just based on the spectral noise, without the need of imposing any prior through the regularisation term.   We applied the wild bootstrapping procedure to both the cases, with fixed ([Z/H]=0.06), and free ([Z/H]=0.0, 0.22) metallicity in the fit, using the same templates and age grid as for the SIMPLEFIT runs. The best-fit results are fully consistent with those from SIMPLEFIT, as shown in Table~\ref{tab:stack_spec}, and Figure~\ref{fig:sfh} of the main text.

\section{Deriving an upper limit for missing stellar mass in the discs}\label{app:med_sed}

As noticed in Section~\ref{sec:seds}, the age and stellar mass of the galaxy discs estimated through SED fitting in the UV/optical rest-frame, could be biased towards lower values, since an older and more massive stellar population may be outshined by the emission of the youngest stars.   
However, since the emission at longer wavelengths (e.g., $\lambda_{rest}>1.5\mu$m) is instead dominated by massive old ($>0.5-1$ Gyr) stars, in Section~\ref{sec:seds} we proved the robustness of our SED fitting results by showing that the observed IRAC fluxes (not used in the fit) are in agreement, within 10\%, with those extracted from our total (bulge+disc ) SED models. 

\begin{figure}
\begin{center}
\includegraphics[width=0.48\textwidth]{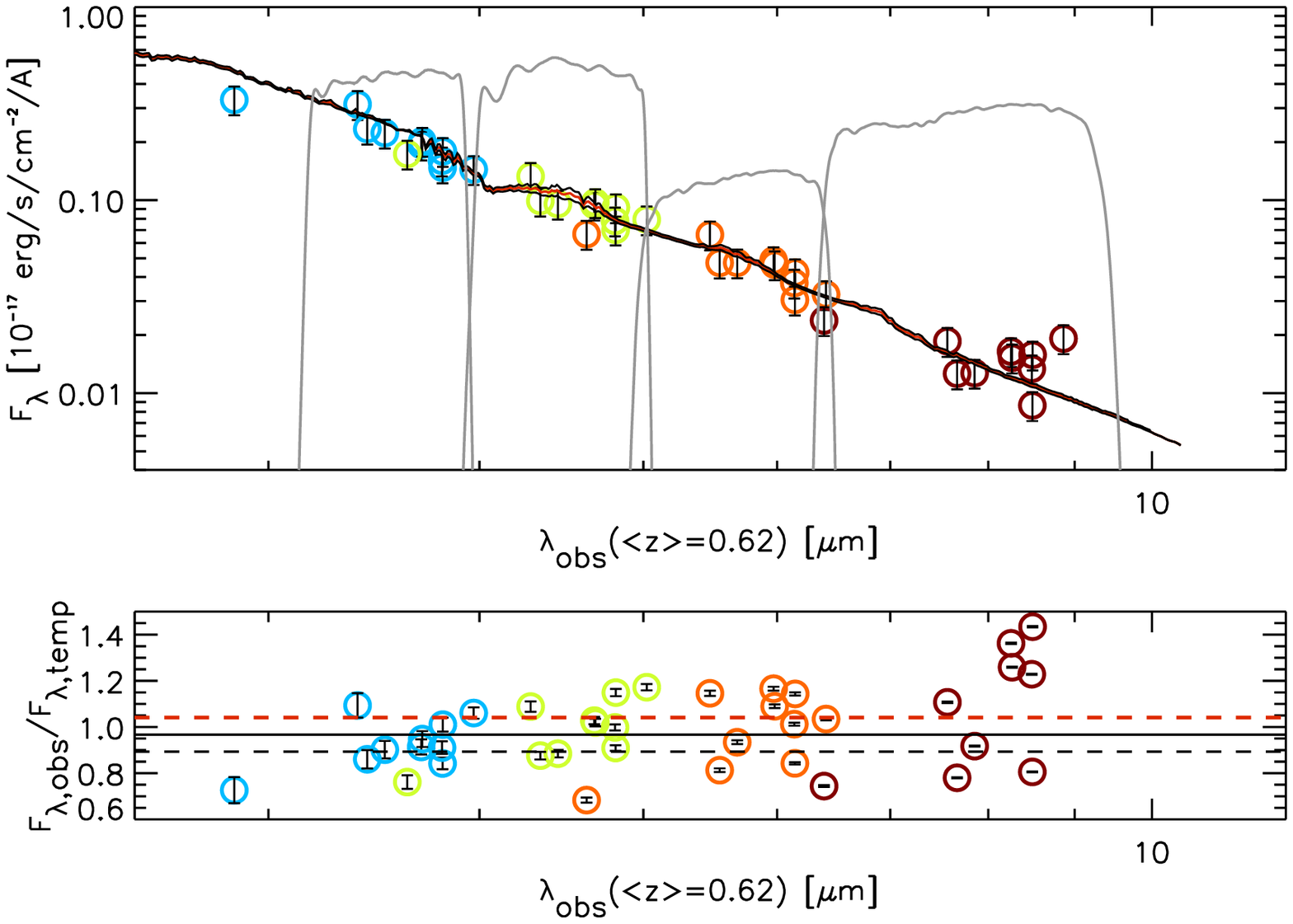}
\caption{{\bf Top panel.} Average SED (red line) and its 1$\sigma$ dispersion (black lines) of the sample at the Spitzer/IRAC wavelengths. The 3.6, 4.5, 5.8, and 8.0~$\mu$m IRAC data points for our 10 galaxies are shown in cyan, green, orange and brown, respectively. {\bf Bottom panel.} Observed to model flux ratio (F$_{\lambda,obs}$/F$_{\lambda,temp}$) as a function of the observed wavelength at $\langle z\rangle=0.62$. The dashed black, and red lines show the lower and upper limits, at $3 \times \sigma_m$, corresponding to three times the standard deviation of the mean.}\label{fig:med_sed}
\end{center}
\end{figure}

\begin{figure}
\begin{center}
\includegraphics[width=0.48\textwidth]{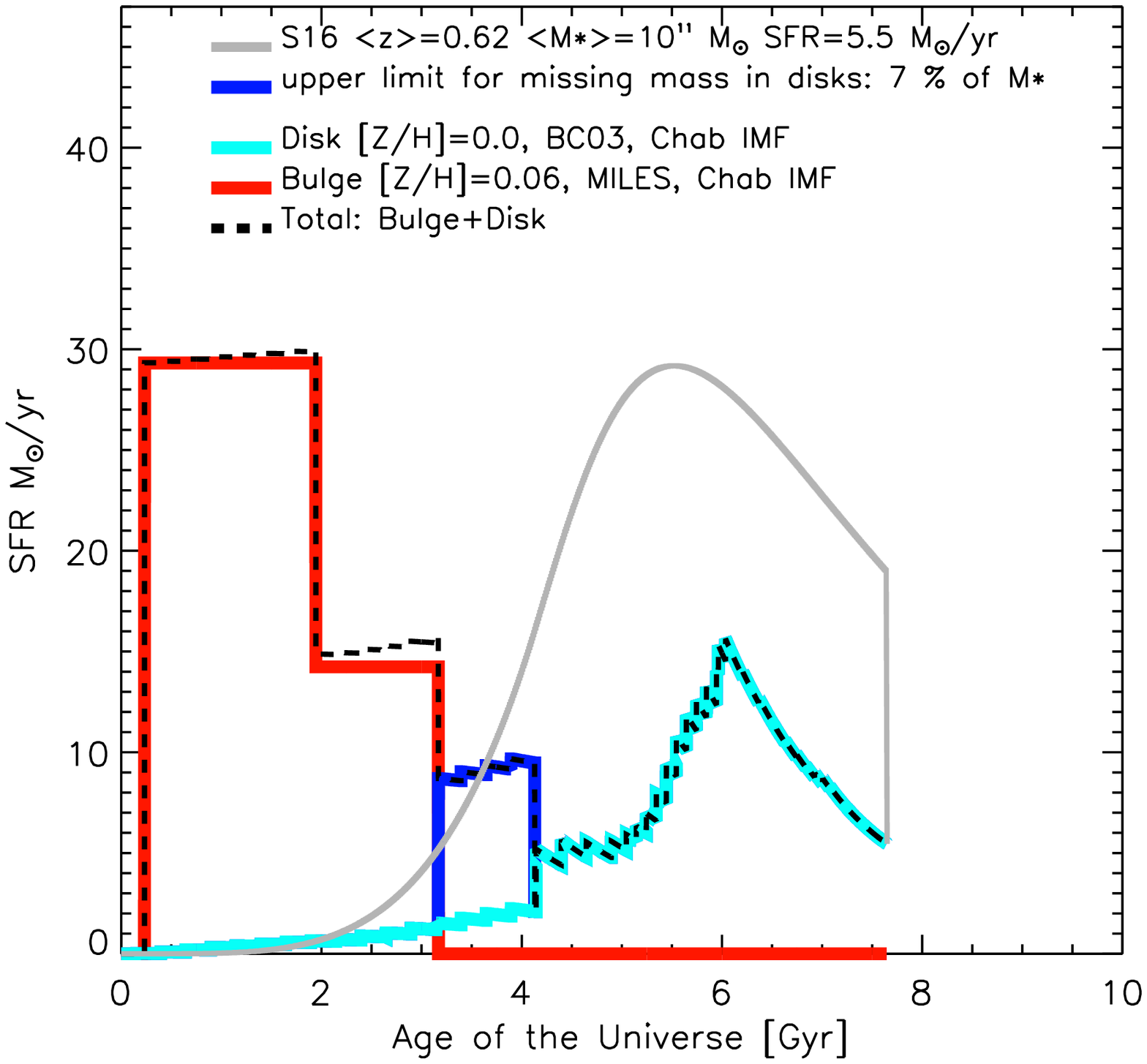}
\caption{Same as Figure~\ref{fig:sfh}, top-left panel, in which the maximum amount of extra stellar mass possibly missed in the disc, estimated as detailed in the text (i.e. 7\% of the total $M_*$, cf. Appendix~\ref{app:med_sed}) has been added in the time gap between the bulge and disc formation (dark blue line).}\label{fig:sfh_extram}
\end{center}
\end{figure}

Nevertheless, in view of the subsequent analysis on the galaxy SFHs, it is also important to derive an upper limit for the stellar mass that could have been missed in the discs, if any, and test how much the extra mass would affect the derived SFH. 
To this purpose, we built an average total SED and measured the residuals between the observed and model photometry in the first three IRAC channels (at 3.6, 4.5, and 5.8~$\mu$m), by excluding IRAC 8.0~$\mu$m fluxes, that could suffer PAH contamination at $z\sim 0.62$. 
The individual total SEDs, obtained by summing the bulge and disc  best-fit models (grey dashed curves in Figure~\ref{fig:seds}), were combined together after they were brought back to the same redshift, $z\sim 0.62$, and normalised to the same flux at $\lambda_{obs}=$ 3.6 $\mu$m. The average SED and its (1$\sigma$) dispersion is shown in the top panel of Figure~\ref{fig:med_sed}. The same normalisation, and shift in redshift were applied to the IRAC photometric data-points, also reported in the figure with different colours for each IRAC channel. This method allows us to have a good sampling of the whole SED in the range $2~\mu$m$ <\lambda_{rest}<5 \mu$m. 
In the bottom panel of Figure~\ref{fig:med_sed} the comparison between the observed and model fluxes is shown. We found that the average ratio is $\langle$F$_{\lambda,\rm obs}$/F$_{\lambda,\rm temp}\rangle=0.97 \pm0.02$ (where the reported uncertainty is the standard deviation of the mean, $\sigma_m$). 
The upper limit to the excess extra mass in the discs was then taken as 3$\times \sigma_m$, which corresponds to $\sim$ 7\% of the total galaxy stellar mass. 
In Figure~\ref{fig:sfh_extram} we show that the excess extra mass possibly missed in the disc , even if assembled at intermediate epochs between the bulk of the bulge and disc  formation, would not strongly affect the average SFH of our sample galaxies.    

\bsp	
\label{lastpage}
\end{document}